\documentclass[acmsmall,nonacm]{acmart}

\renewcommand\footnotetextcopyrightpermission[1]{} % removes footnote with conference information in first column
\usepackage{array}
\newcolumntype{P}[1]{>{\centering\arraybackslash}p{#1}}  

\usepackage{algorithmic}
\usepackage{graphicx}
\usepackage{textcomp}
\usepackage{xcolor}
\usepackage{multirow}
\usepackage{pifont}
\usepackage{colortbl}
\usepackage{wasysym}
\usepackage{fontawesome}
\newcommand{\gc}{\cellcolor{Gray}}
\usepackage{booktabs, siunitx}
\usepackage{color, colortbl}
\definecolor{Gray}{gray}{0.9}
\definecolor{LightCyan}{rgb}{0.88,1,1}
\usepackage{arydshln}
\usepackage{longtable}
\usepackage{caption}
\usepackage{supertabular}
\usepackage{subcaption}
\usepackage{tcolorbox}

\usepackage{listings, xcolor}

\definecolor{verylightgray}{rgb}{.97,.97,.97}

\lstdefinelanguage{Solidity}{
	keywords=[1]{anonymous, assembly, assert, balance, break, call, callcode, case, catch, class, constant, continue, constructor, contract, debugger, default, delegatecall, delete, do, else, emit, event, experimental, export, external, false, finally, for, function, gas, if, implements, import, in, indexed, instanceof, interface, internal, is, length, library, log0, log1, log2, log3, log4, memory, modifier, new, payable, pragma, private, protected, public, pure, push, require, return, returns, revert, selfdestruct, send, solidity, storage, struct, suicide, super, switch, then, this, throw, transfer, true, try, typeof, using, value, view, while, with, addmod, ecrecover, keccak256, mulmod, ripemd160, sha256, sha3}, % generic keywords including crypto operations
	keywordstyle=[1]\color{blue}\bfseries,
	keywords=[2]{address, bool, byte, bytes, bytes1, bytes2, bytes3, bytes4, bytes5, bytes6, bytes7, bytes8, bytes9, bytes10, bytes11, bytes12, bytes13, bytes14, bytes15, bytes16, bytes17, bytes18, bytes19, bytes20, bytes21, bytes22, bytes23, bytes24, bytes25, bytes26, bytes27, bytes28, bytes29, bytes30, bytes31, bytes32, enum, int, int8, int16, int24, int32, int40, int48, int56, int64, int72, int80, int88, int96, int104, int112, int120, int128, int136, int144, int152, int160, int168, int176, int184, int192, int200, int208, int216, int224, int232, int240, int248, int256, mapping, string, uint, uint8, uint16, uint24, uint32, uint40, uint48, uint56, uint64, uint72, uint80, uint88, uint96, uint104, uint112, uint120, uint128, uint136, uint144, uint152, uint160, uint168, uint176, uint184, uint192, uint200, uint208, uint216, uint224, uint232, uint240, uint248, uint256, var, void, ether, finney, szabo, wei, days, hours, minutes, seconds, weeks, years},	% types; money and time units
	keywordstyle=[2]\color{teal}\bfseries,
	keywords=[3]{block, blockhash, coinbase, difficulty, gaslimit, number, timestamp, msg, data, gas, sender, sig, value, now, tx, gasprice, origin},	% environment variables
	keywordstyle=[3]\color{violet}\bfseries,
	identifierstyle=\color{black},
	sensitive=false,
	comment=[l]{//},
	morecomment=[s]{/*}{*/},
	commentstyle=\color{black}\ttfamily,
	stringstyle=\color{red}\ttfamily,
	morestring=[b]',
	morestring=[b]"
}

\definecolor{nickgreen}{HTML}{03A60D}
\definecolor{dgreen}{HTML}{03A60D}
\definecolor{cadmiumorange}{rgb}{0.93, 0.53, 0.18}

\newcommand{\circI}{{\scriptsize\Circle}}
\newcommand{\circII}{{\scriptsize\LEFTcircle}}
\newcommand{\circIII}{{\scriptsize\CIRCLE}}

\newcommand\cbox[1]{\begin{tcolorbox}#1\end{tcolorbox}}

\definecolor{myblue}{HTML}{152e5c}
\newcommand{\rev}[1]{{\color{black} #1}}

\AtBeginDocument{%
  \providecommand\BibTeX{{%
    \normalfont B\kern-0.5em{\scshape i\kern-0.25em b}\kern-0.8em\TeX}}}

\begin{document}

\title{Security Defense For Smart Contracts: A Comprehensive Survey}


\author{Nikolay Ivanov}
\email{ivanovn1@msu.edu}
\affiliation{%
  \institution{Michigan State University}
  \streetaddress{426 Auditorium Road}
  \city{East Lansing}
  \state{Michigan}
  \country{USA}
  \postcode{48824}
}

\author{Chenning Li}
%\authornote{The author is currently affiliated with Massachusetts Institute of Technology, USA.}
\email{lichenni@msu.edu}
\affiliation{%
  \institution{Michigan State University}
  \streetaddress{426 Auditorium Road}
  \city{East Lansing}
  \state{Michigan}
  \country{USA}
  \postcode{48824}
}

\author{Qiben Yan}
    \authornote{Corresponding Author: Qiben Yan (e-mail: qyan@msu.edu)}
\email{qyan@msu.edu}
\affiliation{
  \institution{Michigan State University}
  \streetaddress{426 Auditorium Road}
  \city{East Lansing}
  \state{Michigan}
  \country{USA}
  \postcode{48824}
}

\author{Zhiyuan Sun}
\authornote{The author is also affiliated with Southern University of Science and Technology, China as a joint PhD student.}
\email{zhi-yuan.sun@connect.polyu.hk}
\affiliation{%
  \institution{The Hong Kong Polytechnic University}
  \streetaddress{11 Yuk Choi Road}
  \city{Kowloon}
  \state{Hung Hom}
  \country{Hong Kong}
}

\author{Zhichao Cao}
\email{caozc@msu.edu}
\affiliation{%
\institution{Michigan State University}
  \streetaddress{426 Auditorium Road}
  \city{East Lansing}
  \state{Michigan}
  \country{USA}
  \postcode{48824}
}

\author{Xiapu Luo}
\email{daniel.xiapu.luo@polyu.edu.hk}
\affiliation{%
  \institution{The Hong Kong Polytechnic University}
  \streetaddress{PQ828 Mong Man Wai Building}
  \city{Kowloon}
  \state{Hung Hom}
  \country{Hong Kong}
}

% \author{Qiben Yan}
% \affiliation{%

% \author{Ting Chen}
% \affiliation{%

\renewcommand{\shortauthors}{N. Ivanov, et al.}

\begin{abstract}
The blockchain technology, initially created for cryptocurrency, has been re-purposed for recording state transitions of smart contracts --- decentralized applications that can be invoked through external transactions. Smart contracts gained popularity and accrued hundreds of billions of dollars in market capitalization in recent years. Unfortunately, like all other computer programs, smart contracts are prone to security vulnerabilities that have incurred multibillion-dollar damages over the past decade. As a result, many automated threat mitigation solutions have been proposed to counter the security issues of smart contracts. These threat mitigation solutions include various tools and methods that are challenging to compare. This survey develops a comprehensive classification taxonomy of smart contract threat mitigation solutions within five orthogonal dimensions: defense modality, core method, targeted contracts, input-output data mapping, and threat model. We classify 133 existing threat mitigation solutions using our taxonomy and confirm that the proposed five dimensions allow us to concisely and accurately describe any smart contract threat mitigation solution. In addition to learning \emph{what} the threat mitigation solutions do, we also show \emph{how} these solutions work by synthesizing their actual designs into a set of uniform workflows corresponding to the eight existing defense core methods. We further create an integrated coverage map for the known smart contract vulnerabilities by the existing threat mitigation solutions. Finally, we perform the evidence-based evolutionary analysis, in which we identify trends and future perspectives of threat mitigation in smart contracts and pinpoint major weaknesses of the existing methodologies. For the convenience of smart contract security developers, auditors, users, and researchers, we deploy and maintain a regularly updated comprehensive open-source online registry of threat mitigation solutions, called Security Threat Mitigation (STM) Registry at \url{https://seit.egr.msu.edu/research/stmregistry/}.

\end{abstract}




\maketitle

\section{Introduction}\label{sec:introduction}
%In one decade, the blockchain technology has emerged from a ledger of barely known cryptocurrency to an ever-penetrating industry with hundreds of billions of dollars in market capitalization~\cite{coinmarketcap-total-cap}. 

Blockchain is a decentralized network that sustains distributed records stored in immutable blocks to form an ever-growing chain. In one decade, blockchain technology has evolved from the ledger of cryptocurrency (e.g., Bitcoin, Monero) to the decentralized computing platform (e.g., Ethereum, EOS) that facilitates the deployment and execution of smart contracts. 
\emph{Smart contract} is a decentralized program deployed on a blockchain that enforces the execution of protocols and agreements without involving any third party or establishing a mutual trust~\cite{szabo1996smart}. A smart contract provides a set of functions to be called via transactions and executed by the blockchain's virtual machine (VM). \rev{Most smart contracts are written in high-level special-purpose programming languages, such as Solidity or Vyper. The source code of a smart contract is often compiled into bytecode and deployed on the blockchain.} For example, the Ethereum Virtual Machine (EVM) is the blockchain VM for executing smart contracts on the Ethereum platform\footnote{Although it is primarily associated with Ethereum, EVM has also been adopted by some other blockchain platforms, such as Polygon~\cite{polygon} and RSK~\cite{rsk}.}. An important feature of smart contracts is their ability to perform financial operations with cryptocurrency and valuable custom tokens (e.g., ERC20, ERC721). As of March 2023, the total market capitalization of smart contracts exceeds 300 billion USD~\cite{etherscan-tokens}.

% Integration of smart contracts in the blockchain technology is widely known as Blockchain 2.0, which is the major focus of this survey.
% Smart contracts have been promoted as security safeguards for many industries, including healthcare~\cite{griggs2018healthcare} and Internet-of-Things (IoT)~\cite{huh2017managing,novo2018blockchain}. 
% More importantly, some smart contracts have been used to store and transfer enormous amounts of financial assets. For example, as of February 2022, the market capitalization of the Tether USD Token smart contract is around \$40 billion, with more than 4.4 million of holders and tens of billions of dollars in daily transaction volume~\cite{etherscan-tokens}. 

The large amounts of valued assets stored and transacted by smart contracts made them lucrative targets for attackers. Numerous security vulnerabilities and attacks on Ethereum smart contracts have been hampering their widespread adoption~\cite{expl1, expl2}. In the past few years, exploitations of these vulnerabilities caused billions of dollars in damages. For example, in June 2016, about \$150 million were stolen from the popular DAO contract~\cite{thedao}. In July 2017, about \$30 million were stolen from the Parity multi-signature wallet~\cite{breidenbach2017depth}. Not long after that, a bug in the same multi-signature wallet caused the freeze of nearly \$280 million~\cite{browneaccidental}.

%To prevent and defend against smart contract hacks, 
A large number of approaches and tools have been developed to address different types of smart contract security issues. \rev{
In this work, we use the term \emph{threat mitigation solutions} to describe the full spectrum of smart contract security tools. These efforts include active defense and passive preventative solutions aiming to reduce or eliminate risks of exploitation of vulnerabilities in smart contracts. Academic research efforts as well as commercial and open-source software products are both covered in this survey.}

Existing surveys summarize the vulnerabilities and attacks in smart contracts~\cite{atzei2017survey,li2020survey}. Furthermore, the Smart Contract Weakness Classification and Test Cases database, also known as the SWC Registry~\cite{swcregistry}, identifies and describes 37 classes of known smart contract vulnerabilities (as of March 2023). However, all the existing ways of systematizing smart contract security knowledge focus primarily on vulnerabilities and attacks, paying very little or no attention to the broad swath of defense and prevention mechanisms developed in the past decade. In this work, we bridge the gap in the systematization of the threat mitigation solutions via the following four steps: developing classification taxonomy, synthesizing design workflows of core methods of threat mitigation, creating the map of vulnerability coverage, and conducting an evolutionary analysis. 
%(see Fig.~\ref{fig:methodology}).
%In this work, we bridge this gap in four major steps: developing a classification taxonomy of smart contract threat mitigation solutions; developinghttps://www.overleaf.com/project/61cdfcd88f5dcacf16150d97 synthesized design workflows of the eight core methods used by threat mitigation solutions; creating a map of coverage of known vulnerabilities by the existing solutions; and performing an evolutionary analysis of limitations and trends.
%The major goal of this work is to bridge this gap.
 
% One paragraph for taxonomy
\noindent\textbf{Step I: Taxonomy.}
\rev{Smart contract threat mitigation constitutes a diverse set of efforts. Consequently, finding a uniform organizational methodology for all these solutions poses a major challenge.} These solutions employ a variety of techniques, such as symbolic execution~\cite{mossberg2019manticore,nikolic2018finding}, formal verification~\cite{cecchetti12compositional}, static analysis~\cite{zhou2018erays,brent2018vandal}, etc. Some of these solutions target specific vulnerabilities, such as reentrancy~\cite{rodler2018sereum} or integer overflow~\cite{so2020verismart}, while others are general-purpose~\cite{tsankov2018securify}. Some threat mitigation solutions aim at detecting vulnerabilities~\cite{luu2016making}, while others focus on verifying the safety property of a smart contract~\cite{permenev2020verx}. In other words, all these solutions vary within multiple dimensions. In this survey, we formalize these dimensions and create a comprehensive taxonomy of smart contract threat mitigation  based on five dimensions: defense modality, core method, targeted contracts, data mapping, and threat model.

% One paragraph for workflows
\noindent\textbf{Step II: Design Workflows.} \rev{In addition to learning what the smart contract threat mitigation solutions \emph{do}, we also explore \emph{how} they achieve their aimed goals. Meeting both these objective is challenging due to a wide variety of innovations and novel techniques employed by the existing solutions.} In this work, we study the design workflows of all the 133 smart contract threat mitigation solutions under our investigation, and we subdivide them into eight core methods: \emph{static analysis}, \emph{symbolic execution}, \emph{fuzzing}, \emph{formal analysis}, \emph{machine learning}, \emph{execution tracing}, \emph{code synthesis}, and \emph{transaction interception}. Then, we synthesize the actual designs of the threat mitigation solutions corresponding to each of the eight core methods and build eight \emph{uniform} workflows that summarize the whole variety of threat mitigation solutions for smart contracts.

% One paragraph for coverage
\noindent\textbf{Step III: Vulnerability Coverage.} Next, we raise another important question: which known vulnerabilities have been covered (i.e., prevented, detected, or unmasked) by the existing smart contract threat mitigation solutions? \rev{Answering this question requires overcoming two significant challenges. First, many existing threat mitigation solutions do not explicitly declare the specific vulnerabilities they address. Second, the existing solutions do not adhere to uniform definitions of smart contract vulnerabilities.} To overcome these challenges, we meticulously translate, group, or un-group the vulnerabilities referred to by the authors of the threat mitigation solutions to match the vulnerability classification proposed by the popular SWC Registry. Thus, we develop a unified vulnerability coverage map for these solutions based on the SWC registry.

% One paragraph for Evolution
\noindent\textbf{Step IV: Evolutionary Analysis.} We perform an evidence-based evolutionary analysis of existing smart contract threat mitigation solutions to identify trends and potential future research directions. Specifically, we identify the three most promising vectors of development of smart contract threat mitigation solutions: dynamic transaction interception, AI-driven security, and study of human-machine interaction in smart contracts. \rev{In addition, we identify two deficiencies of existing threat mitigation solutions: \emph{under-representation of non-Ethereum smart contracts}, and \emph{lack of security-related large-scale measurements}.}

% This work is a comprehensive survey of smart contract threat mitigation methods. In addition to classification and assessing vulnerability coverage, we analyze smart contract threat mitigation methods through their architecture (design). Not only do we scrutinize what the threat mitigation methods \emph{do}, but also we are the first to give a systematization of \emph{how} they do what they do. Some of the existing surveys also include brief overviews of defense methods for addressing the vulnerabilities~\cite{singh2020blockchain,chen2020survey}. Unfortunately, none of these surveys focuses on the design of the the threat mitigation (defense) tools.

%\noindent\textbf{Our contribution.} 

In summary, this survey makes the following contributions:

\begin{itemize}
    \item We develop a five-dimensional threat mitigation taxonomy tailored for smart contracts, and we use this taxonomy to classify 133 existing smart contract threat mitigation solutions.
    \item We pinpoint eight core methods adopted by the existing smart contract threat mitigation solutions, and we develop synthesized workflows of these methods to demonstrate the working mechanisms of smart contract threat mitigation.
    \item We identify the threat mitigation solutions that explicitly declare protection against specific vulnerabilities, and we create a smart contract vulnerability coverage map for these solutions.
    \item We identify trends and deficiencies of the existing smart contract mitigation solutions based on the findings of this survey and other solid evidence.
    \item Finally, in the spirit of open research, we develop and publish a constantly updated online registry of threat mitigation solutions, called the STM Registry\footnote{https://seit.egr.msu.edu/research/stmregistry/}.
\end{itemize}

\noindent\textbf{Organization.} The rest of this work is organized as follows. First, we compare our work with previous surveys related to smart contract security (\S\ref{sec:prior-surveys}). 
Then, we describe the methodology employed in this survey (\S\ref{sec:methodology}). After that, we classify 133 threat mitigation solutions based on the developed five-dimensional taxonomy (\S\ref{sec:classification}), followed by a detailed comparative description of designs of the eight core methods of threat mitigation (\S\ref{sec:design-comparison}). Next, we compare the threat mitigation methods by their ability to address specific known smart contract vulnerabilities (\S\ref{sec:coverage}). Then, we discuss trends and future perspectives of threat mitigation in smart contracts (\S\ref{sec:trends}), and finally, we conclude our work %with a brief summary 
(\S\ref{sec:conclusion}).

\section{Prior Surveys}\label{sec:prior-surveys}

Previous surveys on smart contract security focus on different perspectives than this survey. Atzei et al.~\cite{atzei2017survey} propose the first systematic exposition of the Ethereum security vulnerabilities by organizing the vulnerabilities in three levels: Solidity\footnote{Solidity is an object-oriented programming language used mostly for writing Ethereum smart contracts.}, EVM\footnote{The Ethereum Virtual Machine (EVM) is a software platform for executing Ethereum smart contracts. All smart contracts are compiled into bytecode and run on the EVM of all Ethereum nodes.} bytecode, and blockchain. They also illustrate six influential attacks in different application scenarios. In contrast, we primarily target vulnerability mitigation methods rather than the classification of programming pitfalls. Chen et al.~\cite{chen2020defining} propose an empirical survey that provides a systematic study of smart contract defects on the Ethereum platform from five aspects: security, availability, performance, maintainability, and re-usability. They collect and analyze smart contract-related posts on  \textit{Ethereum.StackExchange}\footnote{\url{https://ethereum.stackexchange.com/}} as well as real-world smart contracts to define 20 kinds of contract flaws and 5 relevant impacts. Zou et al.~\cite{zou2019smart} perform an exploratory research to illustrate the current state and potential challenges in smart contract development. \rev{Specifically, they conduct semi-structured interviews with 20 developers and professionals, followed by a survey of 232 practitioners.}
% to confirm the 5 conclusions from the interviews that focus primarily on smart contract development.
In addition, Zhang et al.~\cite{zhang2020framework} present a new classification framework for smart contract bugs and construct a dataset of 176 buggy smart contracts.
Wang et al.~\cite{wang2021ethereum} conduct an analysis of the security of Ethereum smart contracts and categorize these security challenges into abnormal contracts, program vulnerabilities, and unsafe external data.
Vacca et al.~\cite{VACCA2021110891} provide a systematic review of techniques and tools used to address the software engineering-specific challenges of blockchain-based applications by analyzing 96 research papers.
The above surveys summarize smart contract security and development issues, while our survey specifically focuses on vulnerability mitigation solutions.

There are also a number of surveys that investigate the vulnerability mitigation solutions.
Chen et al.~\cite{chen2020survey} present a comprehensive and systematic survey on Ethereum systems security which includes vulnerabilities, attacks, and defenses.
The authors discuss 44 kinds of vulnerabilities based on the layers of the Ethereum architecture and describe the history, cause, tactic, and direct impact of 26 attacks.
As for defenses, the authors enumerate 47 defense mechanisms and provide the best practices to guide contract development.
Although they divide the defenses into proactive and reactive, they are lacking an explanation of how the different tools are designed. Another survey by Wang and He et al.~\cite{wang2021securityenhancement} reviews 6 kinds of vulnerability detection methods and privacy protection techniques in 3 platforms (i.e., Ethereum, Hyperledger fabric and Corda), and summarizes several commonly used tools for each method.
Di Angelo et al.~\cite{di2019survey} investigate 27 analysis tools of Ethereum smart contracts regarding availability, maturity level, methods employed, and detection of security issues. They examine the availability and functionality of the tools and compare their characteristics in a structured manner. In comparison, we carry out a multi-dimensional classification of 133 solutions and take into account different aspects of threat mitigation. Besides, we also analyze different defense mechanisms through their architecture.
Furthermore, Samreen et al.~\cite{samreen2021survey} review some detection tools and discuss eight vulnerabilities by analyzing past exploitation cases.
Ni et al.~\cite{ni2020survey} propose a three-layered threat model for smart contract security and introduce 15 major vulnerabilities of Ethereum at three levels: programming language, virtual machine, and blockchain. They also summarize and compare the three most commonly used vulnerability mitigation techniques, viz., fuzzing, symbolic execution, and formal verification. Li et al.~\cite{li2020survey} survey the security threats of blockchain and enumerate 6 real attack cases. They also review the security enhancement solutions for blockchain by introducing 5 commonly used defense tools. In contrast, we categorize defenses in 5 orthogonal dimensions and compare 133 commonly used solutions. Praitheeshan et al.~\cite{praitheeshan2019security} review the security of Ethereum smart contracts through 16 types of security vulnerabilities, 19 software security issues, and 3 defense methods. For each defense method, they list several common tools but do not compare the different methods and tools.
\rev{
Wan et al.~\cite{wan2021smart} conduct qualitative and quantitative studies involving 13 interviews and
156 survey responses from smart contract security professionals regarding smart contract security in order to understand the role of security in smart contract development lifecycle.
Hu et al.~\cite{hu2021comprehensive} explore three major categories of existing schemes for smart contract construction and execution. They further identify the main challenges preventing wide adoption of smart contracts, and outline further directions. Harz et al.~\cite{harz2018towards} deliver a security survey of blockchain and smart contracts by summarizing paradigms, types, instruction sets, semantics, and metering techniques. They also examine verification tools and methods for smart contracts and blockchains.}
%, and present future research directions.}
In contrast, we summarize 5 more vulnerability core methods and compare them through 5 dimensions. Moreover, we also construct a compact vulnerability map that contains 37 known vulnerabilities to summarize the vulnerability-addressing ability of 38 classes of threat mitigation solutions.

There are several studies that delve into a specific defense method (e.g., formal verification).
Tolmach et al.~\cite{tolmach2021survey} scrutinize formal models and specifications of smart contracts. They categorize the specifications of smart contracts in various application domains and propose a four-layered framework to classify smart contract analysis methods. After that, they summarize the tools for formal verification and group them based on the utilized techniques. In addition, the authors also discuss the difficulties in smart contract verification and development. Similarly, Singh et al.~\cite{singh2020blockchain} conduct a systematic survey about current formalization research on all smart contract-enabled blockchain platforms by summarizing 35 studies between 2015 and 2019. However, these studies focus purely on formal verification without examining other types of threat mitigation. On the contrary, we provide eight commonly used vulnerability mitigation core methods and identify future research trends and directions in smart contract threat mitigation.

\rev{
Smart contract threat mitigation is also partially covered by some books. Antonopoulos et al.~\cite{antonopoulos2018mastering} provide a comprehensive introductory guide to the Ethereum platform, which includes one chapter devoted to smart contract security as of year 2018. Blockchain foundations and applications are described in a book by Yi et al.~\cite{yi2022blockchain}, which includes a chapter on security analysis of smart contract-based self-tallying voting systems.
}

\rev{The above surveys and books either cover a limited set of methods or describe smart contract threat mitigation without sufficient technical depth. To bridge these gaps, our survey explores the topic of smart contract threat mitigation comprehensively and in-depth. Overall, we undertake \textbf{four} major steps to shed light on the ever-evolving threat mitigation landscape of smart contracts. First, we deliver a comprehensive 5-dimensional classification taxonomy. Second, we synthesize design workflows corresponding to the eight core methods. Third, we create a vulnerability coverage map. Fourth, we conduct an evolutionary analysis with trends and perspectives.} The combination of these four steps applied to 133 solutions makes our work the most comprehensive systematization of smart contract threat mitigation to date.

\section{Methodology}\label{sec:methodology}

\begin{figure}
    \centering
    \includegraphics[width=\textwidth]{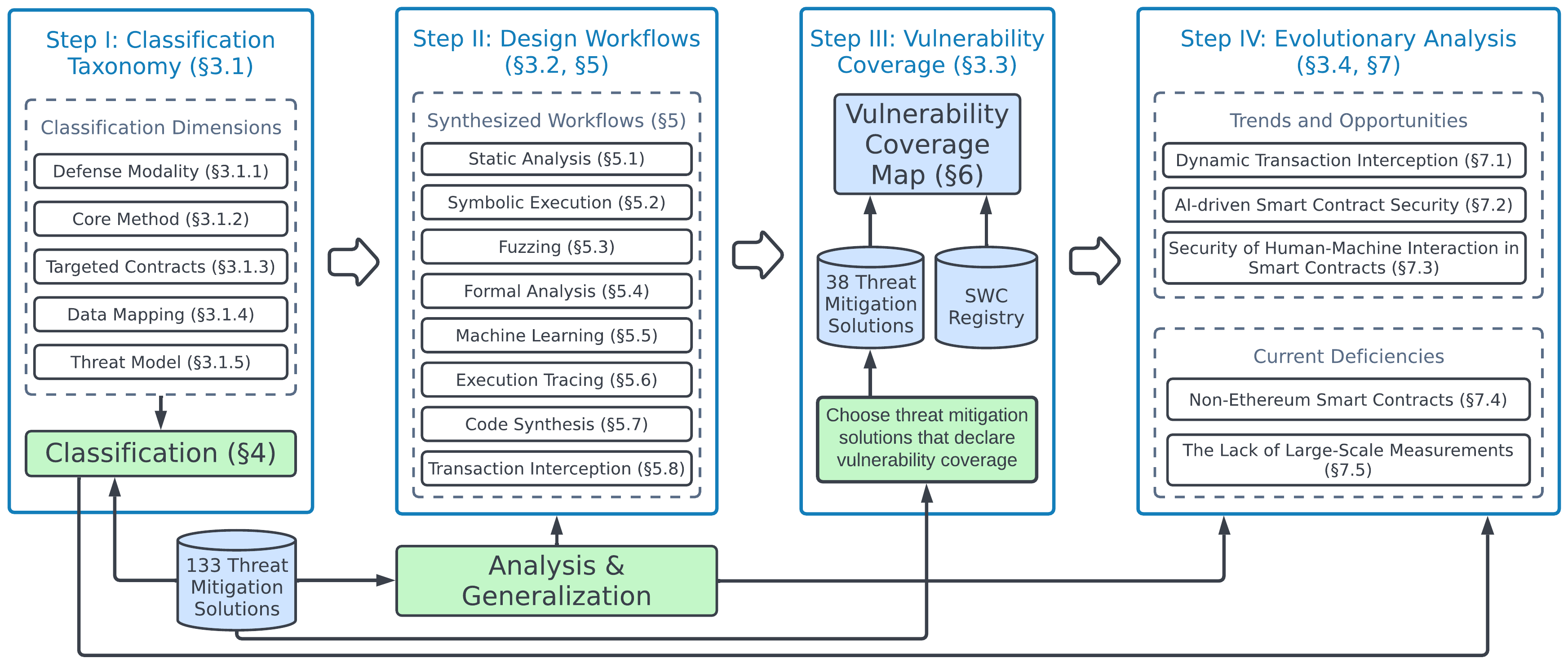}
    \caption{Four-step methodology of this survey.}
    \label{fig:methodology}
\end{figure}

In this section, we describe the details of the 4-step methodology that we use in this survey.
Fig.~\ref{fig:methodology} depicts these steps, which include: \textbf{Step I}: developing the classification taxonomy of smart contract threat mitigation solutions (\S\ref{subsec:taxonomy}); \textbf{Step II}: synthesizing the workflows of the core methods of threat mitigation solutions (\S\ref{subsec:workflow}); \textbf{Step III}: developing the vulnerability coverage map by threat mitigation solutions (\S\ref{subsec:coverage}); and \textbf{Step IV}: investigating the evolutionary trends and deficiencies of threat mitigation in smart contracts (\S\ref{subsec:evolution}). %Let us take a closer look at 
Next, we describe the approaches employed by these four steps in detail.

\subsection{Classification Taxonomy}\label{subsec:taxonomy}

% \begin{figure}
%     \centering
%     \includegraphics[width=0.8\textwidth]{img/taxonomy.png}
%     \caption{Smart contract threat mitigation taxonomy.}
%     \label{fig:taxonomy}
% \end{figure}

\begin{table}
\rev{
\setlength{\tabcolsep}{4pt}
    \centering
    \caption{\rev{Smart contract threat mitigation taxonomy}}
    \label{tab:taxonomy}
    \begin{tabular}{c||c|c}

        \arrayrulecolor{black}\toprule
        
        % ============ UPPER HEADERS =================
        \textbf{Classification Dimension} & \textbf{Possible Values} & \textbf{Short Notation}  \\
        
        \arrayrulecolor{black}\midrule
        
        % ============ DATA =================
        
        \multirow{3}{*}{\textit{Defense Modality}} & prevention & \texttt{PREV} \\
        & detection & \texttt{DET} \\
        & exploration & \texttt{EXPL} \\
        \hline
    
        \multirow{8}{*}{\textit{Core Method}} & static analysis & \texttt{StAn} \\
        & symbolic execution & \texttt{SymEx} \\
        & fuzzing & \texttt{Fuz} \\
        & formal analysis & \texttt{FormAn} \\
        & machine learning & \texttt{ML} \\
        & execution tracing & \texttt{ExTrace} \\
        & code synthesis & \texttt{CodeSyn} \\
        & transaction interception & \texttt{TxInt} \\
        \hline
        
        \multirow{4}{*}{\textit{Targeted Contracts}} & Ethereum & \texttt{ETH} \\
        & EVM-compatible & \texttt{EVM} \\
        & any contract & \texttt{any} \\
        & non-Ethereum & \texttt{non-ETH} \\
        \hline
        
        \multirow{1}{*}{\textit{Data Mapping}} & 
        \begin{tabular}{c:c}
            % \multicolumn{2}{c}{$\;\;\;\;$ \textbf{Input} $\;\;\mapsto\;$ \textbf{Output}} \\
            \textbf{Input} & \textbf{Output} \\
            \hline
            source code  & report \\
            bytecode  & source code \\
            ABI  & bytecode \\
            specifications  & action \\
            chain data  & exploits \\
            assembly code  & metadata
        \end{tabular} &
        \begin{tabular}{c:c}
            \textbf{Input} & \textbf{Output} \\
            \hline
            \texttt{Src} & \texttt{Rep} \\
            \texttt{BCode}  & \texttt{Src} \\
            \texttt{ABI}  & \texttt{BCode} \\
            \texttt{Spec}  & \texttt{Act} \\
            \texttt{CData} & \texttt{Exp} \\
            \texttt{Assemb} & \texttt{Meta}
        \end{tabular} \\
        
        \hline
        
        \multirow{3}{*}{\textit{Threat Model}} & vulnerable contract only & \texttt{VC} \\
        & malicious contract only & \texttt{MC} \\
        & malicious \emph{or} vulnerable contract & \texttt{MVC} \\

        % ============ LEGEND: ICONS =================
        \arrayrulecolor{black}\bottomrule
        %\multicolumn{5}{l}{$^\dagger$ ...}

    \end{tabular}
    }
\end{table}

To classify the smart contract threat mitigation solutions, we build a comprehensive taxonomy of threat mitigation, which includes the following five orthogonal dimensions (see Table~\ref{tab:taxonomy}): 1) defense modality, 2) core method, 3) targeted contracts, 4) data mapping, and 5) threat model. We empirically verify that our taxonomy is not only concise but also allows to describe a threat mitigation solution with high accuracy. For example, using our taxonomy, the popular threat mitigation tool Oyente~\cite{luu2016making} can be accurately described via the following single sentence:

\begin{quote}
\textit{``Oyente is a security tool based on symbolic execution that detects and reports vulnerabilities in the bytecode of malicious or buggy Ethereum smart contracts."}     
\end{quote}

Moreover, our taxonomy is cross-platform and general enough to be applied to the future developments of threat mitigation for smart contracts, even when new methods or platforms emerge. Next, we describe all these five dimensions of the threat mitigation taxonomy in detail (\S\ref{subsec:defense-modality}--\S\ref{subsec:threat-model}).

\subsubsection{Defense Modality}\label{subsec:defense-modality}
\rev{
The \emph{defense modality} is the essential philosophy used by a threat mitigation solution to achieve its goals. There are three types of the defense modality: \emph{prevention}, \emph{detection}, and \emph{exploration}.
}
The \emph{prevention} methods aim at verifying or enforcing certain security properties of a smart contract. For example, the requirement that if a smart contract accepts cryptocurrency deposits, it must also provide the functionality for cryptocurrency withdrawal, can be used by a solution with the \emph{prevention} modality as a property to enforce or verify security. The detection methods look for known vulnerabilities in smart contracts. For instance, defense tools that search for reentrancy vulnerabilities in smart contracts pertain to the \emph{detection} defense modality. The \emph{exploration} approaches enhance the transparency of a smart contract or associated transactions in order to facilitate security audits. For example, an auditing tool that allows demystifying the call stack of a complicated smart contract, thereby exposing the potential security problems, would belong to the \emph{exploration} defense modality.

\subsubsection{Core Method}\label{subsec:core-method}
The \emph{core method} is the technical approach describing the implementation principles of a given threat mitigation solution. \rev{Unlike defense modality, which describes the general philosophy of a solution, the core method describes the implementation methodology utilized by the solution. In other words, the same defense philosophy can be implemented in a number of different core methods.} Threat mitigation solutions belonging to the same core method, despite the diversity of implementations, share the same major workflow with possible minor additions. \rev{For example, all symbolic execution methods take a smart contract and a set of specifications as an input, utilize an SMT solver, and produce a human-readable report as an output. However, many symbolic execution implementations include some additional modules and data units that are not uniform across all solutions.} In this work, we build workflows that demonstrate which items are essential and which of them provide an incremental augmentation.

\subsubsection{Targeted Contracts}\label{subsec:targeted-contracts}
\begin{figure}
    \centering
    \includegraphics[width=0.55\textwidth]{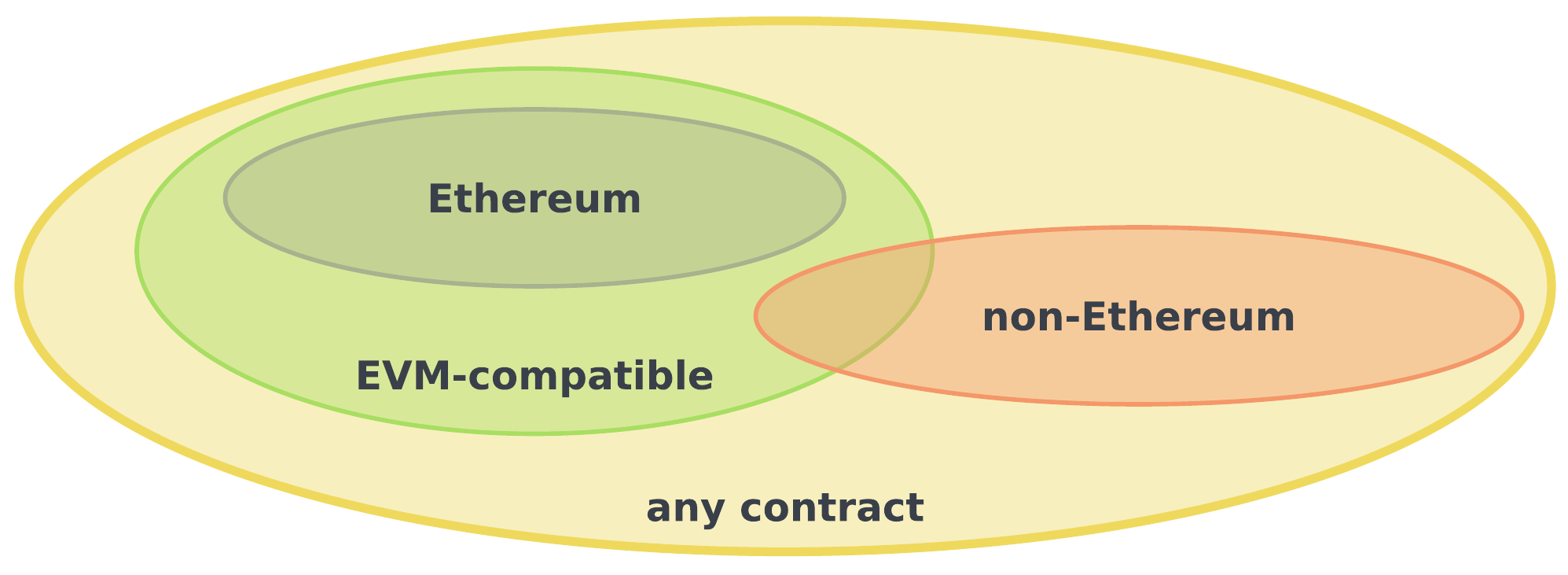}
    \caption{Venn diagram of relationships between different scopes of smart contracts.}
    \label{fig:venn}
\end{figure}

The dimension of \emph{targeted contracts} describes the class of smart contracts that a threat mitigation solution applies to. This dimension is largely shaped by the practical circumstance, in which the vast majority of smart contract threat mitigation solutions target the popular Ethereum platform. Moreover, we notice that within the Ethereum platform, there is very little variety in terms of what kind of Ethereum smart contracts the threat mitigation solutions target. In other words, most solutions target Ethereum, and these Ethereum-based solutions are suitable for any Ethereum contract. Thus, to accurately represent the practical reality of the distribution of smart contract threat mitigation solutions in the dimension of targeted contract, we subdivide this dimension into four classes: \emph{Ethereum smart contracts}, \emph{EVM-compatible smart contracts}, \emph{non-Ethereum smart contracts}, and \emph{any smart contract} (i.e., platform-agnostic). Fig.~\ref{fig:venn} shows the Venn diagram of the relationships between these classes. Specifically, all Ethereum contracts are EVM-compatible, but there are non-Ethereum platforms that may or may not be EVM-compatible. At the same time, the ``any contract'' scope would embrace all the types of smart contracts mentioned above, without prioritizing any of them.

\subsubsection{Data Mapping}\label{subsec:data-mapping}
The \emph{data mapping} dimension describes what the input and output of a given threat mitigation solution are. As shown in Table~\ref{tab:taxonomy}, the input of a threat mitigation solution may be a combination of 1) source code; 2) bytecode; 3) application binary interface (ABI); 4) security specifications; 5) chain data; or 6) assembly code. The output can be represented by any combination of the following six entities: 1) security report; 2) source code; 3) bytecode; 4) defense action; 5) set of exploits; or 6) metadata. In this work, we use the symbol $\mapsto$ as a convention for data mapping. For example, if the input of a threat mitigation solution is a set of specifications with the source code of the smart contract, and the output is a human-readable report, then we denote such a mapping as \mbox{\texttt{Sp,S}$\mapsto$\texttt{R}}. As we can see, the data mapping dimension allows to concisely and informatively describe the requirements for the input and expectations for the output for a smart contract threat mitigation solution.
% Describe each of the input/output types. Figure out how.

\subsubsection{Threat Model}\label{subsec:threat-model}
The dimension of \emph{threat model} describes the vector(s) of potential attacks that the threat mitigation solution aims to prevent, detect, or explore. We empirically observe that all the smart contract threat mitigation solutions belong to either of the three general threat models: 1) the one with the malicious smart contract; 2) the one in which the smart contract is the victim; and 3) the agnostic model, in which the contract may be either malicious or a victim. For example, the threat mitigation solutions capable of preventing exploitations of the reentrancy vulnerability, responsible for the infamous DAO hack~\cite{thedao}, belong to the \texttt{VC} (victim contract) model. Conversely, a tool defending against honeypot smart contracts, which set unexpected traps for hackers attempting to exploit known smart contract vulnerabilities, is a typical example of a threat mitigation tool assuming the malicious contract (\texttt{MC}) threat model. \rev{However, some solutions defend against vulnerabilities that can be used both in a malicious or a victim smart contract. In the latter case, we use the malicious-or-vulnerable contract (\texttt{MVC}) model.} For example, the SWC-123 vulnerability~\cite{swc123}, called \emph{Requirement Violation}, can be both a bug in a vulnerable smart contract or an intentional malicious action of the smart contract developer.

%describes the explicit or implicit ability of a threat mitigation solution to detect, prevent, or mitigate the exploitation of 

\subsection{Workflows of Core Methods}\label{subsec:workflow}
In this survey, not only do we explore \emph{what} the smart contract threat mitigation solutions do, but we also explore, for the first time, \emph{how} these solutions accomplish their goals. \rev{In order to do that, we adopt the following approach. For each of the eight core methods introduced in \S\ref{subsec:core-method}, we synthesize the workflows of all the existing solutions implementing these methods. In these synthesized workflow charts, we showcase the mandatory (common for all solutions) and augmented (observed in some solutions) elements.} Sections~\ref{sec:workflow-static-analysis}---\ref{sec:workflow-tx-interception} describe the synthesized workflows of all the eight core methods of smart contract threat mitigation. In order to embrace the diverse variety of implementations, we use a uniform set of conventions in the eight workflows. Specifically, we use three types of elements connected with flows (arrows): modules (data processors), data entities, and environments (groups).
%Continue along the same lines.

%Refer the reader to the future chapters.

%: 1) static analysis, 2) symbolic execution, 3) fuzzing, 4) formal analysis, 5) machine learning, 6) execution tracing, 7) code synthesis, and 8) transaction interception.

%For each of the eight core methods,we synthesize the workflows of all the existing solutions implementing this method to demonstrate the mandatory and optional components.
%We identify that each of the existing smart contract threat mitigation methods employs one or a combination of the following eight core methods:

% \begin{enumerate}
%     \item static analysis;
%     \item symbolic execution;
%     \item fuzzing;
%     \item formal analysis;
%     \item machine learning;
%     \item execution tracing; 
%     \item code synthesis; and
%     \item transaction interception.
% \end{enumerate}

\subsection{Vulnerability Coverage}\label{subsec:coverage}
% The third step of our survey is scrutinizing the vulnerability coverage, i.e., to determine \emph{which known vulnerabilities are detectable and/or preventable by the existing threat mitigation solutions.} 
\rev{The third step of our survey is scrutinizing the vulnerability coverage of the existing threat mitigation solutions. The vulnerability coverage of a solution determines which known vulnerabilities are detectable and/or preventable by this solution.} To accomplish that, we create a uniform \emph{vulnerability coverage} map using the popular SWC Registry. This task poses two major challenges: i) many threat mitigation solutions do not explicitly or even implicitly declare the set of addressed vulnerabilities; ii) the majority of threat mitigation solutions refer to the existing vulnerabilities using custom names and/or groupings, which often do not correspond to the SWC taxonomy. Here, we select the 38 threat mitigation solutions that explicitly specify the list of targeted vulnerabilities, and then we meticulously translate the declared vulnerability coverage provided by the selected 38 solutions into the SWC conventions.

\subsection{Threat Mitigation Evolution}\label{subsec:evolution}
Our final step explores the evolution of the smart contract threat mitigation solutions, as well as the trends and obstacles observed in this area of computer security. Specifically, we explore the adoption and augmentation of new core methods over time. For each threat mitigation solution, we keep track of the publication date as well as the initial release or announcement date, whenever available. \rev{Additionally, we analyze the ``blind spots'' of the existing body of smart contract mitigation solutions --- the potentially feasible yet unexplored combinations of approaches that can bring more benefits.} As a result, we make five observations supported by data and evidence. First, we identify that dynamic transaction interception methods of smart contract threat mitigation are gaining momentum in the research community. Second, we show that the smart contract threat mitigation solutions utilizing AI and machine learning have started playing an important role in smart contract defense. Third, we identified the emerging trend for studying human-machine interaction in the domain of smart contracts. Fourth, we confirm that Ethereum smart contracts are over-represented by the threat mitigation solutions, and we discuss likely reasons explaining this phenomenon. Finally, we discuss the necessity for more exploration tools and large-scale measurements
for gathering important data about smart contract security, such as the real market value of smart contracts and the traces of choices made by miners and crypto exchanges.

%for answering some important questions about smart contract security.
 % [no papers]
\section{Threat Mitigation Classification}\label{sec:classification}

% TABLE I
\begin{table}
\rev{
\linespread{0.95}
\setlength{\tabcolsep}{4pt}
    \centering
    \small
    \caption{\rev{Classification of threat mitigation solutions based on the proposed taxonomy.}}
    \label{tab:classification}
    \begin{tabular}{P{0.6cm}:P{2.7cm}||P{1.2cm}:P{2.1cm}:P{1.0cm}:P{2.55cm}:P{0.8cm}}

        \arrayrulecolor{black}\toprule
        
        % ============ UPPER HEADERS =================
        \multicolumn{1}{c}{\multirow{2}{*}{\textbf{}}} &
        \multicolumn{1}{c}{\multirow{2}{*}{\textbf{Threat}}} &
        \multicolumn{5}{c}{\textit{Classification Criteria (Dimensions)}$\mathrm{^\dagger}$} \\
        % \cmidrule(lr){4-5} 
        \cmidrule(lr){3-7}

        % ============ LOWER HEADERS =================
        \textbf{STM} & \textbf{mitigation} &
        {\small Defense} & {\small Core} & {\small Target} & {\small Data} & {\small Threat} \\
        \textbf{code
        %$\mathrm{^\dagger}$
        } & \textbf{solution} &
        {\small Modality} & {\small Method} & {\small Contract} & {\small Mapping} & {\small Model} \\
        \arrayrulecolor{black}\midrule
        
        % ============ DATA =================

        001 & Oyente~\cite{luu2016making} & \texttt{DET} & \texttt{SymEx} & \texttt{ETH} & \texttt{BCode} $\mapsto$ \texttt{Rep} & \texttt{MVC} \\
        \hline
        002 & Mythril~\cite{mueller2018smashing} & \texttt{DET} & \texttt{SymEx} & \texttt{ETH} & \texttt{BCode} $\mapsto$ \texttt{Rep} & \texttt{MVC} \\
        \hline
        003 & Securify~\cite{tsankov2018securify} & \texttt{DET+PREV} & \texttt{StAn} & \texttt{ETH} & \texttt{BCode,Src} $\mapsto$ \texttt{R,M} & \texttt{MVC} \\
        \hline
        004 & Maian~\cite{nikolic2018finding} & \texttt{DET} & \texttt{SymEx} & \texttt{ETH} & \texttt{BCode} $\mapsto$ \texttt{Rep} & \texttt{MVC} \\
        \hline
        005 & Manticore~\cite{mossberg2019manticore} & \texttt{DET} & \texttt{SymEx} & \texttt{ETH} & \texttt{BCode} $\mapsto$ \texttt{Rep} & \texttt{MVC} \\
        \hline
        006 & KEVM~\cite{hildenbrandt2018kevm} & \texttt{EXPL} & \texttt{FormAn} & \texttt{ETH} & \texttt{BCode} $\mapsto$ \texttt{Rep} & \texttt{MVC} \\
        \hline
        007 & ZEUS~\cite{kalra2018zeus} & \texttt{PREV} & \texttt{SymEx} & \texttt{ETH} & \texttt{BCode} $\mapsto$ \texttt{Rep} & \texttt{MVC} \\
        \hline
        008 & Sereum~\cite{rodler2018sereum} & \texttt{DET} & \texttt{ExTrace,TxInt} & \texttt{ETH} & \texttt{CData} $\mapsto$ \texttt{Rep} & \texttt{VC} \\
        \hline
        009 & ECFChecker~\cite{grossman2017online} & \texttt{PREV} & \texttt{ExTrace,TxInt} & \texttt{ETH} & \texttt{CData} $\mapsto$ \texttt{Rep} & \texttt{VC} \\
        \hline
        010 & teEther~\cite{krupp2018teether} & \texttt{DET} & \texttt{SymEx} & \texttt{ETH} & \texttt{BCode} $\mapsto$ \texttt{Exp} & \texttt{VC} \\
        \hline
        011 & Hydra~\cite{breidenbach2018enter} & \texttt{PREV} & \texttt{CodeSyn} & \texttt{ETH} & \texttt{Src} $\mapsto$ \texttt{BCode} & \texttt{VC} \\
        \hline
        
        012 & Erays~\cite{zhou2018erays} & \texttt{EXPL} & \texttt{StAn} & \texttt{ETH} & \texttt{BCode} $\mapsto$ \texttt{Meta} & \texttt{MVC} \\
        \hline
        013 & TokenScope~\cite{chen2019tokenscope} & \texttt{DET} & \texttt{ExTrace} & \texttt{ETH} & \texttt{CData} $\mapsto$ \texttt{Rep} & \texttt{MVC} \\ 
        \hline
        014 & Osiris~\cite{torres2018osiris} & \texttt{DET} & \texttt{SymEx} & \texttt{ETH} & \texttt{BCode} $\mapsto$ \texttt{Rep} & \texttt{VC} \\
        \hline
        015 & Vandal~\cite{brent2018vandal} & \texttt{DET} & \texttt{StAn} & \texttt{ETH} & \texttt{BCode} $\mapsto$ \texttt{Rep} & \texttt{MVC} \\
        \hline
        016 & FSolidM~\cite{mavridou2018designing} & \texttt{PREV} & \texttt{CodeSyn} & \texttt{ETH} & \texttt{S,Sp} $\mapsto$ \texttt{Src} & \texttt{VC} \\
        \hline
        017 & ContractFuzzer~\cite{jiang2018contractfuzzer} & \texttt{DET} & \texttt{Fuz} & \texttt{ETH} & \texttt{ABI,B} $\mapsto$ \texttt{Rep} & \texttt{VC} \\
        \hline
        018 & S-GRAM~\cite{liu2018s} & \texttt{DET} & \texttt{StAn} & \texttt{ETH} & \texttt{Src} $\mapsto$ \texttt{Rep} & \texttt{MVC} \\
        \hline
        019 & MadMax~\cite{grech2018madmax} & \texttt{DET} & \texttt{StAn} & \texttt{ETH} & \texttt{BCode} $\mapsto$ \texttt{Rep} & \texttt{MVC} \\
        \hline
        020 & SmartCheck~\cite{tikhomirov2018smartcheck} & \texttt{DET} & \texttt{StAn} & \texttt{ETH} & \texttt{Src} $\mapsto$ \texttt{Rep} & \texttt{MVC} \\
        \hline
        
        021 & ReGuard~\cite{liu2018reguard} & \texttt{DET} & \texttt{Fuz} & \texttt{ETH} & \texttt{Src} $\mapsto$ \texttt{R,E} & \texttt{VC} \\
        \hline
        022 & GASPER~\cite{chen2017under} & \texttt{DET} & \texttt{StAn} & \texttt{ETH} & \texttt{BCode} $\mapsto$ \texttt{Rep} & \texttt{MC} \\
        \hline
        023 & Grishchenko~et~al.~\cite{grishchenko2018semantic} & \texttt{EXPL} & \texttt{FormAn} & \texttt{ETH} & \texttt{BCode} $\mapsto$ \texttt{Meta} & \texttt{MVC} \\
        \hline
        024 & Lolisa~\cite{yang2018lolisa} & \texttt{PREV} & \texttt{FormAn} & \texttt{ETH} & \texttt{Src} $\mapsto$ \texttt{Rep} & \texttt{MVC} \\
        \hline
        025 & SASC~\cite{zhou2018security} & \texttt{EXPL} & \texttt{StAn} & \texttt{ETH} & \texttt{Src} $\mapsto$ \texttt{Rep} & \texttt{MVC} \\
        \hline
           026 & Chen et al.~\cite{chen2018detecting} & \texttt{DET} & \texttt{ExTrace,StAn} & \texttt{ETH} & \texttt{C,S} $\mapsto$ \texttt{Rep} & \texttt{MC} \\
        \hline
        027 & Solidity*~\cite{bhargavan2016formal} & \texttt{PREV} & \texttt{StAn,FormAn} & \texttt{ETH} & \texttt{Src} $\mapsto$ \texttt{Rep} & \texttt{MVC} \\
        \hline
        028 & Amani et al.~\cite{amani2018towards} & \texttt{PREV} & \texttt{StAn,FormAn} & \texttt{ETH} & \texttt{BCode,Spec} $\mapsto$ \texttt{Rep} & \texttt{MVC} \\
        \hline
        029 & Model-Check.~\cite{nehai2018model} & \texttt{PREV} & \texttt{FormAn} & \texttt{ETH} & \texttt{Sp,S} $\mapsto$ \texttt{Rep} & \texttt{MVC} \\
        \hline
        030 & EtherTrust~\cite{grishchenko2018ethertrust} & \texttt{DET} & \texttt{StAn} & \texttt{ETH} & \texttt{BCode} $\mapsto$ \texttt{Rep} & \texttt{MVC} \\
        \hline
        
        031 & Flint~\cite{schrans2018writing} & \texttt{PREV} & \texttt{CodeSyn} & \texttt{ETH} & \texttt{Spec} $\mapsto$ \texttt{Src} & \texttt{VC} \\
        \hline
        032 & HoneyBadger~\cite{torres2019art} & \texttt{DET} & \texttt{StAn,SymEx} & \texttt{ETH} & \texttt{BCode} $\mapsto$ \texttt{Rep} & \texttt{MC} \\
        \hline
        033 & ILF~\cite{he2019learning} & \texttt{DET} & \texttt{Fuz,ML} & \texttt{ETH} & \texttt{BCode,Src} $\mapsto$ \texttt{Rep} & \texttt{MVC} \\
        \hline
        034 & VeriSolid~\cite{mavridou2019verisolid} & \texttt{PREV} & \texttt{FormAn,CodeSyn} & \texttt{ETH} & \texttt{S,Sp} $\mapsto$ \texttt{Src} & \texttt{VC} \\
        \hline
        035 & solc-verify~\cite{hajdu2019solc} & \texttt{PREV} & \texttt{StAn} & \texttt{ETH} & \texttt{S,Sp} $\mapsto$ \texttt{Rep} & \texttt{MVC} \\
        \hline
        036 & Slither~\cite{feist2019slither} & \texttt{DET} & \texttt{StAn} & \texttt{ETH} & \texttt{Src} $\mapsto$ \texttt{Rep} & \texttt{MVC} \\
        \hline
        037 & sCompile~\cite{chang2019scompile} & \texttt{DET} & \texttt{SymEx} & \texttt{ETH} & \texttt{BCode} $\mapsto$ \texttt{Rep} & \texttt{MVC} \\
        \hline
        038 & NPChecker~\cite{wang2019detecting} & \texttt{DET} & \texttt{StAn} & \texttt{ETH} & \texttt{BCode} $\mapsto$ \texttt{Rep} & \texttt{MVC} \\
        \hline
        039 & BitML~\cite{atzei2019developing} & \texttt{PREV} & \texttt{StAn} & \texttt{non-ETH} & \texttt{C,Sp} $\mapsto$ \texttt{R,E} & \texttt{VC} \\
        \hline
        040 & CESC~\cite{li2019finding} & \texttt{DET} & \texttt{StAn} & \texttt{ETH} & \texttt{BCode} $\mapsto$ \texttt{Rep} & \texttt{VC} \\   
        
        % ============ LEGEND: ICONS =================
        \arrayrulecolor{black}\bottomrule
        % \multicolumn{7}{l}{\scriptsize
        % $\mathrm{^\dagger}$ see Table~\ref{tab:taxonomy} for notation
        % } \\
        
        \multicolumn{7}{c}{\scriptsize
        $\mathrm{^\dagger}$ \texttt{DET}--- detection; \texttt{PREV} --- prevention; \texttt{EXPL} --- exploration; \texttt{StAn} --- static analysis; \texttt{SymEx} --- symbolic execution; \texttt{Fuz} --- fuzzing;
        } \\
        
        \multicolumn{7}{c}{\scriptsize 
        \texttt{FormAn} --- formal analysis; \texttt{ML} --- machine learning; \texttt{ExTrace} --- execution tracing; \texttt{CodeSyn} --- code synthesis; \texttt{Act} --- action;
        } \\
        
        \multicolumn{7}{c}{\scriptsize 
         \texttt{TxInt} --- transaction interception;
        \texttt{Src} --- source code; \texttt{BCode} --- bytecode; \texttt{Spec} --- specification; 
        \texttt{CData} --- chain data;
        } \\
        
        \multicolumn{7}{c}{\scriptsize
        \texttt{Assemb} --- assembly code;
        \texttt{Rep} --- report;
        \texttt{Exp} --- exploits; \texttt{Meta} --- metadata;
        \texttt{ETH} --- Ethereum;
        \texttt{MC} --- malicious contract;
        } \\

        \multicolumn{7}{c}{\scriptsize
        \texttt{non-ETH} --- non-Ethereum;
        \texttt{EVM} --- EVM-comp.;
        \texttt{any} --- any contract;
        \texttt{VC} --- vulnerable contract;
        \texttt{MVC} --- mal. or vuln. contract.
        } \\
        
        \arrayrulecolor{black}\hline
        
        % ============ LEGEND: SUPERSCRIPTS =================
        %\multicolumn{4}{l}{$\mathrm{^\dagger}$ https://stmregistry.io/.}
        
    \end{tabular}
    }
\end{table}

% TABLE II
\begin{table}
\rev{
\linespread{0.95}
\setlength{\tabcolsep}{4pt}
    \small
    \centering
    
    %\caption{Classification of threat mitigation solutions based on the developed taxonomy.}
    %\label{tab:retention}
    \begin{tabular}{P{0.6cm}:P{2.7cm}||P{1.2cm}:P{2.1cm}:P{1.0cm}:P{2.55cm}:P{0.8cm}}
        
        \arrayrulecolor{black}\toprule
        
        % ============ UPPER HEADERS =================
        \multicolumn{1}{c}{\multirow{2}{*}{\textbf{}}} &
        \multicolumn{1}{c}{\multirow{2}{*}{\textbf{Threat}}} &
        \multicolumn{5}{c}{\textit{Classification Criteria (Dimensions)}$\mathrm{^\dagger}$} \\
        % \cmidrule(lr){4-5} 
        \cmidrule(lr){3-7}

        % ============ LOWER HEADERS =================
        \textbf{STM} & \textbf{mitigation} &
        {\small Defense} & {\small Core} & {\small Target} & {\small Data} & {\small Threat} \\
        \textbf{code
        %$\mathrm{^\dagger}$
        } & \textbf{solution} &
        {\small Modality} & {\small Method} & {\small Contract} & {\small Mapping} & {\small Model} \\
        \arrayrulecolor{black}\midrule
        
        % ============ DATA =================

        %\hline
        041 & EasyFlow~\cite{gao2019easyflow} & \texttt{DET} & \texttt{StAn} & \texttt{ETH} & \texttt{CData} $\mapsto$ \texttt{Rep} & \texttt{VC} \\
        \hline
        042 & Vultron~\cite{wang2019vultron} & \texttt{DET} & \texttt{CodeSyn} & \texttt{ETH} & \texttt{Src} $\mapsto$ \texttt{Rep} & \texttt{VC} \\          
        
        \hline
        043 & SAFEVM~\cite{albert2019safevm} & \texttt{PREV} & \texttt{StAn} & \texttt{ETH} & \texttt{S,B} $\mapsto$ \texttt{Rep} & \texttt{MVC} \\
        \hline
        044 & EthRacer~\cite{kolluri2019exploiting} & \texttt{DET} & \texttt{Fuz} & \texttt{ETH} & \texttt{BCode,CData}$\mapsto$\texttt{Rep} & \texttt{MVC} \\
        \hline
        045 & SolidityCheck~\cite{zhang2019soliditycheck} & \texttt{DET} & \texttt{StAn} & \texttt{ETH} & \texttt{Src} $\mapsto$ \texttt{Rep} & \texttt{MVC} \\   
        \hline
        046 & EVMFuzz~\cite{fu2019evmfuzz} & \texttt{DET} & \texttt{Fuz} & \texttt{ETH} & \texttt{Src} $\mapsto$ \texttt{Rep} & \texttt{MVC} \\
        \hline
        047 & EVulHunter~\cite{quan2019evulhunter} & \texttt{DET} & \texttt{StAn} & \texttt{ETH} & \texttt{Assemb} $\mapsto$ \texttt{Rep} & \texttt{VC} \\
        \hline
        048 & GasFuzz~\cite{ma2019gasfuzz} & \texttt{DET} & \texttt{Fuz} & \texttt{ETH} & \texttt{BCode} $\mapsto$ \texttt{Rep} & \texttt{MVC} \\
        \hline
        049 & NeuCheck~\cite{lu2019neucheck} & \texttt{DET} & \texttt{StAn} & \texttt{ETH} & \texttt{Src} $\mapsto$ \texttt{Rep} & \texttt{MVC} \\
        \hline
        050 & SolAnalyser~\cite{akca2019solanalyser} & \texttt{DET} & \texttt{StAn} & \texttt{ETH} & \texttt{Src} $\mapsto$ \texttt{Rep} & \texttt{MVC} \\
        \hline
        
        051 & SoliAudit~\cite{liao2019soliaudit} & \texttt{DET} & \texttt{Fuz} & \texttt{ETH} & \texttt{Src} $\mapsto$ \texttt{Rep} & \texttt{MVC} \\
        \hline
        052 & MPro~\cite{zhang2019mpro} & \texttt{DET} & \texttt{StAn,SymEx} & \texttt{ETH} & \texttt{Src} $\mapsto$ \texttt{Rep} & \texttt{MVC} \\
        \hline
        053 & Li et al.~\cite{li2019formal} & \texttt{PREV} & \texttt{FormAn} & \texttt{ETH} & \texttt{Src} $\mapsto$ \texttt{Rep} & \texttt{VC} \\
        \hline
        054 & Gastap~\cite{albert2019running} & \texttt{PREV} & \texttt{StAn} & \texttt{ETH} & \texttt{S,B,As} $\mapsto$ \texttt{Rep} & \texttt{MVC} \\
        \hline
        055 & Momeni et al.~\cite{momeni2019machine} & \texttt{DET} & \texttt{ML} & \texttt{ETH} & \texttt{Src} $\mapsto$ \texttt{Meta} & \texttt{MVC} \\
        \hline
        056 & KSolidity~\cite{jiao2020semantic} & \texttt{EXPL} & \texttt{FormAn} & \texttt{ETH} & \texttt{Src} $\mapsto$ \texttt{Meta} & \texttt{MVC} \\
        \hline
        057 & VerX~\cite{permenev2020verx} & \texttt{PREV} & \texttt{SymEx,CodeSyn} & \texttt{ETH} & \texttt{Src} $\mapsto$ \texttt{Rep} & \texttt{MVC} \\
        \hline
        058 & VeriSmart~\cite{so2020verismart} & \texttt{DET} & \texttt{StAn} & \texttt{ETH} & \texttt{Src} $\mapsto$ \texttt{Rep} & \texttt{MVC} \\
        \hline
        059 & TxSpector~\cite{zhang2020txspector} & \texttt{EXPL} & \texttt{StAn,ExTrace} & \texttt{ETH} & \texttt{CData} $\mapsto$ \texttt{Rep} & \texttt{MVC} \\
        \hline
        060 & Zhou et al.~\cite{zhou2020ever} & \texttt{EXPL} & \texttt{StAn} & \texttt{ETH} & \texttt{CData} $\mapsto$ \texttt{Rep} & \texttt{MVC} \\
        \hline
        
        061 & ETHBMC~\cite{frank2020ethbmc} & \texttt{DET} & \texttt{SymEx} & \texttt{ETH} & \texttt{BCode} $\mapsto$ \texttt{Rep} & \texttt{MVC} \\
        \hline
        062 & SODA~\cite{chen2020soda} & \texttt{DET} & \texttt{TxInt} & \texttt{EVM} & \texttt{CData} $\mapsto$ \texttt{R,Ac} & \texttt{VC} \\
        \hline
        063 & Ethor~\cite{schneidewind2020ethor} & \texttt{PREV} & \texttt{StAn,FormAn} & \texttt{ETH} & \texttt{BCode} $\mapsto$ \texttt{Rep} & \texttt{MVC} \\
        \hline
        064 & \AE GIS~\cite{ferreira2020aegis} & \texttt{DET} & \texttt{TxInt} & \texttt{ETH} & \texttt{CData} $\mapsto$ \texttt{R,Ac} & \texttt{VC} \\
        \hline
        065 & SafePay~\cite{li2020safepay} & \texttt{DET} & \texttt{SymEx} & \texttt{ETH} & \texttt{S,B} $\mapsto$ \texttt{Rep} & \texttt{VC} \\
        \hline
        066 & Solar~\cite{feng2020summary} & \texttt{DET} & \texttt{CodeSyn,SymEx} & \texttt{ETH} & \texttt{Spec} $\mapsto$ \texttt{Rep} & \texttt{VC} \\
        \hline
        067 & EVMFuzzer~\cite{fu2019evmfuzzer} & \texttt{DET} & \texttt{Fuz} & \texttt{EVM} & \texttt{Spec} $\mapsto$ \texttt{Rep} & \texttt{MVC} \\
        \hline
        068 & ModCon~\cite{liu2020modcon} & \texttt{DET+PREV} & \texttt{Fuz} & \texttt{any} & \texttt{Src} $\mapsto$ \texttt{Rep} & \texttt{MVC} \\
        \hline
        069 & Harvey~\cite{wustholz2020harvey} & \texttt{DET} & \texttt{Fuz} & \texttt{ETH} & \texttt{Src} $\mapsto$ \texttt{Rep} & \texttt{MVC} \\
        \hline
        070 & Solythesis~\cite{li2020securing} & \texttt{PREV} & \texttt{CodeSyn} & \texttt{ETH} & \texttt{Src} $\mapsto$ \texttt{Src} & \texttt{VC} \\
        \hline
        
        071 & Ethainter~\cite{brent2020ethainter} & \texttt{DET} & \texttt{StAn} & \texttt{ETH} & \texttt{Src} $\mapsto$ \texttt{Rep} & \texttt{MVC} \\
        \hline
        072 & sFuzz~\cite{nguyen2020sfuzz} & \texttt{DET} & \texttt{Fuz} & \texttt{ETH} & \texttt{BCode,ABI} $\mapsto$ \texttt{Rep} & \texttt{MVC} \\
        \hline
        073 & Seraph~\cite{yang2020seraph} & \texttt{DET} & \texttt{SymEx} & \texttt{any} & \texttt{Src} $\mapsto$ \texttt{Rep} & \texttt{MVC} \\
        \hline
        074 & Clairvoyance~\cite{ye2020clairvoyance} & \texttt{DET} & \texttt{StAn} & \texttt{ETH} & \texttt{Src} $\mapsto$ \texttt{Rep} & \texttt{VC} \\
        \hline
        075 & Artemis~\cite{wang2020artemis} & \texttt{DET} & \texttt{SymEx} & \texttt{ETH} & \texttt{BCode} $\mapsto$ \texttt{Rep} & \texttt{MVC} \\
        \hline
        076 & Echidna~\cite{grieco2020echidna} & \texttt{DET} & \texttt{Fuz} & \texttt{ETH} & \texttt{BCode,Sc} $\mapsto$ \texttt{R,M} & \texttt{MVC} \\
        \hline
        077 & EShield~\cite{yan2020eshield} & \texttt{PREV} & \texttt{CodeSyn} & \texttt{ETH} & \texttt{BCode} $\mapsto$ \texttt{BCode} & \texttt{VC} \\
        \hline
        078 & SmartShield~\cite{zhang2020smartshield} & \texttt{DET} & \texttt{CodeSyn} & \texttt{ETH} & \texttt{BCode} $\mapsto$ \texttt{BCode,R} & \texttt{VC} \\
        \hline
        079 & ETHPLOIT~\cite{zhang2020ethploit} & \texttt{DET} & \texttt{Fuz} & \texttt{ETH} & \texttt{Src} $\mapsto$ \texttt{Exp} & \texttt{MVC} \\
        \hline
        080 & Cecchetti et al.~\cite{cecchetti2020securing} & \texttt{PREV} & \texttt{CodeSyn} & \texttt{any} & \texttt{Src} $\mapsto$ \texttt{Rep} & \texttt{VC} \\
        
        \hline
        081 & EthScope~\cite{wu2020ethscope} & \texttt{DET} & \texttt{ExTrace} & \texttt{ETH} & \texttt{CData} $\mapsto$ \texttt{Rep} & \texttt{MC} \\
        \hline
        082 & ContractWard~\cite{wang2020contractward} & \texttt{DET} & \texttt{ML} & \texttt{ETH} & \texttt{Src} $\mapsto$ \texttt{R,M} & \texttt{MVC} \\
        \hline
        083 & RA~\cite{chinen2020ra} & \texttt{DET} & \texttt{StAn,SymEx} & \texttt{ETH} & \texttt{BCode} $\mapsto$ \texttt{Rep} & \texttt{VC} \\
        \hline
        084 & Camino et al.~\cite{camino2020data} & \texttt{DET} & \texttt{StAn} & \texttt{ETH} & \texttt{BCode} $\mapsto$ \texttt{Rep} & \texttt{MC} \\
        \hline
        085 & OpenBalthazar~\cite{arganaraz2020detection} & \texttt{DET} & \texttt{StAn} & \texttt{ETH} & \texttt{Src} $\mapsto$ \texttt{Rep} & \texttt{MVC} \\
        \hline
        086 & sGUARD~\cite{nguyen2021sguard} & \texttt{PREV} & \texttt{StAn,FormAn} & \texttt{ETH} & \texttt{BCode} $\mapsto$ \texttt{Rep} & \texttt{VC} \\
        \hline
        087 & SmartPulse~\cite{stephens2021smartpulse} & \texttt{PREV} & \texttt{StAn} & \texttt{ETH} & \texttt{S,Sp} $\mapsto$ \texttt{Rep} & \texttt{MVC} \\

        % ============ LEGEND: ICONS =================
        \arrayrulecolor{black}\bottomrule
        
        % \multicolumn{38}{c}{\scriptsize
        % \CIRCLE~--- full support; \LEFTcircle~--- partial support; \Circle~--- no support; \faQuestionCircle~--- possible support (insufficient data); \faExclamationCircle~--- tautological determination.
        % } \\
        
        % \arrayrulecolor{red}\hline
        
        % ============ LEGEND: SUPERSCRIPTS =================
        %\multicolumn{4}{l}{$\mathrm{^\dagger}$ https://swcregistry.io/.}
        
    \end{tabular}
    }
\end{table}

% TABLE III
\begin{table}
\rev{
\linespread{0.95}
\setlength{\tabcolsep}{4pt}
    \small
    \centering
    % \caption{Classification of threat mitigation solutions based on the developed taxonomy.}
    % \label{tab:retention}
    \begin{tabular}{P{0.6cm}:P{2.8cm}||P{1.2cm}:P{2.1cm}:P{1.0cm}:P{2.55cm}:P{0.8cm}}
        
        \arrayrulecolor{black}\toprule
        
        % ============ UPPER HEADERS =================
        \multicolumn{1}{c}{\multirow{2}{*}{\textbf{}}} &
        \multicolumn{1}{c}{\multirow{2}{*}{\textbf{Threat}}} &
        \multicolumn{5}{c}{\textit{Classification Criteria (Dimensions)}$\mathrm{^\dagger}$} \\
        % \cmidrule(lr){4-5} 
        \cmidrule(lr){3-7}

        % ============ LOWER HEADERS =================
        \textbf{STM} & \textbf{mitigation} &
        {\small Defense} & {\small Core} & {\small Target} & {\small Data} & {\small Threat} \\
        \textbf{code
        %$\mathrm{^\dagger}$
        } & \textbf{solution} &
        {\small Modality} & {\small Method} & {\small Contract} & {\small Mapping} & {\small Model} \\
        \arrayrulecolor{black}\midrule
        
        % ============ DATA =================
        
        %\hline
        088 & SeRIF~\cite{cecchetti12compositional} & \texttt{DET} & \texttt{FormAn} & \texttt{any} & \texttt{Src} $\mapsto$ \texttt{Rep} & \texttt{VC} \\        
        \hline
        \multirow{2}{*}{089} & \multirow{2}{*}{EVMPatch~\cite{rodler2021evmpatch}} & \multirow{2}{*}{\texttt{DET}} & \multirow{2}{*}{\texttt{CodeSyn}} & \multirow{2}{*}{\texttt{ETH}} & \texttt{BCode,CData} $\mapsto$ \texttt{BCode} & \multirow{2}{*}{\texttt{VC}} \\
        \hline
        090 & Perez et al.~\cite{perez2021smart} & \texttt{EXPL} & \texttt{ExTrace} & \texttt{ETH} & \texttt{CData} $\mapsto$ \texttt{Rep} & \texttt{VC} \\
        \hline
        
        091 & DEFIER~\cite{su2021evil} & \texttt{DET} & \texttt{ExTrace} & \texttt{ETH} & \texttt{CData} $\mapsto$ \texttt{Rep} & \texttt{MVC} \\
        \hline
        092 & SmarTest~\cite{so2021smartest} & \texttt{DET} & \texttt{SymEx,StAn} & \texttt{ETH} & \texttt{BCode} $\mapsto$ \texttt{R,E} & \texttt{MVC} \\
        \hline
        093 & EOSAFE~\cite{he2021eosafe} & \texttt{DET} & \texttt{StAn} & \texttt{non-ETH} & \texttt{BCode} $\mapsto$ \texttt{Rep} & \texttt{MVC} \\
        \hline
        094 & Ivanov et al.~\cite{ivanov2021targeting} & \texttt{PREV+DET} & \texttt{StAn} & \texttt{ETH} & \texttt{Src} $\mapsto$ \texttt{Rep} & \texttt{MC} \\
        \hline
        095 & ConFuzzius~\cite{ferreira2021confuzzius} & \texttt{DET} & \texttt{Fuz} & \texttt{ETH} & \texttt{BCode,ABI} $\mapsto$ \texttt{Rep} & \texttt{MVC} \\
        \hline
        096 & Huang et al.~\cite{huang2021hunting} & \texttt{DET} & \texttt{StAn} & \texttt{ETH} & \texttt{BCode} $\mapsto$ \texttt{Rep} & \texttt{VC} \\
        \hline
        097 & STC/STV~\cite{hu2021security} & \texttt{PREV} & \texttt{StAn} & \texttt{ETH} & \texttt{Src} $\mapsto$ \texttt{Rep} & \texttt{VC} \\
        \hline
        098 & Horus~\cite{ferreira2021eye} & \texttt{DET} & \texttt{ExTrace} & \texttt{ETH} & \texttt{CData} $\mapsto$ \texttt{Rep} & \texttt{MVC} \\
        \hline
        099 & BlockEye~\cite{wang2021blockeye} & \texttt{DET} & \texttt{ExTrace,TxInt} & \texttt{ETH} & \texttt{CData} $\mapsto$ \texttt{Rep} & \texttt{MVC} \\
        \hline
        100 & Sailfish~\cite{bose2021sailfish} & \texttt{DET} & \texttt{SymEx,StAn} & \texttt{ETH} & \texttt{BCode} $\mapsto$ \texttt{Rep} & \texttt{MVC} \\
        \hline
        
        101 & DeFiRanger~\cite{wu2021defiranger} & \texttt{DET} & \texttt{StAn} & \texttt{ETH} & \texttt{CData} $\mapsto$ \texttt{Rep} & \texttt{VC} \\
        \hline
        102 & ESCORT~\cite{lutz2021escort} & \texttt{DET} & \texttt{ML} & \texttt{ETH} & \texttt{BCode,Spec} $\mapsto$ \texttt{Rep} & \texttt{MVC} \\
        \hline
        103 & DefectChecker~\cite{chen2021defectchecker} & \texttt{DET} & \texttt{SymEx} & \texttt{ETH} & \texttt{BCode} $\mapsto$ \texttt{Rep} & \texttt{MVC} \\
        \hline
        104 & Hu et al.~\cite{hu2021transaction} & \texttt{DET} & \texttt{StAn,ML} & \texttt{ETH} & \texttt{BCode} $\mapsto$ \texttt{Rep} & \texttt{MVC} \\
        \hline
        105 & HFContractFuzzer~\cite{ding2021hfcontractfuzzer} & \texttt{DET} & \texttt{Fuz} & \texttt{non-ETH} & \texttt{Src} $\mapsto$ \texttt{Rep} & \texttt{VC} \\
        \hline
        106 & Solidifier~\cite{antonino2021solidifier} & \texttt{PREV} & \texttt{FormAn} & \texttt{ETH} & \texttt{Src} $\mapsto$ \texttt{Rep} & \texttt{VC} \\
        \hline
        107 & SafelyAdmin.~\cite{ivanov2021rectifying} & \texttt{PREV} & \texttt{CodeSyn,ML} & \texttt{ETH} & \texttt{Src} $\mapsto$ \texttt{Src} & \texttt{MC} \\
        \hline
        108 & EXGEN~\cite{jin2022exgen} & \texttt{DET} & \texttt{SymEx} & \texttt{any} & \texttt{Src} $\mapsto$ \texttt{Rep} & \texttt{VC} \\
        \hline
        109 & EtherProv~\cite{linoyetherprov} & \texttt{DET} & \texttt{StAn} & \texttt{ETH} & \texttt{Src} $\mapsto$ \texttt{BCode,M} & \texttt{MVC} \\
        \hline
        110 & Abdellatif et al.~\cite{abdellatif2018formal} & \texttt{PREV} & \texttt{FormAn} & \texttt{any} & \texttt{CData} $\mapsto$ \texttt{Rep} & \texttt{MVC} \\
        \hline
        
        111 & Bai et al.~\cite{bai2018formal} & \texttt{PREV} & \texttt{FormAn} & \texttt{any} & \texttt{Spec} $\mapsto$ \texttt{Rep} & \texttt{MVC} \\
        \hline
        112 & Bigi et al.~\cite{bigi2015validation} & \texttt{PREV} & \texttt{FormAn} & \texttt{any} & \texttt{Spec} $\mapsto$ \texttt{Rep} & \texttt{MVC} \\
        \hline
        113 & Findel~\cite{biryukov2017findel} & \texttt{PREV} & \texttt{CodeSyn} & \texttt{any} & \texttt{Spec} $\mapsto$ \texttt{Src} & \texttt{VC} \\
        \hline
        114 & ContractLarva~\cite{ellul2018runtime} & \texttt{PREV} & \texttt{CodeSyn} & \texttt{ETH} & \texttt{S,Sp} $\mapsto$ \texttt{Src} & \texttt{MVC} \\
        \hline
        115 & Le et al.~\cite{le2018proving} & \texttt{PREV} & \texttt{FormAn} & \texttt{any} & \texttt{Src} $\mapsto$ \texttt{Rep} & \texttt{MVC} \\
        \hline
        116 & Solicitous~\cite{marescotti2020accurate} & \texttt{PREV} & \texttt{StAn} & \texttt{ETH} & \texttt{Src} $\mapsto$ \texttt{BCode} & \texttt{MVC} \\
        \hline
        117 & VeriSol~\cite{wang2019formal} & \texttt{PREV} & \texttt{FormAn} & \texttt{EVM} & \texttt{Src} $\mapsto$ \texttt{Rep} & \texttt{VC} \\
        \hline
        118 & SmartCopy~\cite{feng2019precise} & \texttt{DET} & \texttt{Fuz} & \texttt{ETH} & \texttt{BCode,ABI} $\mapsto$ \texttt{Rep} & \texttt{VC} \\
        \hline
        119 & WANA~\cite{wang2020wana} & \texttt{DET} & \texttt{SymEx} & \texttt{any} & \texttt{BCode} $\mapsto$ \texttt{Rep} & \texttt{MVC} \\
        \hline
        120 & E-EVM~\cite{norvill2018visual} & \texttt{EXPL} & \texttt{ExTrace} & \texttt{ETH} & \texttt{CData} $\mapsto$ \texttt{Rep} & \texttt{MVC} \\
        
        \hline
        
         121 & AMEVulDetector~\cite{liu2021smart} & \texttt{DET} & \texttt{ML} & \texttt{any} & \texttt{Src} $\mapsto$ \texttt{Rep} & \texttt{MVC} \\
        \hline
        122 & Javadity~\cite{ahrendt2019verification} & \texttt{PREV} & \texttt{CodeSyn} & \texttt{EVM} & \texttt{Src} $\mapsto$ \texttt{Src} & \texttt{MVC} \\
        \hline
        123 & Alqahtani et al.~\cite{alqahtani2020formal} & \texttt{PREV} & \texttt{StAn} & \texttt{any} & \texttt{Src} $\mapsto$ \texttt{BCode} & \texttt{MVC} \\
        \hline
        124 & Bartoletti et al.~\cite{bartoletti2019verifying} & \texttt{PREV} & \texttt{FormAn} & \texttt{non-ETH} & \texttt{Src} $\mapsto$ \texttt{Rep} & \texttt{MVC} \\
        \hline
        125 & Beckert et al.~\cite{beckert2018formal} & \texttt{PREV} & \texttt{FormAn} & \texttt{non-ETH} & \texttt{Src} $\mapsto$ \texttt{Rep} & \texttt{VC} \\
        \hline
        126 & SmartInspect~\cite{bragagnolo2018smartinspect} & \texttt{EXPL} & \texttt{StAn} & \texttt{ETH} & \texttt{S,C} $\mapsto$ \texttt{Rep} & \texttt{MVC} \\
        \hline
        127 & CPN~\cite{duo2020formal} & \texttt{EXPL} & \texttt{StAn} & \texttt{ETH} & \texttt{S,B} $\mapsto$ \texttt{Rep} & \texttt{MVC} \\
        \hline
        128 & Hajdu et al.~\cite{hajdu2020formal} & \texttt{PREV} & \texttt{FormAn,StAn} & \texttt{ETH} & \texttt{Src} $\mapsto$ \texttt{Rep} & \texttt{MVC} \\
        \hline
        129 & Kongmanee~et~al.~\cite{kongmanee2019securing} & \texttt{PREV} & \texttt{FormAn} & \texttt{ETH} & \texttt{Spec} $\mapsto$ \texttt{Meta} & \texttt{MVC} \\
        \hline
        \multirow{2}{*}{130} & \multirow{2}{*}{EVM*~\cite{ma2019evm}} & \multirow{2}{*}{\texttt{PREV}} & \multirow{2}{*}{\texttt{TxInt}} & \multirow{2}{*}{\texttt{ETH}} & \texttt{BCode,CData} $\mapsto$ \texttt{Rep,Act} & \multirow{2}{*}{\texttt{MVC}} \\
        \hline
        
        131 & OpenZeppelin~\cite{openzeppelin-contracts} & \texttt{PREV} & \texttt{CodeSyn} & \texttt{ETH} & \texttt{Src} $\mapsto$ \texttt{Src} & \texttt{VC} \\
        \hline
        132 & MythX~\cite{mythx} & \texttt{DET} & many & \texttt{ETH} & \texttt{BCode} $\mapsto$ \texttt{Rep} & \texttt{MVC} \\
        \hline
        133 & Contract Library~\cite{contract-library} & \texttt{DET+PREV} & \texttt{CodeSyn} & \texttt{ETH} & \texttt{BCode} $\mapsto$ \texttt{Rep} & \texttt{MVC} \\

        % ============ LEGEND: ICONS =================
        \arrayrulecolor{black}\bottomrule
        
        % \multicolumn{38}{c}{\scriptsize
        % \CIRCLE~--- full support; \LEFTcircle~--- partial support; \Circle~--- no support; \faQuestionCircle~--- possible support (insufficient data); \faExclamationCircle~--- tautological determination.
        % } \\
        
        % \arrayrulecolor{red}\hline
        
        % ============ LEGEND: SUPERSCRIPTS =================
        %\multicolumn{4}{l}{$\mathrm{^\dagger}$ https://swcregistry.io/.}
        
    \end{tabular}
}
\end{table}

% % TABLE IV
% \begin{table}
% \setlength{\tabcolsep}{4pt}
%     \centering    
%     % \caption{Classification of threat mitigation solutions based on the developed taxonomy.}
%     % \label{tab:retention}
%     \begin{tabular}{c:c||c:c:c:c:c}
        
%         \arrayrulecolor{red}\toprule
        
%         % ============ UPPER HEADERS =================
%         \multicolumn{1}{c}{\multirow{2}{*}{\textbf{STM}}} &
%         \multicolumn{1}{c}{\multirow{2}{*}{\textbf{Threat}}} &
%         \multicolumn{5}{c}{\textit{Classification criteria}} \\
%         % \cmidrule(lr){4-5} 
%         \cmidrule(lr){3-7}

%         % ============ LOWER HEADERS =================
%         \textbf{registry} & \textbf{mitigation} &
%         Defense & Core & Targeted & Data & Threat \\
%         \textbf{code} & \textbf{solution} &
%         Modality & Method & Contracts & Mapping & Model \\
%         \arrayrulecolor{red}\midrule
        
%         % ============ DATA =================

%         % ============ LEGEND: ICONS =================
%         \arrayrulecolor{red}\bottomrule
        
%         % \multicolumn{38}{c}{\scriptsize
%         % \CIRCLE~--- full support; \LEFTcircle~--- partial support; \Circle~--- no support; \faQuestionCircle~--- possible support (insufficient data); \faExclamationCircle~--- tautological determination.
%         % } \\
        
%         % \arrayrulecolor{red}\hline
        
%         % ============ LEGEND: SUPERSCRIPTS =================
%         \multicolumn{4}{l}{$\mathrm{^\dagger}$ https://stmregistry.io/.}
        
%     \end{tabular}
% \end{table}

\begin{figure}[t]
\captionsetup[subfigure]{skip=5pt,margin={-10pt,0pt}}
\hspace{20pt}\begin{subfigure}{.36\textwidth}
    \includegraphics[height=1.58in]{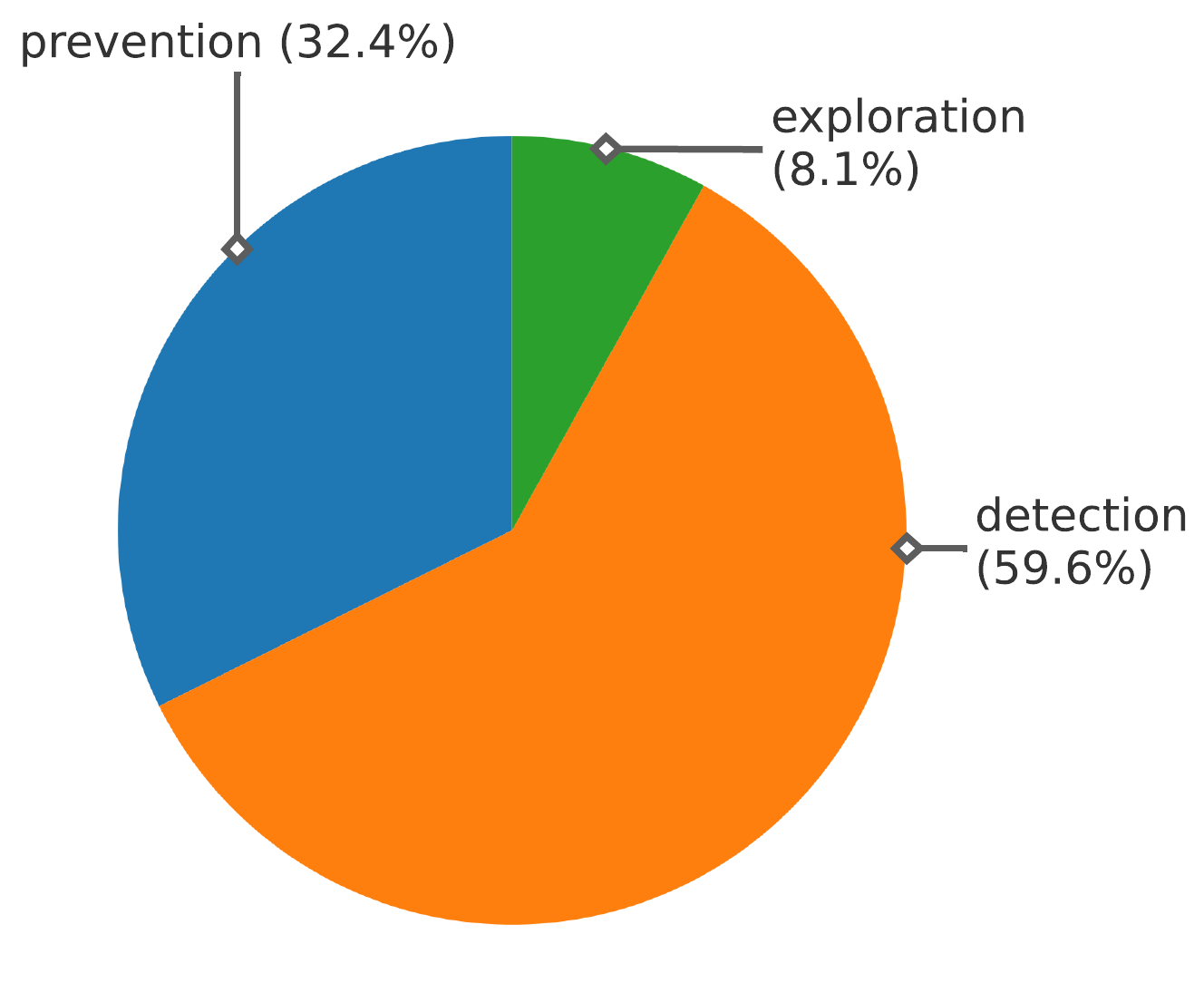}
  \caption{Defense modality}
  \label{fig:defense-modality-pie}
\end{subfigure}\hspace{29pt}\captionsetup[subfigure]{skip=5pt,margin={10pt,0pt}}
\begin{subfigure}{.51\textwidth}
    \vspace{-3pt}\includegraphics[height=1.64in]{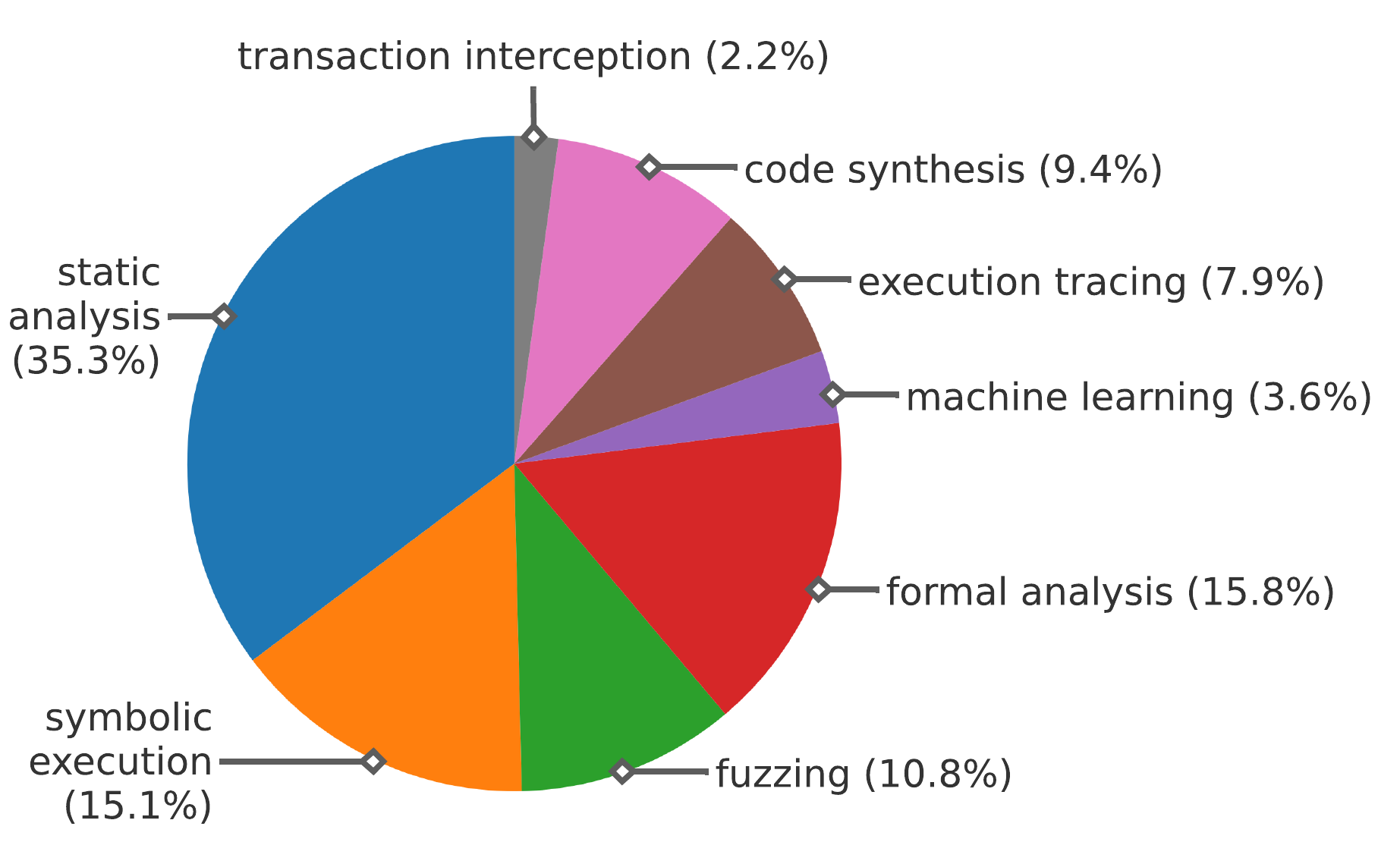}
  \caption{Core method}
  \label{fig:core-method-pie}
\end{subfigure}

\vspace{15pt}

\captionsetup[subfigure]{skip=5pt,margin={-30pt,0pt}}
\hspace{-17pt}\begin{subfigure}{.49\textwidth}
    \includegraphics[height=1.59in]{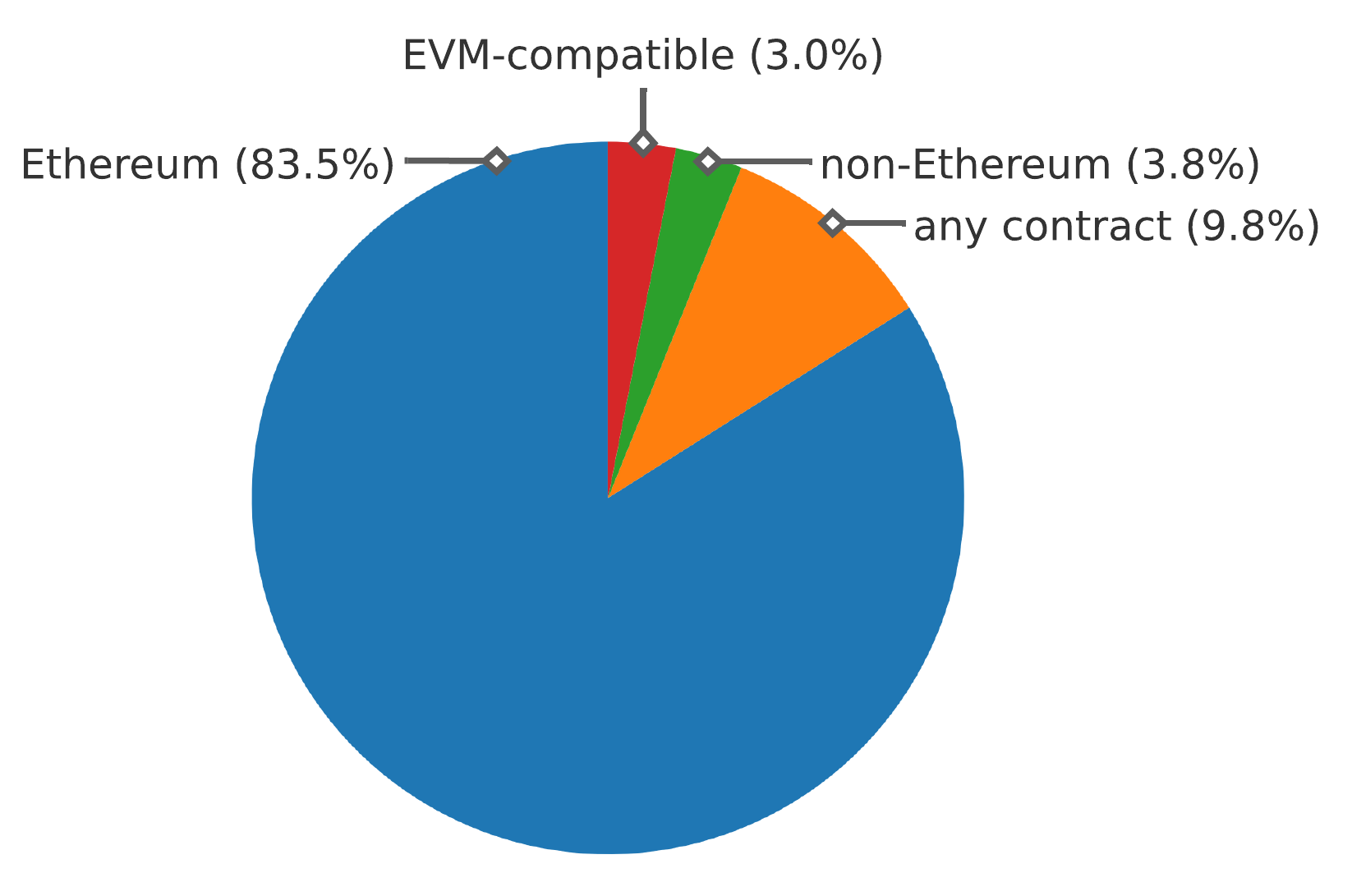}
  \caption{Targeted contracts}
  \label{fig:targeted-contracts-pie}
\end{subfigure}\hspace{-18pt}\captionsetup[subfigure]{skip=5pt,margin={-20pt,0pt}}
\begin{subfigure}{.49\textwidth}
    \vspace{12pt}\includegraphics[height=1.43in]{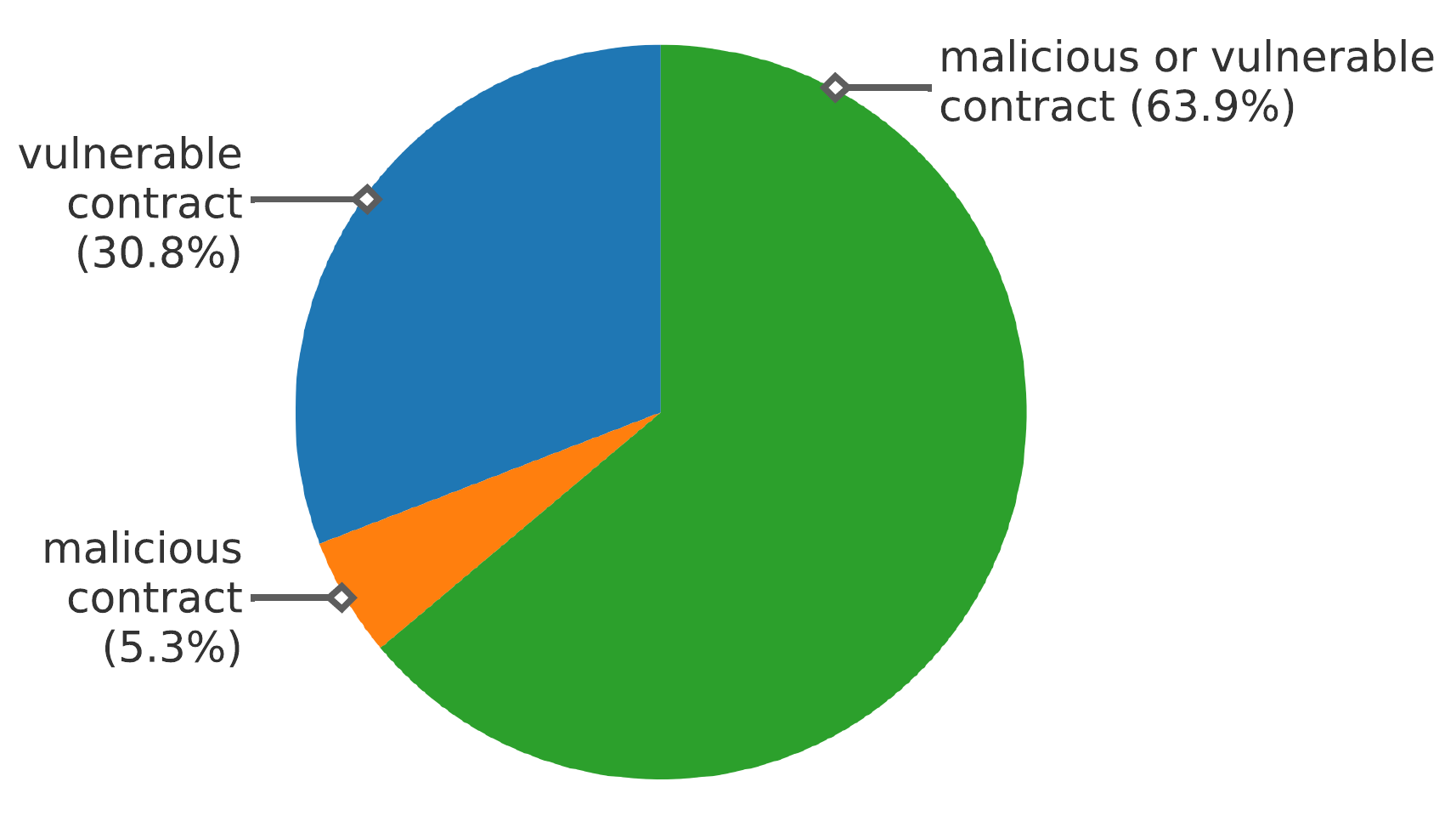}
  \caption{Threat model}
  \label{fig:threat-model-pie}
\end{subfigure}

\caption{Distribution of threat mitigation methods by four criteria: defense modality, core method, targeted contracts, and threat model.}
\label{fig:pies}
\end{figure}

In this section, we apply the taxonomy developed earlier (\S\ref{subsec:taxonomy}) to describe each of the threat mitigation solutions via the five orthogonal dimensions: threat mitigation modality (\S\ref{sec:modalities}), core method (\S\ref{sec:core-methods}), the scope of targeted contracts (\S\ref{sec:targeted-contracts}), the input-output data mapping of the solution (\S\ref{sec:data-mapping}), and the assumed threat model (\S\ref{sec:threat-modeling}). The results of our classification are given in Table~\ref{tab:classification}.
Furthermore, we perform a frequency analysis of the results along the five dimensions, and create a visual representation of the distributions of defense modalities, core methods, targeted contracts, and threat models in Fig.~\ref{fig:pies}. In the first column of the table, we assign to each of the threat mitigation solutions a permanent Security Threat Mitigation (STM) registry identifier. The second column provides the name of the tool implementing the solution along with its reference; if a solution does not have a common name, we refer to the solution by its authors (e.g., Ivanov et al.). In columns 3--7, we provide the values along the five classification dimensions for each of the 133 threat mitigation solutions. Furthermore, to keep the data in this table up to date and handy, we deploy the Smart Contract Threat Mitigation Registry (STM Registry) at \url{https://seit.egr.msu.edu/research/stmregistry/}.

\noindent\textbf{Selection Method of the Threat Mitigation Solutions.} For this survey, we select 133 threat mitigation solutions, encompassing both academic research projects (e.g., Securify~\cite{tsankov2018securify}, Oyente~\cite{luu2016making} and commercial non-academic efforts (e.g., OpenZeppelin Contracts~\cite{openzeppelin-contracts}, MythX~\cite{mythx}). To assure the quality of our study, we use the following four criteria for selecting threat mitigation solutions:

\begin{enumerate}
    \item \textit{Implementation}. We select solutions that are implemented and evaluated, either as a proof-of-concept (PoC) prototype or in the form of a final product.
    \item \textit{Publication}. For academic research projects, we search for the papers published or accepted at a reputable peer-reviewed venue.
    \item \textit{Impact}. We select solutions that deliver specific improvements or other unique qualities compared to the state-of-the-art solutions.
    \item \textit{Novelty}. Not only do we consider the fact of improvement or impact, but  we also consider the presence of technical novelty, i.e., a specific innovation that leads to the improvement.
\end{enumerate}

In some cases, we include threat mitigation solutions that do not meet all the four above criteria, such as the academic project Vandal~\cite{brent2018vandal}, which has never been published at a peer-reviewed venue. However, we include this work in our survey because it is widely adopted and cited.

\cbox{\textbf{Lessons learned:} There are more than 200 claims of smart contract threat mitigation solutions. Yet, our thorough manual examination reveals various problems associated with some of them. For example, we observed that two research papers may refer to the same implementation (e.g., poster or journal extension articles). In the end, 133 instances have been selected to represent the body of smart contract threat mitigation solutions. Therefore, manual scrutiny of each work is required.}
%The lesson we learned from that experience is that selection criteria are necessary but not sufficient, and manual scrutiny of each work is required.}

% Third column
% Classification criteria.

% \begin{figure}
%     \centering
%     \includegraphics[width=1.0\textwidth]{img/pies.png}
%     \caption{Distribution of threat mitigation methods by their threat model, targeted contracts, core methods, and defense modalities.}
%     \label{fig:pies}
% \end{figure}

\subsection{Threat Mitigation Modalities}\label{sec:modalities}
A threat mitigation modality is a philosophy that a smart contract threat mitigation method employs to address security issues of a smart contract. The threat mitigation solutions that employ the detection modality are designed to identify vulnerabilities in smart contracts. Some of them (e.g., Oyente~\cite{luu2016making}, Securify~\cite{tsankov2018securify}, Vandal~\cite{brent2018vandal}, and Mythril~\cite{mueller2018smashing}) target several groups of vulnerabilities. Other detection-based threat mitigation solutions focus on specific classes of vulnerabilities, such as Sereum~\cite{rodler2018sereum}, which detects only reentrancy vulnerabilities (SWC-107~\cite{swc107}). Another narrow-focused detection tool is VeriSmart~\cite{so2020verismart}, which detects arithmetic bugs only. Overall, we note that the detection solutions that focus on specific vulnerabilities tend to deliver improved detection rates compared to the solutions targeting multiple vulnerabilities.

The solutions belonging to the prevention modality validate some safety properties or rules. ZEUS~\cite{kalra2018zeus} provides eight semantic rules that are used as part of an abstract assertion language for specifying safety properties for ensuring that a smart contract is free of certain vulnerabilities (e.g., reentrancy, unchecked send, integer overflow, etc.). Another salient representation of a prevention solution is SmartPulse~\cite{stephens2021smartpulse}, which creates a linear temporal logic (LTL) language, called SmartLTL, for expressing temporal safety properties in smart contracts and enforcing them with the SmartPulse verifier.

\rev{The exploration modality solutions do not detect vulnerabilities or enforce safety properties. Instead, the exploration solutions reveal previously concealed data. This data can further be used for facilitating human-based or automated auditing of a smart contract.} Erays~\cite{zhou2018erays} is a tool for reverse-engineering of smart contracts that converts a bytecode of a smart contract into pseudocode-like metadata. TxSpector~\cite{zhang2020txspector} is another exploration solution, which is a transaction processing framework that identifies the executed attacks in smart contract execution traces.

\rev{Some threat mitigation solutions adhere to a hybrid detection-plus-prevention modality, which means that they can detect existing vulnerabilities, as well as enforce security properties.} Securify~\cite{tsankov2018securify} not only checks the compliance with security patterns but also detects violations of patterns associated with specific vulnerabilities, such as reentrancy and restricted transfer. Another threat mitigation solution with a hybrid detection-plus-prevention modality is ModCon~\cite{liu2020modcon}, which is a smart contract testing tool that generates a list of states and transitions between these states, thereby enabling further identification of vulnerabilities and confirmation of security properties.

Fig.~\ref{fig:defense-modality-pie} shows the breakdown of the three defense modalities among the 133 threat mitigation solutions. As we can see, 81 (59.6\%) of all the threat mitigation solutions employ the detection modality, 44 (32.4\%) use the verification modality, and the remaining 11 (8.1\%) belong to the exploration modality. Some threat mitigation solutions exhibit a hybrid modality (e.g., \texttt{DET+PREV} --- detection combined with prevention), in which case we identify and assume the predominant modality for the statistical analysis, or we count both modalities in cases when it is impossible to identify the predominant one --- which explains the 136 total modalities considered, despite the fact that they correspond to 133 threat mitigation solutions.

\subsection{Core Methods}\label{sec:core-methods}
The core method describes \emph{how} a threat mitigation solution addresses the security issues of a smart contract. In other words, the core method defines the implementation approach, choice of algorithms, and internal data processing model of a threat mitigation solution. \rev{By scrutinizing all the 133 smart contract threat mitigation solutions, we identify eight distinct core methods: static analysis, symbolic execution, fuzzing, formal analysis, machine learning, execution tracing, code synthesis, and transaction interception.}

Static analysis solutions extract data from smart contracts in order to detect vulnerabilities or confirm safety properties. Most static analysis solutions adhere to the detection modality (e.g., Security~\cite{tsankov2018securify}, S-GRAM~\cite{liu2018s}, MadMax~\cite{grech2018madmax}, SmartCheck~\cite{tikhomirov2018smartcheck}). However, some static analysis solutions enforce policies instead of detecting vulnerabilities (e.g., solc-verify~\cite{hajdu2019solc}, BitML~\cite{atzei2019developing}, GasTap~\cite{albert2019running}, Solicitious~\cite{marescotti2020accurate}). Moreover, we notice that the static analysis core method is often coupled with some other methods. Solidity*~\cite{bhargavan2016formal}, Amani et al.~\cite{amani2018towards}, Ethor~\cite{schneidewind2020ethor}, and sGUARD~\cite{nguyen2021sguard} use static analysis together with formal analysis. Also, static analysis is often used together with the symbolic execution core method, as we can see in HoneyBadger~\cite{torres2019art}, MPro~\cite{zhang2019mpro}, SmarTest~\cite{so2021smartest}, and Sailfish~\cite{bose2021sailfish}.

Symbolic execution methods execute a smart contract with symbolic parameters instead of real ones --- in order to make conclusions regarding some security properties of smart contracts (e.g., the range of values that make a certain condition true). Oyente~\cite{luu2016making}, Mythril~\cite{mueller2018smashing}, Maian~\cite{nikolic2018finding}, Manticore~\cite{mossberg2019manticore}, ZEUS~\cite{kalra2018zeus}, Osiris~\cite{torres2018osiris}, teEther~\cite{krupp2018teether} are popular solutions employing the symbolic execution core method. Similar to static analysis, symbolic execution is also often coupled with other core methods. VerX~\cite{permenev2020verx} and Solar~\cite{feng2020summary} use symbolic execution to guide code synthesis. The solution by Hu et al.~\cite{hu2021transaction} takes advantage of both symbolic execution and machine learning for detecting smart contract vulnerabilities.

Fuzzing methods perform smart contract testing by iteratively generating test cases that are likely to reveal vulnerabilities. ContractFuzzer~\cite{jiang2018contractfuzzer} uses the abstract binary interface (ABI) of the smart contract to facilitate the generation of fuzzing inputs. 
Harvey~\cite{wustholz2020harvey} is a smart contract tester based on greybox fuzzing, which is a middle-ground solution between the absence of code analysis (blackbox fuzzing) and full code execution (whitebox fuzzing); specifically, greybox fuzzing assumes a lightweight (compared to symbolic execution) analysis of the code execution paths. Confuzzius~\cite{ferreira2021confuzzius} is a smart contract fuzzer that uses a combination of genetic algorithms and constraint solving. Overall, fuzzing threat mitigation solutions utilize a diverse variety of predictive methods for balancing accuracy and performance.

%ILF~\cite{he2019learning} ...
%ContractFuzzer~\cite{jiang2018contractfuzzer}, ReGuard~\cite{liu2018reguard}, ILF~\cite{he2019learning}, EthRacer~\cite{kolluri2019exploiting}, GasFuzz~\cite{ma2019gasfuzz}, ModCon~\cite{liu2020modcon}, Harvey~\cite{wustholz2020harvey}, ETHPLOIT~\cite{zhang2020ethploit}, and Confuzzius~\cite{ferreira2021confuzzius} are examples of smart contract threat mitigation solutions employing the fuzzing core method.

Formal analysis methods convert a smart contract into a formal representation and run a solver over this representation to prove or disprove some security properties. Most solutions employing the formal analysis core method belong to either the prevention defense modality (e.g., Lolisa~\cite{yang2018lolisa}, Model-Checking~\cite{nehai2018model}, Li et al.~\cite{li2019formal}, Solidifier~\cite{antonino2021solidifier}, VeriSol~\cite{wang2019formal}) or the exploration modality (e.g., KEVM~\cite{hildenbrandt2018kevm}, Grishchenko et al.~\cite{grishchenko2018semantic}). However, SeRIF~\cite{cecchetti12compositional}, which primary purpose is defense against reentrancy, demonstrates that the formal analysis can also be used for targeting vulnerabilities.

Machine learning methods extract features from smart contracts and train models for detecting vulnerabilities. The smart contract threat mitigation solutions utilizing the machine learning core method are ContractWard~\cite{wang2020contractward}, ESCORT~\cite{lutz2021escort}, AMEVulDetector~\cite{liu2021smart}, and the solution by Momeni et al.~\cite{momeni2019machine}. In \S\ref{sec:ai-driven-security}, we conduct an in-depth discussion about the evolutionary perspective of machine learning in smart contract security.

Execution tracing and transaction interception core methods constitute the transaction-based methods of smart contract threat mitigation. The execution tracing methods examine the runtime traces of the actual transactions submitted to a smart contract in order to detect vulnerabilities, verify safety properties, or facilitate manual auditing. TokenScope~\cite{chen2019tokenscope}, EthScope~\cite{wu2020ethscope}, DEFIER~\cite{su2021evil}, Horus~\cite{ferreira2021eye}, BlockEye~\cite{wang2021blockeye}, E-EVM~\cite{norvill2018visual} are instances of ``pure'' execution tracing methods (i.e., not combined with other methods).

Code synthesis threat mitigation solutions aim at generating vulnerability-free smart contract code resistant to attacks. Hydra~\cite{breidenbach2018enter} is a framework that generates bug bounties for smart contracts using the N-of-N version programming (NNVP) principle. FSolidM~\cite{mavridou2018designing} is a framework for designing secure smart contracts as finite state machines (FSMs) and converting them into Solidity code. Solythesis~\cite{li2020securing} is a source-to-source Solidity compiler that instruments the input source code with additional instructions for validation of security-sensitive invariants.

Transaction interception solutions dynamically observe the transaction pool of a blockchain node in order to prevent the execution of malicious or unsafe transactions. These solutions are represented by SODA~\cite{chen2020soda}, and EVM*~\cite{ma2019evm}. However, we observe that execution tracing is often combined with other core methods. Sereum~\cite{rodler2018sereum} and ECFChecker~\cite{grossman2017online} combine execution tracing with transaction interception, while TxSpector~\cite{zhang2020txspector} and the Ponzi scheme detection solution by Chen et al.~\cite{chen2018detecting} utilize trace execution combined with static analysis.

Fig.~\ref{fig:core-method-pie} shows the distribution of the eight core methods among the 133 threat mitigation solutions. Specifically we found 49 (35.3\%) static analysis tools, 21 (15.1\%) symbolic execution methods, 15 (10.8\%) fuzzing tools, 22 (15.8\%) formal analysis tools, 5 (3.6\%) machine learning solutions, 11 (7.9\%) execution tracing tools, 13 (9.4\%) code synthesis tools, and 3 (2.2\%) transaction interceptors. Notably, some threat mitigation solutions employ a combination of the aforementioned core methods; in this case, we recognize all the methods evolved in Table~\ref{tab:classification}, yet for the purpose of counting and frequency analysis, we reduce the combination of core methods to the predominant core method, if there is one. If it is impossible to identify the predominant core method, we count all of them, which explains that the total count of instances of core methods slightly exceeds the number of the threat mitigation solutions surveyed in this work.

\subsection{Targeted Contracts}\label{sec:targeted-contracts}
Each of the threat mitigation solutions assumes a type of targeted smart contract. Some solutions target general groups of smart contracts, such as Ethereum or even all possible contracts, while some other solutions may target a single specific smart contract instance. Oyente~\cite{luu2016making}, Mythril~\cite{mueller2018smashing}, Securify~\cite{tsankov2018securify}, Sereum~\cite{rodler2018sereum}, Vandal~\cite{brent2018vandal}, OpenZeppelin Contracts~\cite{openzeppelin-contracts}, MythX~\cite{mythx}, Contract Library~\cite{contract-library}, and many other popular threat mitigation solutions are strictly Ethereum-based. Some solutions are EVM-compatible, which means that they are compatible with \emph{but not limited by} the Ethereum smart contracts. SODA~\cite{chen2020soda}, VeriSol~\cite{wang2019formal}, and Javadity~\cite{ahrendt2019verification} are EVM-compatible solutions. Some solutions are universal in terms of the scope of targeted contracts; although they might not support \emph{any} type of smart contracts (e.g., the ones that are not Turing-complete), they do not limit their scope to a specific group either. Such solutions are ModCon~\cite{liu2020modcon}, Seraph~\cite{yang2020seraph}, SeRIF~\cite{cecchetti12compositional}, EXGEN~\cite{jin2022exgen}, and the information flow control solution by Cecchetti et al.~\cite{cecchetti2020securing}. Some threat mitigation solutions target a specific non-Ethereum platform. BitML~\cite{atzei2019developing} targets Bitcoin smart contract overlays, EOSAFE~\cite{he2021eosafe} targets the smart contracts on the EOS blockchain~\cite{eos-whitepaper}, and HFContractFuzzer~\cite{ding2021hfcontractfuzzer} targets the Hyperledger Fabric platform~\cite{androulaki2018hyperledger}.

To make sense of this diverse spectrum, we group the targeted smart contracts into four types, as described in \S~\ref{subsec:targeted-contracts}. Fig.~\ref{fig:targeted-contracts-pie} shows the distribution of different groups of targeted contracts among the threat mitigation methods. Specifically, we discover that as many as 111 (83.5\%) solutions target Ethereum contracts, 13 (9.8\%) are suitable for any contract (including Ethereum, but not specifying it), 5 (3.8\%) aim for some non-Ethereum contracts (e.g., Hyperledger Fabric), and 4 (3.0\%) target EVM-compatible contracts (e.g., Polygon~\cite{polygon}, RSK~\cite{rsk}).

\subsection{Data Mapping}\label{sec:data-mapping}
Next, we explore the design-specified inputs and outputs of each of the threat mitigation solutions. Most smart contract threat mitigation solutions assume a smart contract as an input, either as bytecode, source code, or as part of the chain data. Oyente~\cite{luu2016making}, Mythril~\cite{mueller2018smashing}, Vandal~\cite{brent2018vandal}, ZEUS~\cite{kalra2018zeus}, teEther~\cite{krupp2018teether}, and Osiris~\cite{torres2018osiris} are solutions that take bytecode as a smart contract input. Hydra~\cite{breidenbach2018enter}, S-GRAM~\cite{liu2018s}, SmartCheck~\cite{tikhomirov2018smartcheck}, VerX~\cite{permenev2020verx}, VeriSmart~\cite{so2020verismart}, and SeRIF~\cite{cecchetti12compositional} are solutions that assume source code as the input. Sereum~\cite{rodler2018sereum}, ECFChecker~\cite{grossman2017online}, TokenScope~\cite{chen2019tokenscope}, EasyFlow~\cite{gao2019easyflow}, TxSpector~\cite{zhang2020txspector}, and EthScope~\cite{wu2020ethscope} are the threat mitigation solutions that read smart contract information from the chain data, i.e., stored copy of the blockchain.

Some threat mitigation solutions use a combination of bytecode and source code as an input, e.g., Securify\footnote{Source code is optional in Securify.}~\cite{tsankov2018securify}, SAFEVM~\cite{albert2019safevm}, Gastap~\cite{albert2019running}, SafePay~\cite{li2020safepay}, and CPN~\cite{duo2020formal}. Other solutions, in addition to a smart contract, also take a set of manual specifications as an input, as we see it in FSolidM~\cite{mavridou2018designing}, Model-Checking~\cite{nehai2018model}, VeriSolid~\cite{mavridou2019verisolid}, solc-verify~\cite{hajdu2019solc}, BitML~\cite{atzei2019developing}, SmartPulse~\cite{stephens2021smartpulse}, ESCORT~\cite{lutz2021escort}, and ContractLarva~\cite{ellul2018runtime}. Moreover, a smart contract is not always used as an input of a threat mitigation solution. For instance, Flint~\cite{schrans2018writing}, Solar~\cite{feng2020summary}, EVMFuzzer~\cite{fu2019evmfuzzer}, Findel~\cite{biryukov2017findel}, and the solution by Kongmanee et al.~\cite{kongmanee2019securing} assume a set of specifications as the only input.

Most threat mitigation solutions produce a human-readable report as an output, e.g., Oyente~\cite{luu2016making}, Mythril~\cite{mueller2018smashing}, Maian~\cite{nikolic2018finding}, Manticore~\cite{mossberg2019manticore}, ZEUS~\cite{kalra2018zeus}, and Sereum~\cite{rodler2018sereum}. However, some solutions produce machine-readable metadata (e.g., a formal model) in lieu of a human-readable report, which can be observed in Erays~\cite{zhou2018erays}, the solution by Grishchenko et al.~\cite{grishchenko2018semantic}, the solution by Momeni et al.~\cite{momeni2019machine}, KSolidity~\cite{jiao2020semantic}, and the solution by Kongmanee et al.~\cite{kongmanee2019securing}.

Table~\ref{tab:classification} shows that the majority of the threat mitigation solutions (82.7\%) produce a human-readable report as an output, and for 78.19\% of the solutions, the security report is the only output. Notably, only 4 (3.0\%) of all the threat mitigation solutions result in an action (e.g., stopping a malicious transaction), which is indicative of the predominance of the static methodology in the smart contract defense, which is further discussed in \S\ref{sec:dyn-tx-interception}.

One important property of data mapping is that it often provides fine-tuned information that cannot be inferred from the workflow of the corresponding core method. For example, the workflows of smart contract threat mitigation solutions often specify ``smart contract'' as one of the inputs. However, a smart contract can have several representations: source code, bytecode, deployed address, etc. In this work, we extract the specific meaning of the ``smart contract'' and represent it accordingly in the data mapping.

\subsection{Threat Model}\label{sec:threat-modeling}
Finally, we describe all the threat mitigation solutions through the general description of their assumed threat models. In other words, the threat model specifies the source of the threat, identifies the victim(s), and defines the intent. We generalize all the threat models by subdividing them into three major groups: victim contract, malicious contract, and hybrid malicious \emph{or} victim contract. Sereum~\cite{rodler2018sereum}, teEther~\cite{krupp2018teether}, Hydra~\cite{breidenbach2018enter}, Osiris~\cite{torres2018osiris}, SODA~\cite{chen2020soda}, {\AE}GIS~\cite{ferreira2020aegis}, EVMPatch~\cite{rodler2021evmpatch}, SeRIF~\cite{cecchetti12compositional}, and OpenZeppelin Contracts~\cite{openzeppelin-contracts} are threat mitigation solutions with the vulnerable contract threat model. Solutions with malicious contract threat models are the Ethereum honeypot detector HoneyBadger~\cite{torres2019art}, GASPER~\cite{chen2017under}, and the social engineering attack detector by Ivanov et al.~\cite{ivanov2021targeting}. Most threat mitigation solutions, however, are threat vector agnostic, i.e., they are capable of defending against malicious smart contracts, as well as protecting vulnerable contracts. Securify~\cite{tsankov2018securify}, Oyente~\cite{luu2016making}, ZEUS~\cite{kalra2018zeus}, SmartCheck~\cite{tikhomirov2018smartcheck}, SmartPulse~\cite{stephens2021smartpulse}, SmarTest~\cite{so2021smartest}, and MythX~\cite{mythx} are solutions with a bidirectional vector (malicious \emph{or} victim contract) threat model.

Fig.~\ref{fig:threat-model-pie} shows the breakdown of different threat models among the threat mitigation methods. We find that 41 (30.8\%) methods assume vulnerable contracts, 7 (5.3\%) imply the malicious contract model, and 85 (63.9\%) assume both these vectors. \rev{As we can see, the pure malicious smart contract threat model is underrepresented among the threat mitigation solutions. This suggests that attacks on vulnerable smart contracts are generally perceived as more widespread than the cases of malicious contracts attacking users.} This finding is corroborated by the study by Zhou et al.~\cite{zhou2020ever}, which confirms that the popularity of the honeypot vulnerability, associated with the malicious smart contract modality, is fourth after call injection, call-after-destruct, and airdrop-hunting vulnerabilities, which all assume the victim smart contract threat model.

\section{Design Workflows of Threat Mitigation Methods}\label{sec:design-comparison}
In this section, we scrutinize the designs of the threat mitigation solutions by synthesizing the uniform workflows for all the eight core methods, i.e., static analysis (\S\ref{sec:workflow-static-analysis}), symbolic execution (\S\ref{sec:workflow-symbolic-execution}), fuzzing (\S\ref{sec:workflow-fuzzing}), formal analysis (\S\ref{sec:workflow-formal-analysis}), machine learning (\S\ref{sec:workflow-ml}), execution tracing (\S\ref{sec:workflow-execution-tracing}), code synthesis (\S\ref{sec:workflow-code-synthesis}), and transaction interception (\S\ref{sec:workflow-tx-interception}). Figs.~\ref{fig:sa-workflow}---\ref{fig:ti-workflow} depict the workflows of the eight core methods. Each of these eight workflows utilizes a set of uniform elements: modules, data entities, flows (arrows), and environments. This set of elements allows us to concisely summarize and demystify the wide variety of implementations of smart contract threat mitigation solutions using the aforementioned set of uniform conventions.

The modules (green rectangles) represent items that \emph{do} something, i.e., algorithms, data filters, etc. Modules can be mandatory, i.e., pertaining to any solution with the given core method (solid borders) or optional/augmenting, i.e., implemented by some solutions employing the given core method (dashed borders). The data entities (blue rectangles) represent pieces of data or abstract data structures. The flows, depicted as arrows, show data or execution transitions. Environments (red rectangles) allow grouping of certain elements into single logical modules.

\cbox{\textbf{Lessons learned:} By manually examining the workflows of all the 133 threat mitigation solutions, we learned that every component exhibits a certain degree of generalization. For example, an element called ``smart contract'' is a more general form of what could also be denoted as ``source code'' or ``bytecode''. Thus, one of the challenges we face when synthesizing the workflows is to equate the generalizations of similar workflow elements.}

\subsection{Static Analysis Workflow}\label{sec:workflow-static-analysis}

\begin{figure}[ht!]
    \centering
    \includegraphics[width=0.9\textwidth]{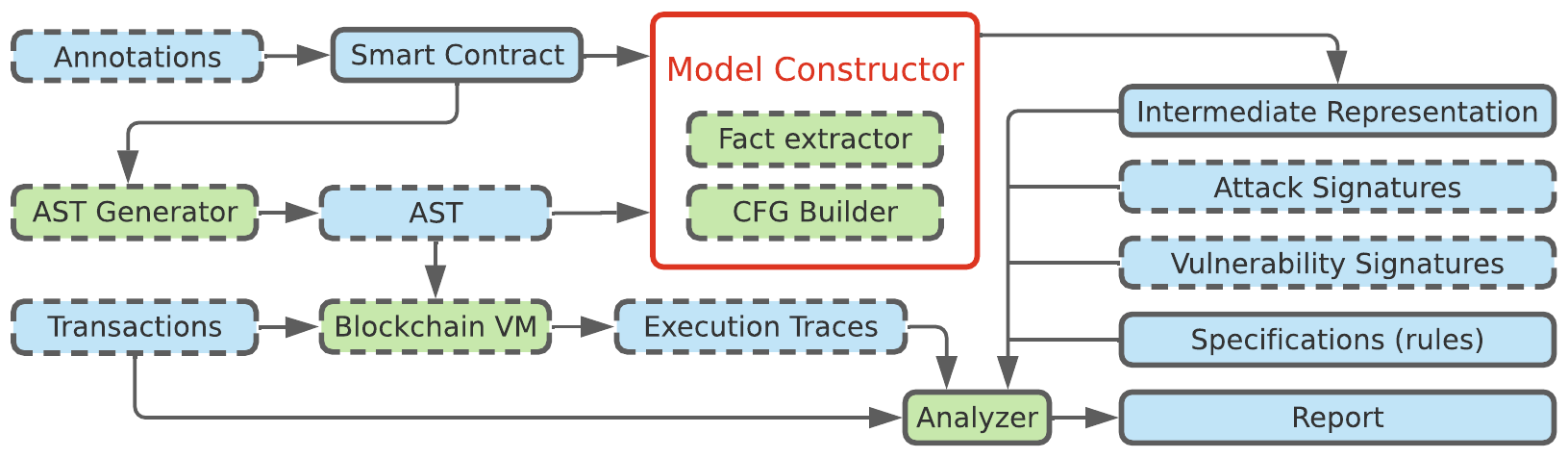}
    \caption{Workflow of the static analysis core method.}
    \label{fig:sa-workflow}
\end{figure}

The static analysis methods apply automated data filtering and syntax analysis techniques to the input. Static analysis methods detect vulnerabilities by extracting information (facts) from the source code or bytecode of a smart contract. Fig.~\ref{fig:sa-workflow} shows the general workflow of static analysis methods.

The static analysis methods take bytecode (e.g., Erays~\cite{zhou2018erays}, Vandal~\cite{brent2018vandal}, MadMax~\cite{grech2018madmax}) or source code (e.g., S-GRAM~\cite{liu2018s}, SmartCheck~\cite{tikhomirov2018smartcheck}, Slither~\cite{feist2019slither}) of a smart contract as an input, while some solutions also analyze previously executed transactions gathered from the chain data (e.g., EasyFlow~\cite{gao2019easyflow}, Zhou et al.~\cite{zhou2020ever}). A large part of the static analysis process is devoted to constructing a model in the form of one or a set of abstract data structures (ADS) that constitute a suitable (and efficient) input for the static analyzer. Control flow graph (CFG) is a popular type of such an ADS, which is utilized by Securify~\cite{tsankov2018securify}, Erays~\cite{zhou2018erays}, and Vandal~\cite{brent2018vandal}, to name a few. The built model, data (in the form of some intermediate representation, e.g., a graph), and a set of pre-defined or user-specified specifications are then directed to the static analyzer, which produces a human-readable security assessment report.

% talk about intermediate representation.

\rev{The input of a static analysis tool is a smart contract. The smart contract may be annotated with some additional metadata for clarifying semantics, such as purpose of a function~\cite{grishchenko2018ethertrust}. The contract is then converted into an intermediate representation (IR) by a model constructor, which sometimes includes manual fact extraction~\cite{brent2018vandal,zhou2018security} (using regular expressions or parsing), followed by construction of a control flow graph (CFG)~\cite{zhou2018erays,feist2019slither}. The goal of the intermediate representation is to create an accurate and predictably structured manifestation of a smart contract suitable for further analysis. The core of a static analysis solution is the analyzer, which often uses multiple data inputs besides intermediate representation, such as blockchain transactions~\cite{torres2019art}, execution traces of a smart contract produced by an instrumented node~\cite{zhang2020txspector}, or signatures of vulnerabilities in a form of formal rules or manual formal security specifications (often called ``security patterns''~\cite{tsankov2018securify}). In all cases, the analyzer produces a human readable report as its output.}

\subsection{Symbolic Execution
Workflow}\label{sec:workflow-symbolic-execution}

\begin{figure}[ht!]
    \centering
    \includegraphics[width=\textwidth]{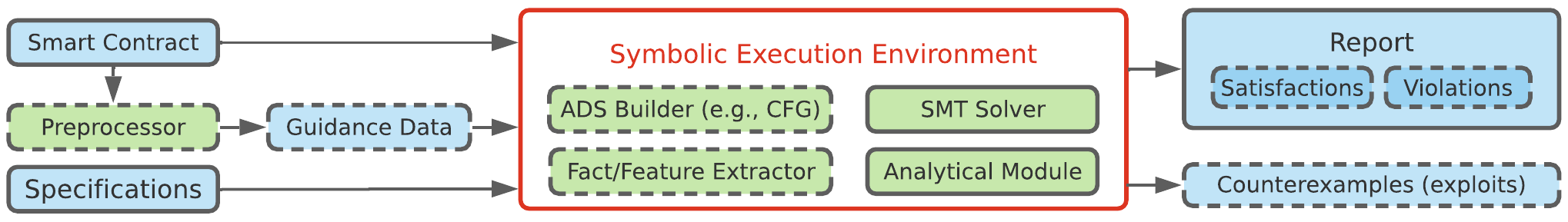}
    \caption{Workflow of the symbolic execution core method.}
    \label{fig:se-workflow}
\end{figure}

Symbolic execution methods~\cite{king1976symbolic} simulate the execution of a smart contract in a way that the actual inputs are replaced with special traceable symbolic parameters. Fig.~\ref{fig:se-workflow} depicts the general workflow of symbolic execution methodology. These methods use smart contract bytecode and a set of specifications as an input. In some cases, the specifications are part of the tool (e.g., Oyente~\cite{luu2016making}, Mythril~\cite{mueller2018smashing}, teEther~\cite{krupp2018teether}, Osiris~\cite{torres2018osiris}), in other cases, the specifications are expected to be provided by the user (e.g., Maian~\cite{nikolic2018finding}). Symbolic execution methods execute smart contracts with traceable (symbolic) parameters in lieu of actual inputs, which allows to prove or disprove some presumptions about smart contracts. Specifically, symbolic execution can answer questions about the possibility of execution of a certain block of code (reachability), the ability to invoke a certain execution path, or the ability to satisfy certain constraints. Similar to static analysis, symbolic execution often involves building a search-efficient data structure, such as CFG, as well as extracting facts and features from the input. However, unlike static analysis, the symbolic execution methods run the code instead of analyzing its syntax. All the existing symbolic execution solutions surveyed in this work employ the Z3~\cite{z3} SMT solver.
% Talk about the analytical module (post-analysis)
% talk about the result.

Some symbolic execution solutions use certain augmentations to the basic design by adding additional features. Oyente~\cite{luu2016making}, teEther~\cite{krupp2018teether}, SafePay~\cite{li2020safepay}, Artemis~\cite{wang2020artemis}, and DEFECTCHECKER~\cite{chen2021defectchecker} process the smart contract to build a CFG. Another augmentation observed in symbolic execution solutions is the production of exploits (sample inputs revealing vulnerabilities), as can be seen in teEther~\cite{krupp2018teether} and EthBMC~\cite{frank2020ethbmc}. Moreover, some symbolic execution methods perform a preliminary analysis (preprocessing) for generating guidance data facilitating the symbolic execution. SmarTest~\cite{so2021smartest} guides symbolic execution with a language-based model in order to achieve higher accuracy and reduce the rate of timeouts.

\rev{
The input of symbolic execution solutions is a combination of smart contract code and specifications to check. Some solutions add a preprocessing stage to the smart contract to produce additional guidance data for symbolic execution, such as results of statistical calculations~\cite{so2021smartest}. The symbolic execution environment includes SMT solver, such as Z3~\cite{z3} or CVC4~\cite{DBLP:conf/cav/BarrettCDHJKRT11}
, as well as an analytical module to formulate the results. Symbolic execution often involves extraction of specific features from the code~\cite{frank2020ethbmc}, as well as building an auxiliary abstract data structure, such as a control flow graph~\cite{luu2016making} or a labeled abstract syntax tree~\cite{jin2022exgen}. Symbolic execution tools produce reports, which often include the list of satisfied rules (satisfactions), and/or the list of unsatisfied rules (violations). In some cases, symbolic execution tools records specific cases of security violations called exploits~\cite{krupp2018teether} or counter-examples~\cite{permenev2020verx}.
}

\subsection{Fuzzing Workflow}\label{sec:workflow-fuzzing}

\begin{figure}[ht!]
    \centering
    \includegraphics[width=\textwidth]{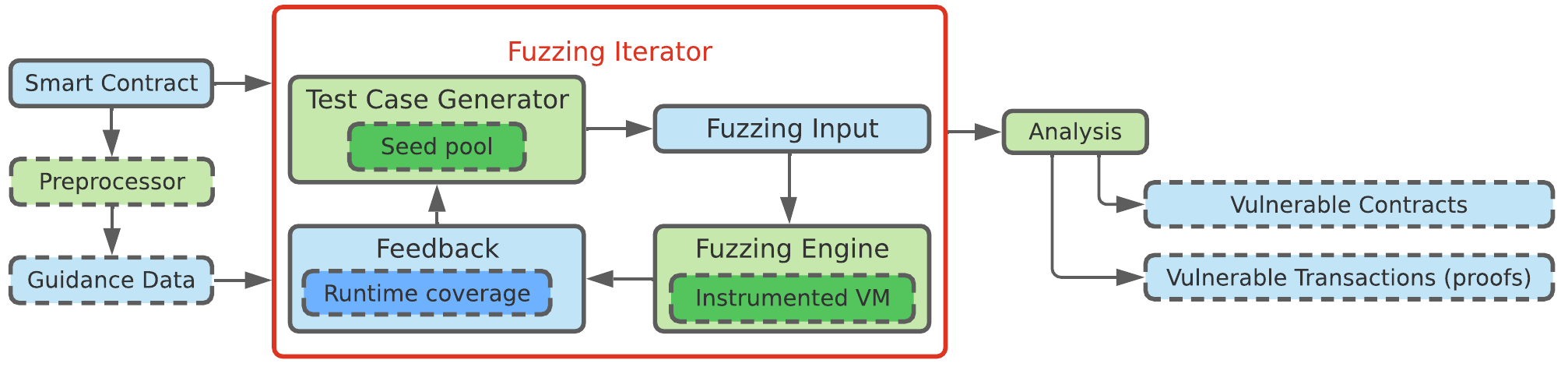}
    \caption{Workflow of the fuzzing core method.}
    \label{fig:f-workflow}
\end{figure}

Fuzzing methods use various techniques for generating subsets of test inputs that could reveal vulnerable execution paths in smart contracts. Fig.~\ref{fig:f-workflow} shows how the fuzzing core method works in smart contracts. Fuzzing tools perform iterative testing of a smart contract by generating test cases and adjusting these cases via a feedback loop. %Fuzzing core methods involve the iterative generation of smart contract test cases with a feedback loop.
The execution of smart contracts is performed by the fuzzing engine, which is either a stand-alone code interpreter or an instrumented (i.e., modified with a custom code) blockchain virtual machine. Fuzzing techniques allow to address the two notorious problems associated with software testing --- input ranges and path explosion. \rev{Even a single parameter of a smart contract function might exhibit a virtually endless range of actual values, e.g., the 256-bit integer in Ethereum. Thus, the goal of a fuzzing method is to pick input samples that are likely to reveal vulnerabilities. The path explosion problem occurs when the user needs to call a sequence of transactions. The number of possible orders of transactions or other variable scenarios might ``explode'' as the number of transactions in the sequence increases. This condition necessitates the use of special techniques, such as pruning of the transaction tree.}

Similar to symbolic execution, some fuzzing methods also utilize guidance data for facilitating test case generation. Confuzzius~\cite{ferreira2021confuzzius} performs a preprocessing in the form of taint analysis in order to guide the fuzzing engine. Also, in addition to identifying a problem in a smart contract, it is common for a fuzzing solution to deliver proof of a vulnerability in the form of a sample malicious transaction or a series thereof, as we see in ReGuard~\cite{liu2018reguard}, SoliAudit~\cite{liao2019soliaudit}, and EthPloit~\cite{zhang2020ethploit}.

\rev{
Fuzzing threat mitigation solutions take a smart contract as an input, and some of them perform preprocessing of the contract to generate additional guidance data, such as indexing tables~\cite{jiang2018contractfuzzer}. The fuzzing iterator implements a test case generator for creating the next input to a fuzzing engine. Fuzzing engines often use a blockchain virtual machine, such as EVM~\cite{wood2014ethereum}, to test transactions (or series thereof) with the inputs produced by the test case generator~\cite{kolluri2019exploiting,fu2019evmfuzzer}. The fuzzing engine executes a test transaction, and sends the corresponding state transition back to the test case generator, possibly with runtime coverage~\cite{ferreira2021confuzzius} --- in order to adjust the generation of the next test case from the seed pool. The multiple fuzzing iterations are then analyzed to generate the list of vulnerable contracts and possibly the lists of test transactions that demonstrate vulnerabilities.
}

\subsection{Formal Analysis Workflow}\label{sec:workflow-formal-analysis}

\begin{figure}[ht!]
    \centering
    \includegraphics[width=0.8\textwidth]{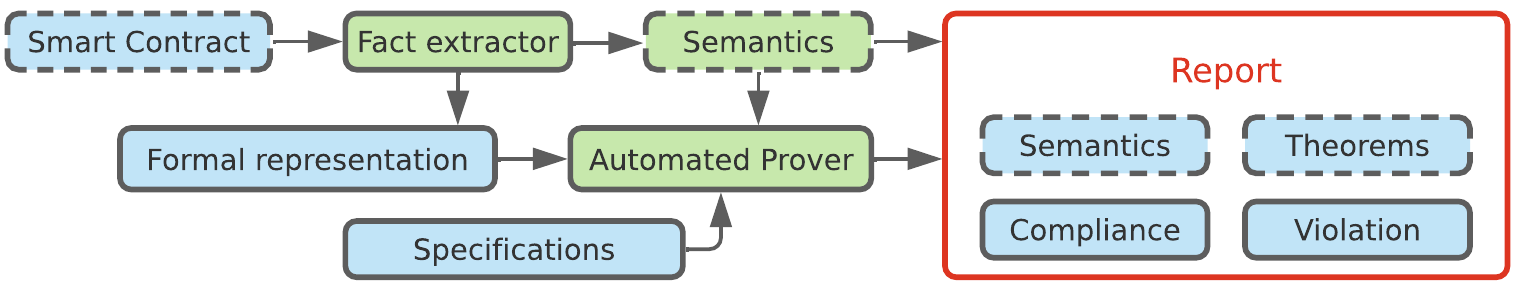}
    \caption{Workflow of the formal analysis core method.}
    \label{fig:fa-workflow}
\end{figure}

Formal analysis methods convert smart contracts into formal representations and use automated provers for deriving deterministic conclusions about the security properties of these smart contracts. Fig.~\ref{fig:fa-workflow} depicts the workflow of the smart contract formal analysis core method. One important component of a formal analysis solution is the fact extractor, which converts a smart contract into a formal representation, usually in a form of a domain-specific language (DSL). The formal representation is then delivered to an automated prover, such as Tamarin~\cite{meier2013tamarin}, along with some specifications representing vulnerabilities or security properties. The prover then juxtaposes the extracted facts with the provided properties to deliver a set of conclusions, which include compliance and violation statements. The output of a formal analysis solution may be supplemented with additional outputs. Specifically, some formal analysis solutions include the intermediate results in the report, e.g., extracted semantics, as seen in KEVM~\cite{hildenbrandt2018kevm}. Also, some solutions not only prove existing theorems, but they also produce theorems based on certain specifications, such as theorems, as we can see in Lolisa~\cite{yang2018lolisa}.

\rev{
The source code or bytecode of a smart contract is not always required for formal analysis. In some cases~\cite{hildenbrandt2018kevm}, instead of smart contract code, a corresponding formal representation is analyzed for compliance or violation of certain rules described in specifications. When a specific smart contract is present, it undergoes preliminary fact extraction procedure for developing formal representation and/or inferring semantics. The core of a formal analysis tool is the automated prover that determines compliance or violation of the formal representation of a smart contract with certain specifications. Some formal methods are capable of generating theorems~\cite{yang2018lolisa} and/or contract semantics~\cite{jiao2020semantic}, which become part of the final report.
}

\subsection{Machine Learning Workflow}\label{sec:workflow-ml}

\begin{figure}[ht!]
    \centering
    \includegraphics[width=0.8\textwidth]{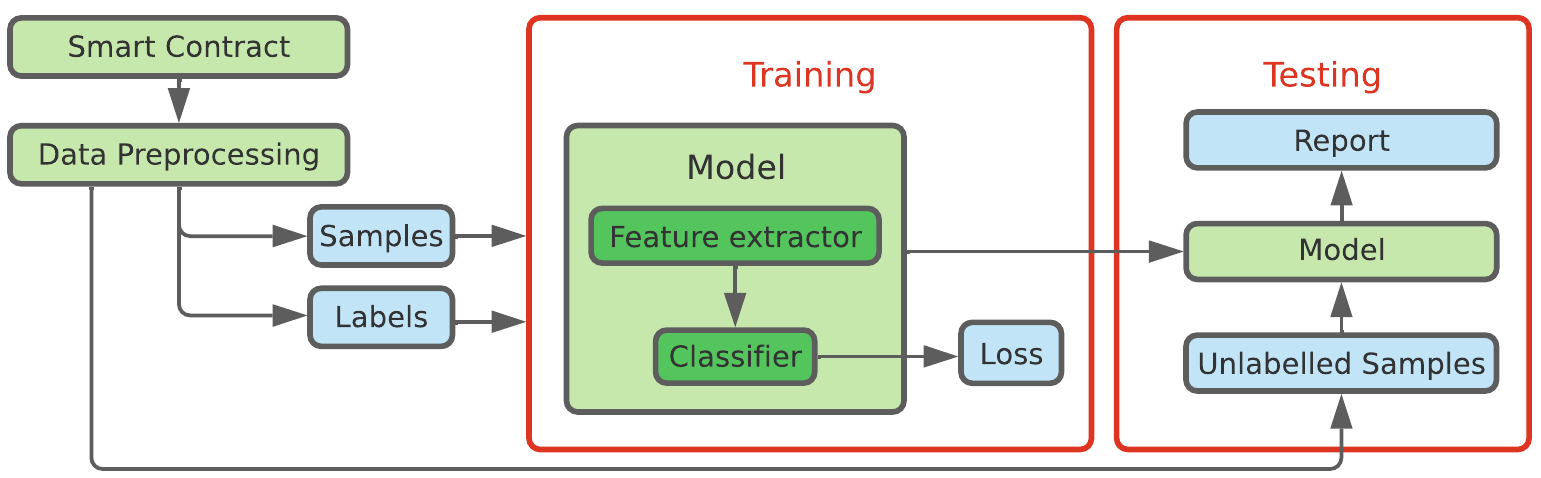}
    \caption{Workflow of the machine learning core method.}
    \label{fig:ml-workflow}
\end{figure}

Machine learning methods extract features from smart contracts or smart contract transactions and train models for classifying smart contracts based on the types of vulnerabilities discovered in them. Fig.~\ref{fig:ml-workflow} shows the general workflow of smart contract machine learning-based threat mitigation solutions. We discover that all the existing machine learning methods of smart contract threat mitigation use supervised models, requiring a subset of labeled smart contract samples. The workflow of a machine learning approach requires the data preprocessing (preparation) step, which includes building a ``clean'' (uniform) dataset, creating training and testing samples, and performing manual labeling (or using an existing one). The primary goal of the training step is to determine the parameters of a chosen model. The goal of the testing step is to verify the robustness of the model candidate. Once the model is trained and properly tested (e.g., using a K-fold method, as observed in the evaluation part of SafelyAdministrated~\cite{ivanov2021rectifying}), the model can detect vulnerabilities or confirm the safety of the unlabeled contracts or smart contract transactions.

Feature extraction and model building are two major characteristics that describe machine learning threat mitigation solutions. Momeni et al.~\cite{momeni2019machine} deliver an ML model for detecting vulnerability patterns in smart contracts, using an abstract syntax tree (AST) and control flow graph (CFG) for feature extraction. ContractWard~\cite{wang2020contractward} approaches an ML-based detection of vulnerabilities in smart contracts based on bigram features. ESCORT~\cite{lutz2021escort} is a machine learning smart contract threat mitigation solution based on a deep neural network (DNN) with a semantic-based feature extractor. AMEVulDetector~\cite{liu2021smart} builds a semantic graph from the source code and applies deep learning to building the vulnerability detection model.

\rev{
Current machine learning solutions for smart contract threat mitigation are all supervised~\cite{momeni2019machine,wang2020contractward,lutz2021escort,liu2021smart}, and some of them are hybrid solutions, used together with fuzzing~\cite{he2019learning}, static analysis~\cite{hu2021transaction}, and code synthesis~\cite{ivanov2021rectifying} workflows. Machine learning approaches, both pure and hybrid, involve processing smart contracts to generate labeled samples, followed by training and testing phases. The training involves developing a model based on features extracted from the code of smart contracts. The testing phase uses the trained model to classify the new unlabeled smart contract samples and produce the final report.
}

\subsection{Execution Tracing Workflow}\label{sec:workflow-execution-tracing}

\begin{figure}[ht!]
    \centering
    \includegraphics[width=0.9\textwidth]{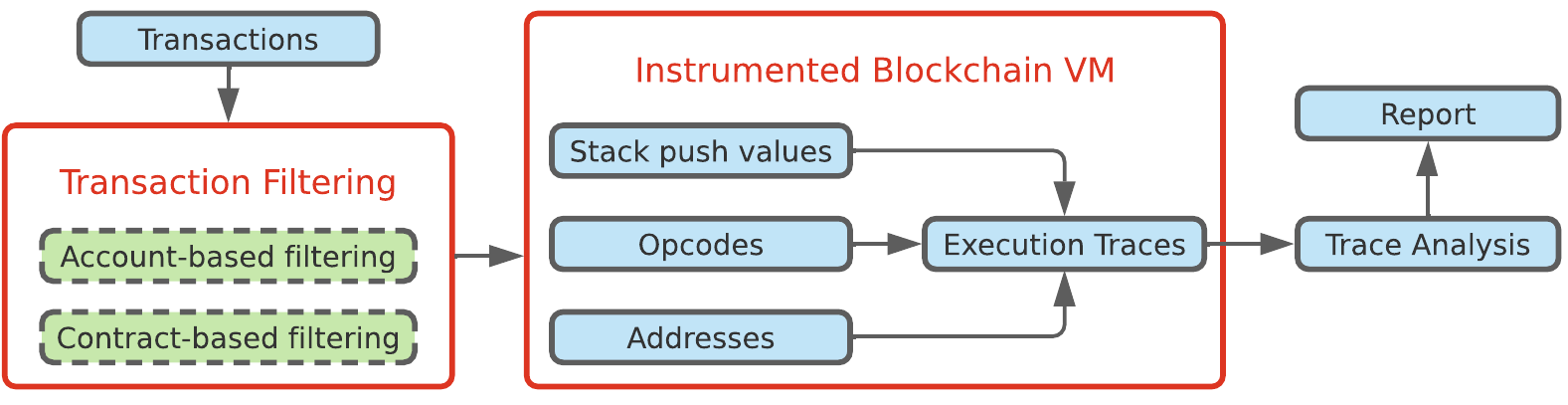}
    \caption{Workflow of the execution tracing core method.}
    \label{fig:et-workflow}
\end{figure}

Execution tracing methods assess the security properties of smart contracts by exploring the execution of transactions sent to a given smart contract or an externally owned account (in cases when the Ethereum platform is targeted\footnote{Ethereum has two types of accounts: smart contract account and externally owned account (EOA). Both EOAs and smart contract accounts can be referenced by their 160-bit public addresses.}). Fig.~\ref{fig:et-workflow} depicts the workflow of execution tracing methods. These solutions use transactions as their input. After that, the transactions are filtered to keep only the ones associated with a specific account, specific smart contract, or a concrete action (e.g., attack). Next, the filtered transactions are executed by the instrumented blockchain virtual machine (e.g., EVM). The instrumented code passively observes the execution of the given transactions and produces a special data structure called \emph{execution traces}. Formally, an execution trace is a path in a control flow graph (CFG) of a smart contract that describes the execution of a specific transaction (or a sequence of transactions). The execution traces are then analyzed to produce a human-readable report.

EthScope~\cite{wu2020ethscope} is a security analysis framework that detects suspicious smart contracts in three steps: collecting related blockchain states, replaying transactions, and reporting data for manual introspective analysis. Perez et al.~\cite{perez2021smart} propose an automated execution tracing framework for Ethereum for detecting both vulnerabilities and actual attacks exploiting these vulnerabilities. DEFIER~\cite{su2021evil} is a tool for the investigation of attack instances associated with Ethereum decentralized applications (DApps), which use Ethereum transaction tracing. Horus~\cite{ferreira2021eye} is an execution tracing framework for the detection and investigation of attacks on smart contracts that use logic-based and graph-based analyses of Ethereum transactions. Another execution tracing solution is \mbox{E-EVM}~\cite{norvill2018visual} that performs emulation and visualization of smart contracts.

%BlockEye~\cite{wang2021blockeye} 

\rev{
Execution tracing tools use historic~\cite{chen2018detecting} or online (run-time) transactions~\cite{rodler2018sereum} for tracing on an instrumented (i.e., manually modified) blockchain virtual machine. The input transactions are typically filtered by certain criteria, such as sender account address~\cite{ferreira2021eye} or destination contract~\cite{chen2019tokenscope}. The transactions are then executed, one opcode at a time, and the results of execution are recorded in execution traces. After obtaining the execution traces, they undergo the final analysis (e.g., checking for signatures of vulnerability exploitation), followed by a final report. 
}

\subsection{Code Synthesis Workflow}\label{sec:workflow-code-synthesis}

\begin{figure}[ht!]
    \centering
    \includegraphics[width=0.9\textwidth]{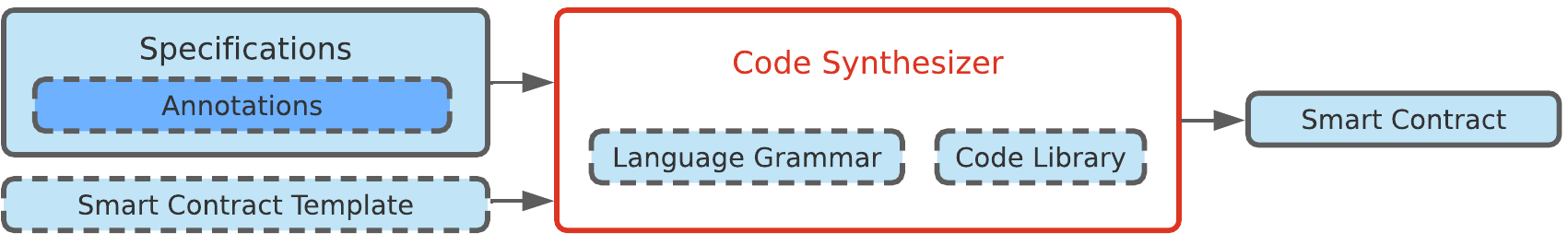}
    \caption{Workflow of the code synthesis core method.}
    \label{fig:cs-workflow}
\end{figure}

The code synthesis methods produce the source code or bytecode of a smart contract with or without a template. The objective of code synthesis methods is to produce a smart contract resistant to specific attacks or vulnerabilities. Fig.~\ref{fig:cs-workflow} shows the workflow of the code synthesis core method. We observe that some code synthesis solutions produce code from specifications only; others require a template to apply specifications to (e.g., ContractLarva~\cite{ellul2018runtime}). Custom source code annotations are an example of specifications, as we can see in Cecchetti et al.~\cite{cecchetti2020securing}.

Some code synthesis solutions utilize language Backus–Naur Form (BNF) grammars or custom code libraries (e.g., SafelyAdministrated~\cite{ivanov2021rectifying} and OpenZeppelin Contracts~\cite{openzeppelin-contracts}) to aid the process. The result of code synthesis is a source code or a bytecode of a smart contract with specific security properties. In addition, some threat mitigation solutions utilize the code synthesis core method to patch vulnerable smart contracts on the bytecode level (e.g., SmartShield~\cite{zhang2020smartshield}).

\rev{
Code synthesis methods produce smart contracts that meet certain security specifications. In some cases, security specifications are provided as annotations inside (or alongside) a template smart contract~\cite{breidenbach2017depth,ivanov2021rectifying}. The specifications are processed by the code specifier, which typically includes a BNF grammar of the target language~\cite{biryukov2017findel}, as well as libraries of coding patterns (e.g., OpenZeppelin Contracts~\cite{openzeppelin-contracts}).
}

\subsection{Transaction Interception Workflow}\label{sec:workflow-tx-interception}

\begin{figure}[ht!]
    \centering
    \includegraphics[width=0.8\textwidth]{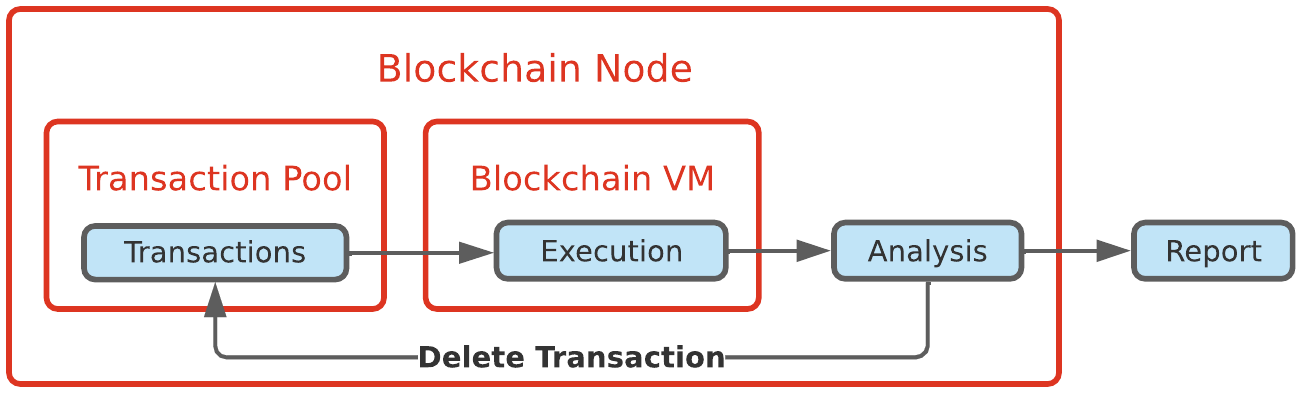}
    \caption{Workflow of the transaction interception core method.}
    \label{fig:ti-workflow}
\end{figure}

A blockchain network is a set of peer-to-peer (P2P) nodes. In this type of workflow, we assume that each node sustains the entire copy of the blockchain, i.e., we assume that the blockchain node is a \emph{full node}. Furthermore, each node has a \emph{transaction pool}, which is a queue of transactions-candidates for addition to the blockchain. Transaction interception methods are dynamic approaches that read submitted transactions from the transaction pool of the blockchain node and prevent the node from including unsafe transactions in the blockchain. Fig.~\ref{fig:ti-workflow} shows the general workflow of the transaction interception core method. Transaction interception methods employ the blockchain P2P node instrumentation, which means that there is a custom code injected into the routines responsible for transaction ordering or smart contract execution. \rev{All the transaction interception solutions surveyed in this work also produce a human-readable report of their operation. This measure is reasonable because deleting transactions from the pool is a deep intervention into the blockchain network protocol, so it must leave a log of the action.}

Transaction interception solutions, although not numerous, exhibit a diverse spectrum of approaches. SODA~\cite{chen2020soda} is a transaction-interception framework for EVM-compatible platforms that allows users to develop custom apps for dynamic defense against attacks. {\AE}GIS~\cite{ferreira2020aegis} is another transaction interception solution that uses a committee of voting security experts to create and approve attack patterns that steer transaction interception by instrumented nodes. Another transaction interception solution is EVM*~\cite{ma2019evm}, which monitors overflows and timestamp bugs.

\rev{
The input of a transaction interception solution is the metadata obtained from the transaction pool~\cite{mempool} of a blockchain node. Typically, a blockchain virtual machine (e.g., EVM) retrieves a transaction from the pool and attempts to execute it~\cite{chen2020soda,ferreira2020aegis}. The transaction interception solutions instrument the node by adding a transaction security analysis component, which can remove an unsafe transaction from the pool, if needed. The report is produced upon transaction removal to ensure transparency.
}
\section{Vulnerability Coverage}\label{sec:coverage}
In this section, we compare threat mitigation solutions from the perspective of their ability to address the known smart contract vulnerabilities. \rev{First, we select 38 solutions that explicitly declare the list of vulnerabilities they cover. Next, we translate the information about these vulnerabilities into the model adopted by the popular SWC Registry~\cite{swcregistry}.} Then we build the vulnerability map, presented in Table~\ref{table:coverage}, which juxtaposes the threat mitigation methods by their ability to address the 37 known smart contract vulnerabilities. The first column of the table has the names of the threat mitigation solutions and corresponding references; if the names are not available, we use the author names instead. The next 37 columns each correspond to the numbered SWC Registry vulnerabilities. Thus, the table constitutes a compact map showing which vulnerabilities are supported (i.e., defended against), which ones are partially supported, and which ones are not supported at all for each of the 38 threat mitigation methods.

The challenge of this approach lies in the fact that different threat mitigation solutions refer to the same vulnerabilities using different names. Moreover, some solutions refer to a group of SWC vulnerabilities as a single weakness. Rodler et al.~\cite{rodler2018sereum} declare the coverage of three vulnerabilities, which correspond to the single reentrancy vulnerability in the SWC Registry, viz., SWC-107~\cite{swc107}. Some other solutions do the opposite: they break down a single SWC vulnerability into several fine-grained subgroups. For instance, the SWC-100~\cite{swc100} and SWC-108~\cite{swc108} vulnerabilities are often treated as a single vulnerability called the ``\texttt{private} modifier'', as we can see in SmartCheck~\cite{tikhomirov2018smartcheck} and in SolidityCheck~\cite{zhang2019soliditycheck}.
%continue & cite.

Table~\ref{table:coverage} unambiguously demonstrates that different vulnerabilities exhibit unequal attention from different threat mitigation solutions. For example, 24 solutions declare defense against reentrancy (SWC-107~\cite{swc107}), whereas none of the solutions declare defense against shadowing the state variables (SWC-119~\cite{swc119}) and RTL-override control character (SWC-130~\cite{swc130}). Remarkably, we observe that both of the vulnerabilities exhibiting close attention by the existing threat mitigation solutions as well as the ones overlooked by these solutions
%the threat mitigation solutions 
are often particularly challenging to pinpoint.

\cbox{\textbf{Lessons learned:} By studying the vulnerability coverage by smart contract threat mitigation solutions, we discovered that some vulnerabilities are covered by multiple threat mitigation solutions. In contrast, many vulnerabilities are not covered by any solutions.}

\begin{table}
\setlength{\tabcolsep}{2.7pt}
    \centering
    \caption{Summary of the defense tools against smart contract vulnerabilities.}\label{table:coverage}
    
\begin{tabular}{c||p{1.0mm}p{1.0mm}p{1.0mm}p{1.0mm}p{1.0mm}p{1.0mm}p{1.0mm}p{1.0mm}p{1.0mm}p{1.0mm}p{1.0mm}p{1.0mm}p{1.0mm}p{1.0mm}p{1.0mm}p{1.0mm}p{1.0mm}p{1.0mm}p{1.0mm}p{1.0mm}p{1.0mm}p{1.0mm}p{1.0mm}p{1.0mm}p{1.0mm}p{1.0mm}p{1.0mm}p{1.0mm}p{1.0mm}p{1.0mm}p{1.0mm}p{1.0mm}p{1.0mm}p{1.0mm}p{1.0mm}p{1.0mm}p{1.0mm}}
        
        \arrayrulecolor{black}\toprule
        
        % ============ UPPER HEADERS =================
        \multicolumn{1}{c}{} &
        \multicolumn{37}{c}{\textit{Vulnerability (SWC Registry Number)}$\mathrm{^\dagger}$} \\
        % \cmidrule(lr){4-5} 
        \cmidrule(lr){2-38}

        % ============ LOWER HEADERS =================
        \textbf{\small Threat Mitigation} &
        \scriptsize \parbox[t]{1mm}{\multirow{2}{*}{\rotatebox[origin=c]{90}{ 100 }}} &
        \scriptsize \parbox[t]{1mm}{\multirow{2}{*}{\rotatebox[origin=c]{90}{ 101 }}} &
        \scriptsize \parbox[t]{1mm}{\multirow{2}{*}{\rotatebox[origin=c]{90}{ 102 }}} &
        \scriptsize \parbox[t]{1mm}{\multirow{2}{*}{\rotatebox[origin=c]{90}{ 103 }}} &
        \scriptsize \parbox[t]{1mm}{\multirow{2}{*}{\rotatebox[origin=c]{90}{ 104 }}} &
        \scriptsize \parbox[t]{1mm}{\multirow{2}{*}{\rotatebox[origin=c]{90}{ 105 }}} &
        \scriptsize \parbox[t]{1mm}{\multirow{2}{*}{\rotatebox[origin=c]{90}{ 106 }}} &
        \scriptsize \parbox[t]{1mm}{\multirow{2}{*}{\rotatebox[origin=c]{90}{ 107 }}} &
        \scriptsize \parbox[t]{1mm}{\multirow{2}{*}{\rotatebox[origin=c]{90}{ 108 }}} &
        \scriptsize \parbox[t]{1mm}{\multirow{2}{*}{\rotatebox[origin=c]{90}{ 109 }}} &
        \scriptsize \parbox[t]{1mm}{\multirow{2}{*}{\rotatebox[origin=c]{90}{ 110 }}} &
        \scriptsize \parbox[t]{1mm}{\multirow{2}{*}{\rotatebox[origin=c]{90}{ 111 }}} &
        \scriptsize \parbox[t]{1mm}{\multirow{2}{*}{\rotatebox[origin=c]{90}{ 112 }}} &
        \scriptsize \parbox[t]{1mm}{\multirow{2}{*}{\rotatebox[origin=c]{90}{ 113 }}} &
        \scriptsize \parbox[t]{1mm}{\multirow{2}{*}{\rotatebox[origin=c]{90}{ 114 }}} &
        \scriptsize \parbox[t]{1mm}{\multirow{2}{*}{\rotatebox[origin=c]{90}{ 115 }}} &
        \scriptsize \parbox[t]{1mm}{\multirow{2}{*}{\rotatebox[origin=c]{90}{ 116 }}} &
        \scriptsize \parbox[t]{1mm}{\multirow{2}{*}{\rotatebox[origin=c]{90}{ 117 }}} &
        \scriptsize \parbox[t]{1mm}{\multirow{2}{*}{\rotatebox[origin=c]{90}{ 118 }}} &
        \scriptsize \parbox[t]{1mm}{\multirow{2}{*}{\rotatebox[origin=c]{90}{ 119 }}} &
        \scriptsize \parbox[t]{1mm}{\multirow{2}{*}{\rotatebox[origin=c]{90}{ 120 }}} &
        \scriptsize \parbox[t]{1mm}{\multirow{2}{*}{\rotatebox[origin=c]{90}{ 121 }}} &
        \scriptsize \parbox[t]{1mm}{\multirow{2}{*}{\rotatebox[origin=c]{90}{ 122 }}} &
        \scriptsize \parbox[t]{1mm}{\multirow{2}{*}{\rotatebox[origin=c]{90}{ 123 }}} &
        \scriptsize \parbox[t]{1mm}{\multirow{2}{*}{\rotatebox[origin=c]{90}{ 124 }}} &
        \scriptsize \parbox[t]{1mm}{\multirow{2}{*}{\rotatebox[origin=c]{90}{ 125 }}} &
        \scriptsize \parbox[t]{1mm}{\multirow{2}{*}{\rotatebox[origin=c]{90}{ 126 }}} &
        \scriptsize \parbox[t]{1mm}{\multirow{2}{*}{\rotatebox[origin=c]{90}{ 127 }}} &
        \scriptsize \parbox[t]{1mm}{\multirow{2}{*}{\rotatebox[origin=c]{90}{ 128 }}} &
        \scriptsize \parbox[t]{1mm}{\multirow{2}{*}{\rotatebox[origin=c]{90}{ 129 }}} &
        \scriptsize \parbox[t]{1mm}{\multirow{2}{*}{\rotatebox[origin=c]{90}{ 130 }}} &
        \scriptsize \parbox[t]{1mm}{\multirow{2}{*}{\rotatebox[origin=c]{90}{ 131 }}} &
        \scriptsize \parbox[t]{1mm}{\multirow{2}{*}{\rotatebox[origin=c]{90}{ 132 }}} &
        \scriptsize \parbox[t]{1mm}{\multirow{2}{*}{\rotatebox[origin=c]{90}{ 133 }}} &
        \scriptsize \parbox[t]{1mm}{\multirow{2}{*}{\rotatebox[origin=c]{90}{ 134 }}} &
        \scriptsize \parbox[t]{1mm}{\multirow{2}{*}{\rotatebox[origin=c]{90}{ 135 }}} &
        \scriptsize \parbox[t]{1mm}{\multirow{2}{*}{\rotatebox[origin=c]{90}{ 136 }}} \\
        \textbf{\small Solution} & & & & & & & & & & & & & & & & & & & & & & & & & & & & & & & & & & & & & \\
        \arrayrulecolor{black}\midrule
        
        % ============ DATA =================

        \gc Oyente~\cite{luu2016making} & \gc \circI & \gc \circI & \gc \circI & \gc \circI & \gc \circI & \gc \circI &
        \gc \circI & \gc \circIII & \gc \circI & \gc \circI & \gc \circI & \gc \circI & \gc \circI & \gc \circIII & \gc \circIII & \gc \circI & \gc \circIII & \gc \circI & \gc \circI & \gc \circI & \gc \circI & \gc \circI & \gc \circI & \gc \circI & \gc \circI & \gc \circI & \gc \circI & \gc \circI & \gc \circI & \gc \circI & \gc \circI & \gc \circI & \gc \circI & \gc \circI & \gc \circI & \gc \circI & \gc \circI ~ \\
        
        Securify~\cite{tsankov2018securify} & \circI & \circI & \circI & \circI & \circIII & \circII & \circIII & \circIII & \circI & \circI & \circI & \circI & \circI & \circI & \circIII & \circI & \circI & \circI & \circI & \circI & \circI & \circI & \circI & \circIII & \circIII & \circI & \circI & \circI & \circI & \circI & \circI & \circI & \circI & \circI & \circI & \circI & \circI ~ \\
        
        \gc Mythril~\cite{mueller2018smashing} & \gc \circI & \gc \circIII & \gc \circI & \gc \circI & \gc \circIII & \gc \circI & 
        \gc \circIII & \gc \circIII & \gc \circI & \gc \circI & \gc \circIII & \gc \circI & \gc \circIII & \gc \circI & \gc \circIII & \gc \circI & \gc \circIII & \gc \circI & \gc \circII & \gc \circI & \gc \circI & \gc \circI & \gc \circI & \gc \circI & \gc \circI & \gc \circI & \gc \circI & \gc \circI & \gc \circI & \gc \circI & \gc \circI & \gc \circI & \gc \circI & \gc \circI & \gc \circI & \gc \circI & \gc \circI ~
        \\
        
        Sereum~\cite{rodler2018sereum} & \circI & \circI & \circI & \circI & \circI & \circI &
        \circI & \circIII & \circI & \circI & \circI & \circI & \circI & \circI & \circI & \circI & \circI & \circI & \circI & \circI & \circI & \circI & \circI & \circI & \circI & \circI & \circI & \circI & \circI & \circI & \circI & \circI & \circI & \circI & \circI & \circI & \circI ~
        \\        
        
        \gc Vandal~\cite{brent2018vandal} & \gc \circI & \gc \circI & \gc \circI & \gc \circI & \gc \circI & \gc \circIII &
        \gc \circIII & \gc \circIII & \gc \circI & \gc \circI & \gc \circI & \gc \circI & \gc \circI & \gc \circI & \gc \circI & \gc \circIII & \gc \circI & \gc \circI & \gc \circIII & \gc \circI & \gc \circI & \gc \circI & \gc \circI & \gc \circI & \gc \circI & \gc \circI & \gc \circI & \gc \circI & \gc \circI & \gc \circI & \gc \circI & \gc \circI & \gc \circI & \gc \circI & \gc \circII & \gc \circI & \gc \circI ~
        \\
        
        sGuard~\cite{nguyen2021sguard} & \circI & \circIII & \circI & \circI & \circI & \circI &
        \circI & \circIII & \circI & \circI & \circI & \circI & \circI & \circI & \circI & \circIII & \circI & \circI & \circI & \circI & \circI & \circI & \circI & \circI & \circI & \circI & \circI & \circI & \circI & \circI & \circI & \circI & \circI & \circI & \circI & \circI & \circI ~
        \\        
        
        \gc ZEUS~\cite{kalra2018zeus} & \gc \circI & \gc \circIII & \gc \circI & \gc \circI & \gc \circI & \gc \circI &
        \gc \circI & \gc \circIII & \gc \circI & \gc \circI & \gc \circI & \gc \circI & \gc \circI & \gc \circI & \gc \circIII & \gc \circIII & \gc \circIII & \gc \circI & \gc \circI & \gc \circI & \gc \circI & \gc \circI & \gc \circI & \gc \circI & \gc \circI & \gc \circI & \gc \circI & \gc \circI & \gc \circI & \gc \circI & \gc \circI & \gc \circI & \gc \circI & \gc \circI & \gc \circIII & \gc \circI & \gc \circI ~
        \\
        
        ConFuzzius~\cite{ferreira2021confuzzius} &  \circI &  \circIII &  \circI &  \circI &  \circIII &  \circI & 
         \circIII &  \circIII &  \circI &  \circI &  \circIII &  \circI &  \circIII &  \circI &  \circIII &  \circI &  \circIII &  \circI &  \circII &  \circI &  \circI &  \circI &  \circI &  \circI &  \circI &  \circI &  \circI &  \circI &  \circI &  \circI &  \circI &  \circI &  \circI &  \circI &  \circI &  \circI &  \circI ~
        \\        
        
        \gc VeriSmart~\cite{so2020verismart} & \gc \circI & \gc \circIII & \gc \circI & \gc \circI & \gc \circI & \gc \circI &
        \gc \circI & \gc \circI & \gc \circI & \gc \circI & \gc \circI & \gc \circI & \gc \circI & \gc \circI & \gc \circI & \gc \circI & \gc \circI & \gc \circI & \gc \circI & \gc \circI & \gc \circI & \gc \circI & \gc \circI & \gc \circI & \gc \circI & \gc \circI & \gc \circI & \gc \circI & \gc \circI & \gc \circI & \gc \circI & \gc \circI & \gc \circI & \gc \circI & \gc \circI & \gc \circI & \gc \circI ~
        \\
        
        SmarTest~\cite{so2021smartest} & \circI & \circIII & \circI & \circI & \circI & \circIII &
        \circIII & \circI & \circI & \circI & \circIII & \circI & \circI & \circI & \circI & \circI & \circI & \circI & \circI & \circI & \circI & \circI & \circI & \circII & \circI & \circI & \circI & \circI & \circI & \circI & \circI & \circI & \circI & \circI & \circI & \circI & \circI ~
        \\

        \gc Maian~\cite{nikolic2018finding} & \gc \circI & \gc \circI & \gc \circI & \gc \circI & \gc \circI & \gc \circIII &
        \gc \circIII & \gc \circI & \gc \circI & \gc \circI & \gc \circI & \gc \circI & \gc \circI & \gc \circI & \gc \circI & \gc \circI & \gc \circI & \gc \circI & \gc \circI & \gc \circI & \gc \circI & \gc \circI & \gc \circI & \gc \circI & \gc \circI & \gc \circI & \gc \circI & \gc \circI & \gc \circI & \gc \circI & \gc \circI & \gc \circI & \gc \circI & \gc \circI & \gc \circI & \gc \circI & \gc \circI ~
        \\

        ECFChecker~\cite{grossman2017online} & \circI & \circI & \circI & \circI & \circI & \circI &
        \circI & \circIII & \circI & \circI & \circI & \circI & \circI & \circI & \circI & \circI & \circI & \circI & \circI & \circI & \circI & \circI & \circI & \circI & \circI & \circI & \circI & \circI & \circI & \circI & \circI & \circI & \circI & \circI & \circI & \circI & \circI ~
        \\       
        
        \gc Osiris~\cite{torres2018osiris} & \gc \circI & \gc \circIII & \gc \circI & \gc \circI & \gc \circI & \gc \circI &
        \gc \circI & \gc \circI & \gc \circI & \gc \circI & \gc \circI & \gc \circI & \gc \circI & \gc \circI & \gc \circI & \gc \circI & \gc \circI & \gc \circI & \gc \circI & \gc \circI & \gc \circI & \gc \circI & \gc \circI & \gc \circI & \gc \circI & \gc \circI & \gc \circI & \gc \circI & \gc \circI & \gc \circI & \gc \circI & \gc \circI & \gc \circI & \gc \circI & \gc \circI & \gc \circI & \gc \circI ~
        \\

        FSolidM~\cite{mavridou2018designing} & \circI & \circI & \circI & \circI & \circI & \circI &
        \circI & \circIII & \circI & \circI & \circI & \circI & \circI & \circI & \circIII & \circI & \circI & \circI & \circI & \circI & \circI & \circI & \circI & \circI & \circI & \circI & \circI & \circI & \circI & \circI & \circI & \circI & \circI & \circI & \circI & \circI & \circI ~
        \\       
                
        \gc ContractFuzzer~\cite{jiang2018contractfuzzer} & \gc \circI & \gc \circI & \gc \circI & \gc \circI & \gc \circIII & \gc \circI &
        \gc \circI & \gc \circIII & \gc \circI & \gc \circI & \gc \circI & \gc \circI & \gc \circIII & \gc \circI & \gc \circI & \gc \circI & \gc \circIII & \gc \circI & \gc \circI & \gc \circI & \gc \circI & \gc \circI & \gc \circI & \gc \circI & \gc \circI & \gc \circI & \gc \circI & \gc \circI & \gc \circI & \gc \circI & \gc \circI & \gc \circI & \gc \circI & \gc \circI & \gc \circIII & \gc \circI & \gc \circI ~
        \\

        MadMax~\cite{grech2018madmax} & \circI & \circI & \circI & \circI & \circI & \circI &
        \circI & \circI & \circI & \circI & \circI & \circI & \circI & \circI & \circI & \circI & \circI & \circI & \circI & \circI & \circI & \circI & \circI & \circI & \circI & \circI & \circIII & \circI & \circIII & \circI & \circI & \circI & \circI & \circI & \circIII & \circI & \circI ~
        \\       

        \gc SmartCheck~\cite{tikhomirov2018smartcheck} & \gc \circIII & \gc \circIII & \gc \circIII & \gc \circI & \gc \circIII & \gc \circI &
        \gc \circI & \gc \circIII & \gc \circIII & \gc \circI & \gc \circI & \gc \circI & \gc \circI & \gc \circIII & \gc \circI & \gc \circIII & \gc \circIII & \gc \circI & \gc \circI & \gc \circI & \gc \circI & \gc \circI & \gc \circI & \gc \circI & \gc \circI & \gc \circI & \gc \circI & \gc \circI & \gc \circI & \gc \circI & \gc \circI & \gc \circI & \gc \circIII & \gc \circI & \gc \circI & \gc \circI & \gc \circI ~
        \\

        ReGuard~\cite{liu2018reguard} & \circI & \circI & \circI & \circI & \circI & \circI &
        \circI & \circIII & \circI & \circI & \circI & \circI & \circI & \circI & \circI & \circI & \circI & \circI & \circI & \circI & \circI & \circI & \circI & \circI & \circI & \circI & \circI & \circI & \circI & \circI & \circI & \circI & \circI & \circI & \circI & \circI & \circI ~
        \\       
        
        \gc ILF~\cite{he2019learning} & \gc \circI & \gc \circI & \gc \circI & \gc \circI & \gc \circIII & \gc \circIII &
        \gc \circIII & \gc \circI & \gc \circI & \gc \circI & \gc \circI & \gc \circI & \gc \circIII & \gc \circI & \gc \circI & \gc \circI & \gc \circIII & \gc \circI & \gc \circI & \gc \circI & \gc \circI & \gc \circI & \gc \circI & \gc \circI & \gc \circI & \gc \circI & \gc \circI & \gc \circI & \gc \circI & \gc \circI & \gc \circI & \gc \circI & \gc \circI & \gc \circI & \gc \circI & \gc \circI & \gc \circI ~
        \\

        NPChecker~\cite{wang2019detecting} & \circI & \circI & \circI & \circI & \circIII & \circI &
        \circI & \circIII & \circI & \circI & \circI & \circI & \circI & \circI & \circIII & \circI & \circIII & \circI & \circI & \circI & \circI & \circI & \circI & \circI & \circI & \circI & \circI & \circI & \circI & \circI & \circI & \circI & \circI & \circI & \circI & \circI & \circI ~
        \\       
                
        \gc EasyFlow~\cite{gao2019easyflow} & \gc \circI & \gc \circIII & \gc \circI & \gc \circI & \gc \circI & \gc \circI &
        \gc \circI & \gc \circI & \gc \circI & \gc \circI & \gc \circI & \gc \circI & \gc \circI & \gc \circI & \gc \circI & \gc \circI & \gc \circI & \gc \circI & \gc \circI & \gc \circI & \gc \circI & \gc \circI & \gc \circI & \gc \circI & \gc \circI & \gc \circI & \gc \circI & \gc \circI & \gc \circI & \gc \circI & \gc \circI & \gc \circI & \gc \circI & \gc \circI & \gc \circI & \gc \circI & \gc \circI ~
        \\

        Vultron~\cite{wang2019vultron} & \circI & \circIII & \circI & \circI & \circIII & \circI &
        \circI & \circIII & \circI & \circI & \circI & \circI & \circI & \circI & \circI & \circI & \circI & \circI & \circI & \circI & \circI & \circI & \circI & \circI & \circI & \circI & \circIII & \circI & \circI & \circI & \circI & \circI & \circI & \circI & \circI & \circI & \circI ~
        \\       
        
        \gc SoidityCheck~\cite{zhang2019soliditycheck} & \gc \circIII & \gc \circIII & \gc \circI & \gc \circI & \gc \circIII & \gc \circI &
        \gc \circI & \gc \circI & \gc \circIII & \gc \circI & \gc \circI & \gc \circI & \gc \circI & \gc \circIII & \gc \circI & \gc \circIII & \gc \circIII & \gc \circI & \gc \circIII & \gc \circI & \gc \circI & \gc \circI & \gc \circI & \gc \circI & \gc \circI & \gc \circI & \gc \circI & \gc \circI & \gc \circI & \gc \circI & \gc \circI & \gc \circI & \gc \circI & \gc \circI & \gc \circI & \gc \circI & \gc \circI ~
        \\

        GasFuzz~\cite{ma2019gasfuzz} & \circIII & \circIII & \circI & \circI & \circI & \circI &
        \circI & \circI & \circIII & \circI & \circI & \circI & \circI & \circI & \circI & \circI & \circI & \circI & \circI & \circI & \circI & \circI & \circI & \circI & \circI & \circI & \circIII & \circI & \circIII & \circI & \circI & \circI & \circI & \circI & \circI & \circI & \circI ~
        \\       
        
        \gc SolAnalyzer~\cite{akca2019solanalyser} & \gc \circI & \gc \circIII & \gc \circI & \gc \circI & \gc \circIII & \gc \circI &
        \gc \circI & \gc \circI & \gc \circI & \gc \circI & \gc \circI & \gc \circI & \gc \circI & \gc \circI & \gc \circI & \gc \circIII & \gc \circIII & \gc \circI & \gc \circI & \gc \circI & \gc \circI & \gc \circI & \gc \circI & \gc \circI & \gc \circI & \gc \circI & \gc \circIII & \gc \circI & \gc \circIII & \gc \circI & \gc \circI & \gc \circI & \gc \circI & \gc \circI & \gc \circI & \gc \circI & \gc \circI ~
        \\

        GasTap~\cite{albert2019running} & \circI & \circI & \circI & \circI & \circI & \circI &
        \circI & \circI & \circI & \circI & \circI & \circI & \circI & \circI & \circI & \circI & \circI & \circI & \circI & \circI & \circI & \circI & \circI & \circI & \circI & \circI & \circIII & \circI & \circIII & \circI & \circI & \circI & \circI & \circI & \circIII & \circI & \circI ~
        \\       
                
        \gc Momeni et al.~\cite{momeni2019machine} & \gc \circI & \gc \circIII & \gc \circI & \gc \circI & \gc \circIII & \gc \circI &
        \gc \circI & \gc \circIII & \gc \circIII & \gc \circIII & \gc \circI & \gc \circI & \gc \circI & \gc \circI & \gc \circI & \gc \circIII & \gc \circI & \gc \circI & \gc \circI & \gc \circI & \gc \circI & \gc \circI & \gc \circI & \gc \circI & \gc \circI & \gc \circI & \gc \circI & \gc \circI & \gc \circI & \gc \circI & \gc \circI & \gc \circI & \gc \circI & \gc \circI & \gc \circI & \gc \circI & \gc \circI ~
        \\

        Harvey~\cite{wustholz2020harvey} & \circI & \circI & \circI & \circI & \circI & \circI &
        \circI & \circI & \circI & \circI & \circIII & \circI & \circI & \circI & \circI & \circI & \circI & \circI & \circI & \circI & \circI & \circI & \circI & \circI & \circIII & \circI & \circI & \circI & \circI & \circI & \circI & \circI & \circI & \circI & \circI & \circI & \circI ~
        \\       
        
        \gc sFuzz~\cite{nguyen2020sfuzz} & \gc \circI & \gc \circIII & \gc \circI & \gc \circI & \gc \circIII & \gc \circI &
        \gc \circI & \gc \circIII & \gc \circI & \gc \circI & \gc \circI & \gc \circI & \gc \circIII & \gc \circI & \gc \circI & \gc \circI & \gc \circIII & \gc \circI & \gc \circI & \gc \circI & \gc \circI & \gc \circI & \gc \circI & \gc \circI & \gc \circI & \gc \circI & \gc \circIII & \gc \circI & \gc \circI & \gc \circI & \gc \circI & \gc \circI & \gc \circI & \gc \circI & \gc \circI & \gc \circI & \gc \circI ~
        \\

        Artemis~\cite{wang2020artemis} & \circI & \circI & \circI & \circI & \circI & \circI &
        \circI & \circI & \circI & \circI & \circI & \circI & \circIII & \circI & \circI & \circI & \circIII & \circI & \circI & \circI & \circI & \circI & \circI & \circI & \circI & \circI & \circIII & \circI & \circI & \circI & \circI & \circI & \circI & \circI & \circI & \circI & \circI ~
        \\       
        
        \gc EthPloit~\cite{zhang2020ethploit} & \gc \circI & \gc \circI & \gc \circI & \gc \circI & \gc \circII & \gc \circII &
        \gc \circI & \gc \circI & \gc \circI & \gc \circI & \gc \circI & \gc \circI & \gc \circI & \gc \circI & \gc \circI & \gc \circI & \gc \circI & \gc \circI & \gc \circI & \gc \circI & \gc \circI & \gc \circI & \gc \circI & \gc \circI & \gc \circI & \gc \circI & \gc \circI & \gc \circI & \gc \circI & \gc \circI & \gc \circI & \gc \circI & \gc \circI & \gc \circI & \gc \circI & \gc \circI & \gc \circIII ~
        \\

        EthScope~\cite{wu2020ethscope} & \circI & \circIII & \circI & \circI & \circI & \circI &
        \circI & \circIII & \circI & \circI & \circI & \circI & \circI & \circI & \circI & \circI & \circI & \circI & \circI & \circI & \circIII & \circI & \circI & \circI & \circI & \circI & \circI & \circI & \circI & \circI & \circI & \circI & \circI & \circI & \circI & \circI & \circI ~
        \\       
                
        \gc RA~\cite{chinen2020ra} & \gc \circI & \gc \circI & \gc \circI & \gc \circI & \gc \circI & \gc \circI &
        \gc \circI & \gc \circIII & \gc \circI & \gc \circI & \gc \circI & \gc \circI & \gc \circI & \gc \circI & \gc \circI & \gc \circI & \gc \circI & \gc \circI & \gc \circI & \gc \circI & \gc \circI & \gc \circI & \gc \circI & \gc \circI & \gc \circI & \gc \circI & \gc \circI & \gc \circI & \gc \circI & \gc \circI & \gc \circI & \gc \circI & \gc \circI & \gc \circI & \gc \circI & \gc \circI & \gc \circI ~
        \\

        SeRIF~\cite{cecchetti12compositional} & \circI & \circI & \circI & \circI & \circI & \circI &
        \circI & \circIII & \circI & \circI & \circI & \circI & \circI & \circI & \circI & \circI & \circI & \circI & \circI & \circI & \circI & \circI & \circI & \circI & \circI & \circI & \circI & \circI & \circI & \circI & \circI & \circI & \circI & \circI & \circI & \circI & \circI ~
        \\       
        
        \gc Huang et al.~\cite{huang2021hunting} & \gc \circI & \gc \circIII & \gc \circI & \gc \circI & \gc \circIII & \gc \circI &
        \gc \circI & \gc \circIII & \gc \circI & \gc \circI & \gc \circI & \gc \circI & \gc \circI & \gc \circI & \gc \circI & \gc \circI & \gc \circI & \gc \circI & \gc \circI & \gc \circI & \gc \circIII & \gc \circI & \gc \circI & \gc \circI & \gc \circI & \gc \circI & \gc \circI & \gc \circI & \gc \circI & \gc \circI & \gc \circI & \gc \circI & \gc \circI & \gc \circI & \gc \circI & \gc \circI & \gc \circI ~
        \\

        DefectChecker~\cite{chen2021defectchecker} & \circI & \circI & \circI & \circI & \circIII & \circIII &
        \circI & \circIII & \circI & \circI & \circI & \circI & \circI & \circII & \circIII & \circI & \circIII & \circI & \circI & \circI & \circI & \circI & \circI & \circI & \circI & \circI & \circI & \circI & \circI & \circI & \circI & \circI & \circI & \circI & \circI & \circI & \circI ~
        \\       
        
        \gc ExGen~\cite{jin2022exgen} & \gc \circI & \gc \circIII & \gc \circI & \gc \circI & \gc \circI & \gc \circI &
        \gc \circIII & \gc \circII & \gc \circI & \gc \circI & \gc \circI & \gc \circI & \gc \circIII & \gc \circI & \gc \circI & \gc \circI & \gc \circI & \gc \circI & \gc \circI & \gc \circI & \gc \circI & \gc \circI & \gc \circI & \gc \circI & \gc \circI & \gc \circI & \gc \circI & \gc \circI & \gc \circI & \gc \circI & \gc \circI & \gc \circI & \gc \circI & \gc \circI & \gc \circI & \gc \circI & \gc \circI ~
        \\

        MythX~\cite{mythx} & \circI & \circIII & \circI & \circI & \circIII & \circIII &
        \circIII & \circIII & \circI & \circIII & \circIII & \circI & \circIII & \circIII & \circI & \circIII & \circIII & \circI & \circI & \circI & \circIII & \circI & \circI & \circI & \circIII & \circI & \circI & \circIII & \circIII & \circI & \circI & \circI & \circI & \circI & \circIII & \circI & \circI ~
        \\       
                
        % \gc Foo~\cite{} & \gc \circI & \gc \circI & \gc \circI & \gc \circI & \gc \circI & \gc \circI &
        % \gc \circI & \gc \circI & \gc \circI & \gc \circI & \gc \circI & \gc \circI & \gc \circI & \gc \circI & \gc \circI & \gc \circI & \gc \circI & \gc \circI & \gc \circI & \gc \circI & \gc \circI & \gc \circI & \gc \circI & \gc \circI & \gc \circI & \gc \circI & \gc \circI & \gc \circI & \gc \circI & \gc \circI & \gc \circI & \gc \circI & \gc \circI & \gc \circI & \gc \circI & \gc \circI & \gc \circI ~
        % \\

        % Bar~\cite{} & \circI & \circI & \circI & \circI & \circI & \circI &
        % \circI & \circI & \circI & \circI & \circI & \circI & \circI & \circI & \circI & \circI & \circI & \circI & \circI & \circI & \circI & \circI & \circI & \circI & \circI & \circI & \circI & \circI & \circI & \circI & \circI & \circI & \circI & \circI & \circI & \circI & \circI ~
        % \\  

        % \gc TxT & \gc \circIII & \gc \circIII & \gc \circIII & \gc \circIII & \gc \circIII & \gc \circI &
        % \gc \circIII & \gc \circIII & \gc \circIII & \gc \circIII & \gc \circIII & \gc \circIII & \gc \circIII & \gc \circIII & \gc \circI & \gc \circIII & \gc \circII & \gc \circIII & \gc \circIII & \gc \circIII & \gc \circII & \gc \circIII & \gc \circIII & \gc \circIII & \gc \circIII & \gc \circIII & \gc \circIII & \gc \circIII & \gc \circIII & \gc \circIII & \gc \circIII & \gc \circIII & \gc \circIII & \gc \circIII & \gc \circI & \gc \circIII & \gc \circI ~
        % \\ % Remove partial support for SWC-136 if no case created.
                        
        % ============ LEGEND: ICONS =================
        \arrayrulecolor{black}\bottomrule
        
        \multicolumn{38}{c}{\scriptsize
        \circI~--- full support; \LEFTcircle~--- partial support; \circI~--- no support.
        } \\
        
        \arrayrulecolor{black}\hline
        
        % ============ LEGEND: SUPERSCRIPTS =================
        \multicolumn{38}{l}{$\mathrm{^\dagger}$ {\scriptsize Available at \url{https://swcregistry.io/} and \url{https://github.com/SmartContractSecurity/SWC-registry}}}
        
    \end{tabular}
\end{table}
\section{Trends and Perspectives}\label{sec:trends}
In this section, we discuss the emerging trends in smart contract threat mitigation (\S\ref{sec:dyn-tx-interception}, \S\ref{sec:ai-driven-security}, \S\ref{sec:human-machine}), the overlooked types of smart contracts (\S\ref{sec:non-ethereum-contracts}), and the necessity for data-driven studies in smart contract security (\S\ref{sec:large-scale-measurements}). To avoid speculations and opinion-based statements, we only make inferences based on our survey data and other strong evidence.

\cbox{\textbf{Lessons learned:} By exploring trends and perspectives associated with smart contract threat mitigation solutions, we discovered that there is a substantial room for future work despite the abundance of existing studies.}

\subsection{Dynamic Transaction Interception}\label{sec:dyn-tx-interception}
Most smart contract threat mitigation solutions use predominantly static code-based detection approaches. However, we note that the focus of the research community is shifting in three major directions:
\begin{enumerate}
    \item static approaches are shifting into the dynamic paradigm;
    \item the code based methods are shifting into the transaction-based ones; and
    \item the detection methods are shifting towards verification.
\end{enumerate}
Following these observations, it would be reasonable to suppose that the next generation of smart contract threat mitigation solutions will likely continue exploring the primarily overlooked area of vulnerability-agnostic dynamic transaction interception. We believe that there are two significant reasons these methods are particularly promising: they are blockchain state-aware and can address zero-day attacks.

% To demonstrate the blockchain state awareness, consider the Ethereum smart contract \texttt{Foo} in Fig.~\ref{fig:motivation}, which transfers cryptocurrency funds to a smart contract \texttt{Bar} (Fig.~\ref{fig:case21}). \texttt{Bar} is deployed on Ethereum Mainnet\footnote{Ethereum Mainnet is the major production Ethereum network supporting the Ether cryptocurrency.}, but not on Ropsten testnet\footnote{Testnets are alternative blockchain networks utilized for development and experiments. Testnets normally execute the same protocols as production networks, but the test cryptocurrency on the testnet does not have any market value.}. Moreover, \texttt{Bar} does not have any payable functions\footnote{A payable function allows to transfer (deposit) cryptocurrency to the smart contract.}, and therefore it cannot accept incoming Ether. As a result, the transfer in line 6 (Fig.~\ref{fig:motivation}) will fail, reverting the entire transaction --- but only on Mainnet, not on Ropsten. Even if the states of all the variables of contract \texttt{Foo} on Ropsten are identical to their counterparts on Mainnet, the behavior of the \texttt{withdraw()} function will be different. This example demonstrates that the state of blockchain is an important factor that determines the outcome of smart contract execution. Unlike the static ones, dynamic transaction interception methods consider the current state of the blockchain, thereby preventing situations such as those illustrated in this example.

A recent study by Zhou et al.~\cite{zhou2020ever} reveals that novel (zero-day) smart contract attacks constantly appear on Ethereum. This trend creates a major challenge: how to defend against attacks we do not yet know about? One way to address this problem is to utilize the prevention methods that enforce security properties instead of searching for flaws, attacks, and vulnerabilities. Unfortunately, the security properties in static prevention solutions are tightly associated with known attacks and vulnerabilities. ECFChecker (STM-009)~\cite{grossman2017online} is a prevention method that verifies the ``callback-free'' property that ensures the safety of a smart contract from the family of reentrancy vulnerabilities. These properties, however, might not be universal enough to protect the smart contract from new vulnerabilities. One possible way to fill this gap is to verify the properties associated with expected outcomes of smart contract functions instead of vulnerability-related properties~\cite{ivanov2023txt}, which requires an extensive research effort.
%TBD % continue.

% \begin{figure}[t]
% \centering
% \begin{subfigure}{.59\textwidth}
%   \centering
%     \lstinputlisting[language=Solidity]{listings/motivation.txt}
%     \caption{smart contract Foo}
%     \label{fig:motivation}
% \end{subfigure}
% \begin{subfigure}{.39\textwidth}
%   \centering
%     \lstinputlisting[language=Solidity]{listings/case21.txt}
%     \vspace{88pt}
%     \caption{smart contract Bar}
%     \label{fig:case21}
% \end{subfigure}

% \caption{A pair of smart contracts demonstrating the importance of the block state.}
% \end{figure}

%\caption{A smart contract that fails only on Mainnet.}

%\caption{A non-payable smart contract deployed on Mainnet at \texttt{0xEc125A03C6F9E75BEB1A420e94d655B2f1352584}. The same address on Ropsten testnet is an 
%externally owned account (EOA).}

% \begin{figure}[t]
%     \centering
%     \lstinputlisting[language=Solidity]{listings/motivation.txt}
%     \caption{A smart contract that fails only on Mainnet.}
%     \label{fig:motivation}
% \end{figure}

% \begin{figure}[t]
%     \centering
%     \lstinputlisting[language=Solidity]{listings/case21.txt}
%     \caption{A non-payable smart contract deployed on Mainnet at \texttt{0xEc125A03C6F9E75BEB1A420e94d655B2f1352584}. The same address on Ropsten testnet is an 
% externally owned account (EOA).}
%     \label{fig:case21}
% \end{figure}

% Talk about zero-day attacks

\rev{
\noindent\textbf{Cross-Layer Attacks~~~~} Blockchains, like many other computer networks, are multi-layered systems. The exact delineation of blockchain layers depends on the platform and specific abstraction model. However, intuitively, the layers responsible for networking and consensus are below the layers of smart contracts, crypto accounts, and decentralized applications. Our assessment of recent security trends in smart contracts reveals a potential emergence of attacks that span several blockchain layers. For example, Qin et al.~\cite{qin2022quantifying} study the blockchain extractable value (BEV) schemes, as well as dangers associated with them. Daian et al.~\cite{daian2020flash} study deployments of arbitrage bots in blockchain systems and decentralized
exchanges (DEXes), including bots’ profit-making strategies. The study reveals systemic risks to consensus-layer security caused by the phenomenon of miner extractable value (MEV). Schwarz-Schilling et al.~\cite{schwarz2022three} explore two recent attacks on proof-of-stake (PoS) Ethereum consensus and create a third attack against the consensus. Although these attacks and risks do not directly involve smart contracts, they inevitably compromise the safety of the deployed smart contracts because smart contracts rely on the security of the underlying consensus. The recently proposed \emph{Transaction Encapsulation} approach, TxT~\cite{ivanov2023txt}, demonstrates that some cross-layer attacks can be alleviated or prevented by dynamic transaction testing, which requires further research.
}

\subsection{AI-driven Security}\label{sec:ai-driven-security}
We identify another recent salient trend in smart contract threat mitigation solutions --- AI-driven approaches involving machine learning. \rev{There are two major reasons why these approaches are capable of making a significant contribution. First, the AI solutions allow to embrace the expressiveness of modern smart contracts. Second, these approaches have been proven successful in securing other domains of computing~\cite{ibm-ai-security,bertino2021ai}}.

The expressiveness of smart contracts limits the capacity of static and formal analytical methods. \rev{Most modern smart contracts are Turing-complete, which allows them to implement sophisticated algorithms using high-level programming languages, such as Solidity and Vyper}. However, the smart contract expressiveness is a double-edged sword, as it creates a virtually infinite number of coding possibilities, which are very hard to embrace by static methods that predominantly rely upon patterns. Although machine learning methods also rely upon some patterns, recent machine learning models (e.g., deep neural network based) could explore much higher-dimensional feature spaces than 
%they allow for a much larger swath of possibilities than 
static approaches.

In the past few years, we have been observing a growing trend of using AI and machine learning for security purposes, such as malware detection~\cite{sahs2012machine}. Although the machine learning methods for smart contract threat mitigation have not yet gained considerable traction, the flexibility and universality of these methods will likely play an increasingly important role in smart contract defense.

\subsection{Human-machine Interaction in Smart Contracts}\label{sec:human-machine}

Smart contracts are often opposed to traditional user software based on the idea of replacing human-based decisions with a deterministic algorithm. However, such a vision is overly idealistic because a human is an integral part of a smart contract lifecycle. Specifically, humans write the source code of smart contracts. Even in the case of automatically synthesized smart contracts, we still require sufficient human intervention for developing templates and specifications. Testing a smart contract also requires a human, even for unit tests, which are developed by a human developer too. The security audit of a smart contract is also impossible without human judgment despite a wide variety of auditing tools available. Finally, interaction with smart contracts is always initiated by a user, regardless of the degree of automation. However, the impact of a human on the security of smart contracts is not sufficiently studied.

The study of human-machine interaction in smart contracts is limited by exploring honeypots and revealing a potential for some social engineering attacks. \rev{Social engineering attacks are attacks targeting humans as the major attack vector. Honeypots are malicious smart contracts that entrap naive attackers who try to exploit a known vulnerability in a smart contract. This allows us to categorize honeypots as a class of social engineering attacks.} HoneyBadger~\cite{torres2019art} is the automated tool that identifies such honeypots. Ivanov et al.~\cite{ivanov2021targeting} expand the scope of social engineering attacks with two more categories: address manipulation and homograph. EthClipper~\cite{ivanov2021ethclipper} is a partially social engineering attack targeting hardware crypto wallets by substituting the recipient address with a visually similar one. \rev{However, the above efforts do not embrace the entire complexity of human-smart contract interaction.}

One unexplored area of human-smart contract interaction is the security implication of the growing population of smart contract users who do not have a deep knowledge of the working mechanics of the blockchain and smart contracts. Another security-sensitive aspect of human-smart contract interaction is the assumption that the decentralization of blockchain implies decentralized applications (i.e., smart contracts) enabled by that blockchain. Specifically, many smart contracts implement routines (e.g, the \texttt{Ownable} parent class in OpenZeppelin Contracts~\cite{openzeppelin-contracts}) that grant excessive power to specified accounts. This excessive power may be abused by the owner or stolen by the attacker~\cite{ivanov2021rectifying} with potentially detrimental consequences. These two examples show the importance of studying human-smart contract interaction from the security perspective, and we envision many future studies in this area.

%TBD~\cite{ivanov2021targeting,torres2019art,ivanov2021rectifying}.

\subsection{Non-Ethereum Contracts}\label{sec:non-ethereum-contracts}
As it is revealed in \S\ref{sec:classification}, the vast majority of the existing smart contract threat mitigation methods target the smart contracts on the Ethereum platform. However, in recent years, the world has been experiencing major growth in the popularity of non-Ethereum smart contract platforms, such as NEO~\cite{neo-whitepaper}, Hyperledger Fabric~\cite{androulaki2018hyperledger}, EOS~\cite{eos-whitepaper}, and many others~\cite{panda2023}. Our analysis of the evolution of smart contract threat mitigation solutions clearly shows the growing attention by the research community to the security of non-Ethereum smart contracts. One reason for such disproportional attention to Ethereum, compared to other platforms, is that Ethereum is an open-data environment with the second-largest market capitalization after Bitcoin, so it is both convenient and important to study~\cite{perez2019smart}. However, these choices come at the expense of overlooking other major smart contract platforms. At the same time, our analysis shows that it is often impossible to extrapolate the lessons learned in Ethereum to the other platforms. \rev{Some existing vulnerabilities and other security issues are directly related to the design of the Ethereum platform or the syntax of Solidity.} Therefore, we expect increased attention to non-Ethereum platforms in the future development of smart contract threat mitigation research.

\subsection{Large-scale Measurements}\label{sec:large-scale-measurements}
Although blockchain is an open-data environment, there are multiple facts and statistics that we are unaware of. One reason is that a large amount of blockchain-related data, such as failed transactions and ERC20 token prices, is stored outside of the blockchain. Moreover, the growing popularity of Decentralized Finance (DeFi) further intensified the exchange of off-chain data~\cite{mev,mev-geth}. As a result, we have seen the growing amounts of on-chain and off-chain data that have not been analyzed from a security perspective.

\rev{
Yet, the existing security-related measurement studies~\cite{zhou2020ever,torres2019art,ferreira2020aegis,perez2021smart} of smart contracts do not provide answers to all the important questions. The smart contract security community is still lacking important data. For example, we have limited information on the fiat value of some blockchain assets, such as fungible and non-fungible tokens, especially the tokens that are not publicly traded. Moreover, we do not have a complete picture of value transfers between blockchain and non-blockchain assets, such as mining, crypto exchanges, and dark web transactions. This data is important because it would assign a specific weight value to the set of smart contracts which are currently weighted only by their native cryptocurrency balances (e.g., Ether). Determining real values and value flows in smart contracts would create a better assessment of threat, allowing to establish evidence-based priorities in threat mitigation. Specifically, we identify the following two areas important for the security of smart contracts in which there is no systematic data.
\begin{enumerate}
    \item \textbf{Measurement and flow of the market value of non-cryptocurrency blockchain assets:} ERC20 tokens, ERC-721 tokens, blockchain oracles, and emerging assets, such as ERC-1155 tokens.
    \item \textbf{Study of purchases and sales of cryptocurrency and tokens:} crypto exchanges, mining rewards, crypto money laundering, and valuable assets that are not publicly traded but can be exchanged into fiat value, e.g., private Initial Coin Offerings (ICOs).
\end{enumerate}
Such data would be very helpful for applying weights to attacks and vulnerabilities to quantify their real impacts based on the actual value flow of the smart contract assets.
}

\section{Conclusion}\label{sec:conclusion}
\rev{This survey covers the full spectrum of smart contract threat mitigation solutions. We presented a general taxonomy for classification of such solutions. The taxonomy applies to today's methods and is suitable for future methods as well, even if new paradigms, blockchain platforms, or vulnerabilities emerge.} Using this taxonomy, we classified 133 existing smart contract threat mitigation solutions. \rev{We identified eight distinct core defense methods employed by the existing solutions. Furthermore, we developed synthesized workflows of the eight core methods.} We studied the ability of the existing smart contract threat mitigation solutions to address the known vulnerabilities. We conducted an evidence-based evolutionary study of smart contract threat mitigation solutions to outline trends and perspectives. To further benefit the community of smart contract security researchers, users, and developers, we deployed an open-source, regularly updated online registry for smart contract threat mitigation at \url{https://seit.egr.msu.edu/research/stmregistry/}.

\bibliographystyle{ACM-Reference-Format}
%\bibliography{bibliography}

\begin{thebibliography}{190}

%%% the Web address.

\ifx \showCODEN    \undefined \def \showCODEN     #1{\unskip}     \fi
\ifx \showDOI      \undefined \def \showDOI       #1{#1}\fi
\ifx \showISBNx    \undefined \def \showISBNx     #1{\unskip}     \fi
\ifx \showISBNxiii \undefined \def \showISBNxiii  #1{\unskip}     \fi
\ifx \showISSN     \undefined \def \showISSN      #1{\unskip}     \fi
\ifx \showLCCN     \undefined \def \showLCCN      #1{\unskip}     \fi
\ifx \shownote     \undefined \def \shownote      #1{#1}          \fi
\ifx \showarticletitle \undefined \def \showarticletitle #1{#1}   \fi
\ifx \showURL      \undefined \def \showURL       {\relax}        \fi
% invisible to TeX
\providecommand\bibfield[2]{#2}
\providecommand\bibinfo[2]{#2}
\providecommand\natexlab[1]{#1}
\providecommand\showeprint[2][]{arXiv:#2}

\bibitem[\protect\citeauthoryear{??}{mem}{2020}]%
        {mempool}
 \bibinfo{year}{2020}\natexlab{}.
\newblock \bibinfo{title}{Exploring the methods of looking into Ethereum’s
  transaction pool (mempool)}.
\newblock
  \bibinfo{howpublished}{{https://chainstack.com/exploring-the-methods-of-looking-into-ethereums-transaction-pool/}}.
\newblock
\newblock
\shownote{Accessed: 2023-03-01}.


\bibitem[\protect\citeauthoryear{??}{ibm}{2022}]%
        {ibm-ai-security}
 \bibinfo{year}{2022}\natexlab{}.
\newblock \bibinfo{title}{{Artificial intelligence (AI) for cybersecurity}}.
\newblock
  \bibinfo{howpublished}{{https://www.ibm.com/security/artificial-intelligence}}.
\newblock
\newblock
\shownote{Accessed: 2022-03-07}.


\bibitem[\protect\citeauthoryear{??}{con}{2022}]%
        {contract-library}
 \bibinfo{year}{2022}\natexlab{}.
\newblock \bibinfo{title}{{Dedadub Contract Library}}.
\newblock \bibinfo{howpublished}{{https://dedaub.com/contract-library}}.
\newblock
\newblock
\shownote{Accessed: 2022-02-28}.


\bibitem[\protect\citeauthoryear{??}{eos}{2022}]%
        {eos-whitepaper}
 \bibinfo{year}{2022}\natexlab{}.
\newblock \bibinfo{title}{{EOS.IO Technical White Paper v2}}.
\newblock
  \bibinfo{howpublished}{{https://github.com/EOSIO/Documentation/blob/master/TechnicalWhitePaper.md}}.
\newblock
\newblock
\shownote{Accessed: 2022-03-07}.


\bibitem[\protect\citeauthoryear{??}{eth}{2022}]%
        {etherscan-tokens}
 \bibinfo{year}{2022}\natexlab{}.
\newblock \bibinfo{title}{{Etherscan Token Tracker}}.
\newblock \bibinfo{howpublished}{{https://etherscan.io/tokens}}.
\newblock
\newblock
\shownote{Accessed: 2022-03-05}.


\bibitem[\protect\citeauthoryear{??}{mev}{2022a}]%
        {mev-geth}
 \bibinfo{year}{2022}\natexlab{a}.
\newblock \bibinfo{title}{Go implementation of MEV-Auction for Ethereum}.
\newblock \bibinfo{howpublished}{{https://github.com/flashbots/mev-geth}}.
\newblock


\bibitem[\protect\citeauthoryear{??}{mev}{2022b}]%
        {mev}
 \bibinfo{year}{2022}\natexlab{b}.
\newblock \bibinfo{title}{Miner Extractable Value (MEV)}.
\newblock
  \bibinfo{howpublished}{{https://ethereum.org/en/developers/docs/mev/}}.
\newblock


\bibitem[\protect\citeauthoryear{??}{myt}{2022}]%
        {mythx}
 \bibinfo{year}{2022}\natexlab{}.
\newblock \bibinfo{title}{{MythX}}.
\newblock \bibinfo{howpublished}{{https://mythx.io/}}.
\newblock
\newblock
\shownote{Accessed: 2022-02-26}.


\bibitem[\protect\citeauthoryear{??}{neo}{2022}]%
        {neo-whitepaper}
 \bibinfo{year}{2022}\natexlab{}.
\newblock \bibinfo{title}{{Neo White Paper}}.
\newblock
  \bibinfo{howpublished}{{https://docs.neo.org/v2/docs/en-us/basic/whitepaper.html}}.
\newblock
\newblock
\shownote{Accessed: 2022-03-07}.


\bibitem[\protect\citeauthoryear{??}{ope}{2022}]%
        {openzeppelin-contracts}
 \bibinfo{year}{2022}\natexlab{}.
\newblock \bibinfo{title}{{OpenZeppelin Contracts}}.
\newblock \bibinfo{howpublished}{{https://openzeppelin.com/contracts/}}.
\newblock
\newblock
\shownote{Accessed: 2022-02-28}.


\bibitem[\protect\citeauthoryear{??}{pol}{2022}]%
        {polygon}
 \bibinfo{year}{2022}\natexlab{}.
\newblock \bibinfo{title}{Polygon}.
\newblock \bibinfo{howpublished}{{https://polygon.technology/}}.
\newblock


\bibitem[\protect\citeauthoryear{??}{rsk}{2022}]%
        {rsk}
 \bibinfo{year}{2022}\natexlab{}.
\newblock \bibinfo{title}{RSK Whitepaper}.
\newblock
  \bibinfo{howpublished}{{https://www.rsk.co/Whitepapers/RSK\_White\_Paper-ORIGINAL.pdf}}.
\newblock


\bibitem[\protect\citeauthoryear{??}{swc}{2022a}]%
        {swc100}
 \bibinfo{year}{2022}\natexlab{a}.
\newblock \bibinfo{title}{SWC-100: Function Default Visibility}.
\newblock \bibinfo{howpublished}{{https://swcregistry.io/docs/SWC-100}}.
\newblock
\newblock
\shownote{Accessed: 2022-03-21}.


\bibitem[\protect\citeauthoryear{??}{swc}{2022b}]%
        {swc107}
 \bibinfo{year}{2022}\natexlab{b}.
\newblock \bibinfo{title}{SWC-107: Reentrancy}.
\newblock \bibinfo{howpublished}{{https://swcregistry.io/docs/SWC-107}}.
\newblock
\newblock
\shownote{Accessed: 2022-03-21}.


\bibitem[\protect\citeauthoryear{??}{swc}{2022c}]%
        {swc108}
 \bibinfo{year}{2022}\natexlab{c}.
\newblock \bibinfo{title}{SWC-108: State Variable Default Visibility}.
\newblock \bibinfo{howpublished}{{https://swcregistry.io/docs/SWC-108}}.
\newblock
\newblock
\shownote{Accessed: 2022-03-21}.


\bibitem[\protect\citeauthoryear{??}{swc}{2022d}]%
        {swc119}
 \bibinfo{year}{2022}\natexlab{d}.
\newblock \bibinfo{title}{SWC-119: Shadowing State Variables}.
\newblock \bibinfo{howpublished}{{https://swcregistry.io/docs/SWC-119}}.
\newblock
\newblock
\shownote{Accessed: 2022-04-17}.


\bibitem[\protect\citeauthoryear{??}{swc}{2022e}]%
        {swc123}
 \bibinfo{year}{2022}\natexlab{e}.
\newblock \bibinfo{title}{SWC-123: Requirement Violation}.
\newblock \bibinfo{howpublished}{{https://swcregistry.io/docs/SWC-123}}.
\newblock
\newblock
\shownote{Accessed: 2022-03-10}.


\bibitem[\protect\citeauthoryear{??}{swc}{2022f}]%
        {swc130}
 \bibinfo{year}{2022}\natexlab{f}.
\newblock \bibinfo{title}{SWC-130: Right-To-Left-Override control character
  (U+202E)}.
\newblock \bibinfo{howpublished}{{https://swcregistry.io/docs/SWC-130}}.
\newblock
\newblock
\shownote{Accessed: 2022-04-17}.


\bibitem[\protect\citeauthoryear{??}{swc}{2022g}]%
        {swcregistry}
 \bibinfo{year}{2022}\natexlab{g}.
\newblock \bibinfo{title}{SWC Registry}.
\newblock \bibinfo{howpublished}{{https://swcregistry.io/}}.
\newblock
\newblock
\shownote{{Accessed: 2022-03-15}}.


\bibitem[\protect\citeauthoryear{??}{z3}{2022}]%
        {z3}
 \bibinfo{year}{2022}\natexlab{}.
\newblock \bibinfo{title}{Z3Prover/z3}.
\newblock \bibinfo{howpublished}{{https://github.com/Z3Prover/z3}}.
\newblock


\bibitem[\protect\citeauthoryear{Abdellatif and Brousmiche}{Abdellatif and
  Brousmiche}{2018}]%
        {abdellatif2018formal}
\bibfield{author}{\bibinfo{person}{Tesnim Abdellatif} {and}
  \bibinfo{person}{Kei-L{\'e}o Brousmiche}.} \bibinfo{year}{2018}\natexlab{}.
\newblock \showarticletitle{Formal verification of smart contracts based on
  users and blockchain behaviors models}. In \bibinfo{booktitle}{\emph{2018 9th
  IFIP International Conference on New Technologies, Mobility and Security
  (NTMS)}}. IEEE, \bibinfo{pages}{1--5}.
\newblock


\bibitem[\protect\citeauthoryear{Ahrendt, Bubel, Ellul, Pace, Pardo, Rebiscoul,
  and Schneider}{Ahrendt et~al\mbox{.}}{2019}]%
        {ahrendt2019verification}
\bibfield{author}{\bibinfo{person}{Wolfgang Ahrendt}, \bibinfo{person}{Richard
  Bubel}, \bibinfo{person}{Joshua Ellul}, \bibinfo{person}{Gordon~J Pace},
  \bibinfo{person}{Ra{\'u}l Pardo}, \bibinfo{person}{Vincent Rebiscoul}, {and}
  \bibinfo{person}{Gerardo Schneider}.} \bibinfo{year}{2019}\natexlab{}.
\newblock \showarticletitle{Verification of smart contract business logic}. In
  \bibinfo{booktitle}{\emph{International Conference on Fundamentals of
  Software Engineering}}. Springer, \bibinfo{pages}{228--243}.
\newblock


\bibitem[\protect\citeauthoryear{Akca, Rajan, and Peng}{Akca
  et~al\mbox{.}}{2019}]%
        {akca2019solanalyser}
\bibfield{author}{\bibinfo{person}{Sefa Akca}, \bibinfo{person}{Ajitha Rajan},
  {and} \bibinfo{person}{Chao Peng}.} \bibinfo{year}{2019}\natexlab{}.
\newblock \showarticletitle{SolAnalyser: A Framework for Analysing and Testing
  Smart Contracts}. In \bibinfo{booktitle}{\emph{2019 26th Asia-Pacific
  Software Engineering Conference (APSEC)}}. IEEE, \bibinfo{pages}{482--489}.
\newblock


\bibitem[\protect\citeauthoryear{Albert, Correas, Gordillo,
  Rom{\'a}n-D{\'\i}ez, and Rubio}{Albert et~al\mbox{.}}{2019a}]%
        {albert2019safevm}
\bibfield{author}{\bibinfo{person}{Elvira Albert}, \bibinfo{person}{Jes{\'u}s
  Correas}, \bibinfo{person}{Pablo Gordillo}, \bibinfo{person}{Guillermo
  Rom{\'a}n-D{\'\i}ez}, {and} \bibinfo{person}{Albert Rubio}.}
  \bibinfo{year}{2019}\natexlab{a}.
\newblock \showarticletitle{SAFEVM: a safety verifier for Ethereum smart
  contracts}. In \bibinfo{booktitle}{\emph{Proceedings of the 28th ACM SIGSOFT
  International Symposium on Software Testing and Analysis}}.
  \bibinfo{pages}{386--389}.
\newblock


\bibitem[\protect\citeauthoryear{Albert, Gordillo, Rubio, and Sergey}{Albert
  et~al\mbox{.}}{2019b}]%
        {albert2019running}
\bibfield{author}{\bibinfo{person}{Elvira Albert}, \bibinfo{person}{Pablo
  Gordillo}, \bibinfo{person}{Albert Rubio}, {and} \bibinfo{person}{Ilya
  Sergey}.} \bibinfo{year}{2019}\natexlab{b}.
\newblock \showarticletitle{Running on fumes}. In
  \bibinfo{booktitle}{\emph{International Conference on Verification and
  Evaluation of Computer and Communication Systems}}. Springer,
  \bibinfo{pages}{63--78}.
\newblock


\bibitem[\protect\citeauthoryear{Alqahtani, He, Gamble, and Mauricio}{Alqahtani
  et~al\mbox{.}}{2020}]%
        {alqahtani2020formal}
\bibfield{author}{\bibinfo{person}{Sarra Alqahtani}, \bibinfo{person}{Xinchi
  He}, \bibinfo{person}{Rose Gamble}, {and} \bibinfo{person}{Papa Mauricio}.}
  \bibinfo{year}{2020}\natexlab{}.
\newblock \showarticletitle{Formal verification of functional requirements for
  smart contract compositions in supply chain management systems}. In
  \bibinfo{booktitle}{\emph{Proceedings of the 53rd Hawaii International
  Conference on System Sciences}}.
\newblock


\bibitem[\protect\citeauthoryear{Amani, B{\'e}gel, Bortin, and Staples}{Amani
  et~al\mbox{.}}{2018}]%
        {amani2018towards}
\bibfield{author}{\bibinfo{person}{Sidney Amani}, \bibinfo{person}{Myriam
  B{\'e}gel}, \bibinfo{person}{Maksym Bortin}, {and} \bibinfo{person}{Mark
  Staples}.} \bibinfo{year}{2018}\natexlab{}.
\newblock \showarticletitle{Towards verifying ethereum smart contract bytecode
  in Isabelle/HOL}. In \bibinfo{booktitle}{\emph{Proceedings of the 7th ACM
  SIGPLAN International Conference on Certified Programs and Proofs}}.
  \bibinfo{pages}{66--77}.
\newblock


\bibitem[\protect\citeauthoryear{Androulaki, Barger, Bortnikov, Cachin,
  Christidis, De~Caro, Enyeart, Ferris, Laventman, Manevich,
  et~al\mbox{.}}{Androulaki et~al\mbox{.}}{2018}]%
        {androulaki2018hyperledger}
\bibfield{author}{\bibinfo{person}{Elli Androulaki}, \bibinfo{person}{Artem
  Barger}, \bibinfo{person}{Vita Bortnikov}, \bibinfo{person}{Christian
  Cachin}, \bibinfo{person}{Konstantinos Christidis}, \bibinfo{person}{Angelo
  De~Caro}, \bibinfo{person}{David Enyeart}, \bibinfo{person}{Christopher
  Ferris}, \bibinfo{person}{Gennady Laventman}, \bibinfo{person}{Yacov
  Manevich}, {et~al\mbox{.}}} \bibinfo{year}{2018}\natexlab{}.
\newblock \showarticletitle{Hyperledger fabric: a distributed operating system
  for permissioned blockchains}. In \bibinfo{booktitle}{\emph{Proceedings of
  the thirteenth EuroSys conference}}. \bibinfo{pages}{1--15}.
\newblock


\bibitem[\protect\citeauthoryear{Antonino and Roscoe}{Antonino and
  Roscoe}{2021}]%
        {antonino2021solidifier}
\bibfield{author}{\bibinfo{person}{Pedro Antonino} {and} \bibinfo{person}{AW
  Roscoe}.} \bibinfo{year}{2021}\natexlab{}.
\newblock \showarticletitle{Solidifier: bounded model checking Solidity using
  lazy contract deployment and precise memory modelling}. In
  \bibinfo{booktitle}{\emph{Proceedings of the 36th Annual ACM Symposium on
  Applied Computing}}. \bibinfo{pages}{1788--1797}.
\newblock


\bibitem[\protect\citeauthoryear{Antonopoulos and Wood}{Antonopoulos and
  Wood}{2018}]%
        {antonopoulos2018mastering}
\bibfield{author}{\bibinfo{person}{Andreas~M Antonopoulos} {and}
  \bibinfo{person}{Gavin Wood}.} \bibinfo{year}{2018}\natexlab{}.
\newblock \bibinfo{booktitle}{\emph{Mastering ethereum: building smart
  contracts and dapps}}.
\newblock \bibinfo{publisher}{O'reilly Media}.
\newblock


\bibitem[\protect\citeauthoryear{Arga{\~n}araz, Ber{\'o}n, Pereira, and
  Henriques}{Arga{\~n}araz et~al\mbox{.}}{2020}]%
        {arganaraz2020detection}
\bibfield{author}{\bibinfo{person}{Mauro Arga{\~n}araz}, \bibinfo{person}{Mario
  Ber{\'o}n}, \bibinfo{person}{Maria~Jo{\~a}o Pereira}, {and}
  \bibinfo{person}{Pedro Henriques}.} \bibinfo{year}{2020}\natexlab{}.
\newblock \showarticletitle{Detection of vulnerabilities in smart contracts
  specifications in ethereum platforms}. In \bibinfo{booktitle}{\emph{9th
  Symposium on Languages, Applications and Technologies (SLATE 2020)}},
  Vol.~\bibinfo{volume}{83}. Schloss Dagstuhl--Leibniz-Zentrum fuer Informatik,
  \bibinfo{pages}{1--16}.
\newblock


\bibitem[\protect\citeauthoryear{Atzei, Bartoletti, and Cimoli}{Atzei
  et~al\mbox{.}}{2017}]%
        {atzei2017survey}
\bibfield{author}{\bibinfo{person}{Nicola Atzei}, \bibinfo{person}{Massimo
  Bartoletti}, {and} \bibinfo{person}{Tiziana Cimoli}.}
  \bibinfo{year}{2017}\natexlab{}.
\newblock \showarticletitle{A survey of attacks on ethereum smart contracts
  (sok)}. In \bibinfo{booktitle}{\emph{International conference on principles
  of security and trust}}. Springer, \bibinfo{pages}{164--186}.
\newblock


\bibitem[\protect\citeauthoryear{Atzei, Bartoletti, Lande, Yoshida, and
  Zunino}{Atzei et~al\mbox{.}}{2019}]%
        {atzei2019developing}
\bibfield{author}{\bibinfo{person}{Nicola Atzei}, \bibinfo{person}{Massimo
  Bartoletti}, \bibinfo{person}{Stefano Lande}, \bibinfo{person}{Nobuko
  Yoshida}, {and} \bibinfo{person}{Roberto Zunino}.}
  \bibinfo{year}{2019}\natexlab{}.
\newblock \showarticletitle{Developing secure Bitcoin contracts with BitML}. In
  \bibinfo{booktitle}{\emph{Proceedings of the 2019 27th ACM Joint Meeting on
  European Software Engineering Conference and Symposium on the Foundations of
  Software Engineering}}. \bibinfo{pages}{1124--1128}.
\newblock


\bibitem[\protect\citeauthoryear{Bai, Cheng, Duan, and Hu}{Bai
  et~al\mbox{.}}{2018}]%
        {bai2018formal}
\bibfield{author}{\bibinfo{person}{Xiaomin Bai}, \bibinfo{person}{Zijing
  Cheng}, \bibinfo{person}{Zhangbo Duan}, {and} \bibinfo{person}{Kai Hu}.}
  \bibinfo{year}{2018}\natexlab{}.
\newblock \showarticletitle{Formal modeling and verification of smart
  contracts}. In \bibinfo{booktitle}{\emph{Proceedings of the 2018 7th
  international conference on software and computer applications}}.
  \bibinfo{pages}{322--326}.
\newblock


\bibitem[\protect\citeauthoryear{Barrett, Conway, Deters, Hadarean, Jovanovic,
  King, Reynolds, and Tinelli}{Barrett et~al\mbox{.}}{2011}]%
        {DBLP:conf/cav/BarrettCDHJKRT11}
\bibfield{author}{\bibinfo{person}{Clark~W. Barrett},
  \bibinfo{person}{Christopher~L. Conway}, \bibinfo{person}{Morgan Deters},
  \bibinfo{person}{Liana Hadarean}, \bibinfo{person}{Dejan Jovanovic},
  \bibinfo{person}{Tim King}, \bibinfo{person}{Andrew Reynolds}, {and}
  \bibinfo{person}{Cesare Tinelli}.} \bibinfo{year}{2011}\natexlab{}.
\newblock \showarticletitle{{CVC4}}. In \bibinfo{booktitle}{\emph{Computer
  Aided Verification - 23rd International Conference, {CAV} 2011, Snowbird, UT,
  USA, July 14-20, 2011. Proceedings}} \emph{(\bibinfo{series}{Lecture Notes in
  Computer Science}, Vol.~\bibinfo{volume}{6806})},
  \bibfield{editor}{\bibinfo{person}{Ganesh Gopalakrishnan} {and}
  \bibinfo{person}{Shaz Qadeer}} (Eds.). \bibinfo{publisher}{Springer},
  \bibinfo{pages}{171--177}.
\newblock
\urldef\tempurl%
\url{https://doi.org/10.1007/978-3-642-22110-1\_14}
\showDOI{\tempurl}


\bibitem[\protect\citeauthoryear{Bartoletti and Zunino}{Bartoletti and
  Zunino}{2019}]%
        {bartoletti2019verifying}
\bibfield{author}{\bibinfo{person}{Massimo Bartoletti} {and}
  \bibinfo{person}{Roberto Zunino}.} \bibinfo{year}{2019}\natexlab{}.
\newblock \showarticletitle{Verifying liquidity of Bitcoin contracts}. In
  \bibinfo{booktitle}{\emph{8th International Conference on Principles of
  Security and Trust, POST 2019 Held as Part of the European Joint Conferences
  on Theory and Practice of Software, ETAPS 2019}},
  Vol.~\bibinfo{volume}{11426}. Springer, \bibinfo{pages}{222--247}.
\newblock


\bibitem[\protect\citeauthoryear{Beckert, Herda, Kirsten, and Schiffl}{Beckert
  et~al\mbox{.}}{2018}]%
        {beckert2018formal}
\bibfield{author}{\bibinfo{person}{Bernhard Beckert}, \bibinfo{person}{Mihai
  Herda}, \bibinfo{person}{Michael Kirsten}, {and} \bibinfo{person}{Jonas
  Schiffl}.} \bibinfo{year}{2018}\natexlab{}.
\newblock \showarticletitle{Formal specification and verification of
  hyperledger fabric chaincode}. In \bibinfo{booktitle}{\emph{Proc. Int. Conf.
  Formal Eng. Methods}}. \bibinfo{pages}{44--48}.
\newblock


\bibitem[\protect\citeauthoryear{Bertino, Kantarcioglu, Akcora, Samtani,
  Mittal, and Gupta}{Bertino et~al\mbox{.}}{2021}]%
        {bertino2021ai}
\bibfield{author}{\bibinfo{person}{Elisa Bertino}, \bibinfo{person}{Murat
  Kantarcioglu}, \bibinfo{person}{Cuneyt~Gurcan Akcora}, \bibinfo{person}{Sagar
  Samtani}, \bibinfo{person}{Sudip Mittal}, {and} \bibinfo{person}{Maanak
  Gupta}.} \bibinfo{year}{2021}\natexlab{}.
\newblock \showarticletitle{AI for Security and Security for AI}. In
  \bibinfo{booktitle}{\emph{Proceedings of the Eleventh ACM Conference on Data
  and Application Security and Privacy}}. \bibinfo{pages}{333--334}.
\newblock


\bibitem[\protect\citeauthoryear{Bhargavan, Delignat-Lavaud, Fournet,
  Gollamudi, Gonthier, Kobeissi, Kulatova, Rastogi, Sibut-Pinote, Swamy,
  et~al\mbox{.}}{Bhargavan et~al\mbox{.}}{2016}]%
        {bhargavan2016formal}
\bibfield{author}{\bibinfo{person}{Karthikeyan Bhargavan},
  \bibinfo{person}{Antoine Delignat-Lavaud}, \bibinfo{person}{C{\'e}dric
  Fournet}, \bibinfo{person}{Anitha Gollamudi}, \bibinfo{person}{Georges
  Gonthier}, \bibinfo{person}{Nadim Kobeissi}, \bibinfo{person}{Natalia
  Kulatova}, \bibinfo{person}{Aseem Rastogi}, \bibinfo{person}{Thomas
  Sibut-Pinote}, \bibinfo{person}{Nikhil Swamy}, {et~al\mbox{.}}}
  \bibinfo{year}{2016}\natexlab{}.
\newblock \showarticletitle{Formal verification of smart contracts: Short
  paper}. In \bibinfo{booktitle}{\emph{Proceedings of the 2016 ACM workshop on
  programming languages and analysis for security}}. \bibinfo{pages}{91--96}.
\newblock


\bibitem[\protect\citeauthoryear{Bigi, Bracciali, Meacci, and Tuosto}{Bigi
  et~al\mbox{.}}{2015}]%
        {bigi2015validation}
\bibfield{author}{\bibinfo{person}{Giancarlo Bigi}, \bibinfo{person}{Andrea
  Bracciali}, \bibinfo{person}{Giovanni Meacci}, {and} \bibinfo{person}{Emilio
  Tuosto}.} \bibinfo{year}{2015}\natexlab{}.
\newblock \showarticletitle{Validation of decentralised smart contracts through
  game theory and formal methods}.
\newblock In \bibinfo{booktitle}{\emph{Programming Languages with Applications
  to Biology and Security}}. \bibinfo{publisher}{Springer},
  \bibinfo{pages}{142--161}.
\newblock


\bibitem[\protect\citeauthoryear{Biryukov, Khovratovich, and
  Tikhomirov}{Biryukov et~al\mbox{.}}{2017}]%
        {biryukov2017findel}
\bibfield{author}{\bibinfo{person}{Alex Biryukov}, \bibinfo{person}{Dmitry
  Khovratovich}, {and} \bibinfo{person}{Sergei Tikhomirov}.}
  \bibinfo{year}{2017}\natexlab{}.
\newblock \showarticletitle{Findel: Secure derivative contracts for Ethereum}.
  In \bibinfo{booktitle}{\emph{International Conference on Financial
  Cryptography and Data Security}}. Springer, \bibinfo{pages}{453--467}.
\newblock


\bibitem[\protect\citeauthoryear{Bose, Das, Chen, Feng, Kruegel, and
  Vigna}{Bose et~al\mbox{.}}{2021}]%
        {bose2021sailfish}
\bibfield{author}{\bibinfo{person}{Priyanka Bose}, \bibinfo{person}{Dipanjan
  Das}, \bibinfo{person}{Yanju Chen}, \bibinfo{person}{Yu Feng},
  \bibinfo{person}{Christopher Kruegel}, {and} \bibinfo{person}{Giovanni
  Vigna}.} \bibinfo{year}{2021}\natexlab{}.
\newblock \showarticletitle{SAILFISH: Vetting Smart Contract
  State-Inconsistency Bugs in Seconds}.
\newblock \bibinfo{journal}{\emph{arXiv preprint arXiv:2104.08638}}
  (\bibinfo{year}{2021}).
\newblock


\bibitem[\protect\citeauthoryear{Bragagnolo, Rocha, Denker, and
  Ducasse}{Bragagnolo et~al\mbox{.}}{2018}]%
        {bragagnolo2018smartinspect}
\bibfield{author}{\bibinfo{person}{Santiago Bragagnolo},
  \bibinfo{person}{Henrique Rocha}, \bibinfo{person}{Marcus Denker}, {and}
  \bibinfo{person}{St{\'e}phane Ducasse}.} \bibinfo{year}{2018}\natexlab{}.
\newblock \showarticletitle{SmartInspect: solidity smart contract inspector}.
  In \bibinfo{booktitle}{\emph{2018 International workshop on blockchain
  oriented software engineering (IWBOSE)}}. IEEE, \bibinfo{pages}{9--18}.
\newblock


\bibitem[\protect\citeauthoryear{Breidenbach, Daian, Juels, and
  Sirer}{Breidenbach et~al\mbox{.}}{2017}]%
        {breidenbach2017depth}
\bibfield{author}{\bibinfo{person}{Lorenz Breidenbach}, \bibinfo{person}{Phil
  Daian}, \bibinfo{person}{Ari Juels}, {and} \bibinfo{person}{Emin~G{\"u}n
  Sirer}.} \bibinfo{year}{2017}\natexlab{}.
\newblock \showarticletitle{An in-depth look at the parity multisig bug}.
\newblock \bibinfo{journal}{\emph{Hacking, Distributed, July}}
  (\bibinfo{year}{2017}).
\newblock


\bibitem[\protect\citeauthoryear{Breidenbach, Daian, Tram{\`e}r, and
  Juels}{Breidenbach et~al\mbox{.}}{2018}]%
        {breidenbach2018enter}
\bibfield{author}{\bibinfo{person}{Lorenz Breidenbach}, \bibinfo{person}{Phil
  Daian}, \bibinfo{person}{Florian Tram{\`e}r}, {and} \bibinfo{person}{Ari
  Juels}.} \bibinfo{year}{2018}\natexlab{}.
\newblock \showarticletitle{Enter the hydra: Towards principled bug bounties
  and exploit-resistant smart contracts}. In \bibinfo{booktitle}{\emph{27th
  {USENIX} Security Symposium ({USENIX} Security 18)}}.
  \bibinfo{pages}{1335--1352}.
\newblock


\bibitem[\protect\citeauthoryear{Brent, Grech, Lagouvardos, Scholz, and
  Smaragdakis}{Brent et~al\mbox{.}}{2020}]%
        {brent2020ethainter}
\bibfield{author}{\bibinfo{person}{Lexi Brent}, \bibinfo{person}{Neville
  Grech}, \bibinfo{person}{Sifis Lagouvardos}, \bibinfo{person}{Bernhard
  Scholz}, {and} \bibinfo{person}{Yannis Smaragdakis}.}
  \bibinfo{year}{2020}\natexlab{}.
\newblock \showarticletitle{Ethainter: A smart contract security analyzer for
  composite vulnerabilities}. In \bibinfo{booktitle}{\emph{Proceedings of the
  41st ACM SIGPLAN Conference on Programming Language Design and
  Implementation}}. \bibinfo{pages}{454--469}.
\newblock


\bibitem[\protect\citeauthoryear{Brent, Jurisevic, Kong, Liu, Gauthier,
  Gramoli, Holz, and Scholz}{Brent et~al\mbox{.}}{2018}]%
        {brent2018vandal}
\bibfield{author}{\bibinfo{person}{Lexi Brent}, \bibinfo{person}{Anton
  Jurisevic}, \bibinfo{person}{Michael Kong}, \bibinfo{person}{Eric Liu},
  \bibinfo{person}{Francois Gauthier}, \bibinfo{person}{Vincent Gramoli},
  \bibinfo{person}{Ralph Holz}, {and} \bibinfo{person}{Bernhard Scholz}.}
  \bibinfo{year}{2018}\natexlab{}.
\newblock \showarticletitle{Vandal: A scalable security analysis framework for
  smart contracts}.
\newblock \bibinfo{journal}{\emph{arXiv preprint arXiv:1809.03981}}
  (\bibinfo{year}{2018}).
\newblock


\bibitem[\protect\citeauthoryear{Browne}{Browne}{2017}]%
        {browneaccidental}
\bibfield{author}{\bibinfo{person}{R Browne}.} \bibinfo{year}{2017}\natexlab{}.
\newblock \bibinfo{title}{Accidental bug may have frozen 280 million worth of
  digital coin ether in a cryptocurrency wallet}.
\newblock
  \bibinfo{howpublished}{{https://www.cnbc.com/2017/11/08/accidental-bug-may-have-frozen-280-worth-of-ether-on-parity-wallet.html}}.
\newblock


\bibitem[\protect\citeauthoryear{Camino, Torres, Baden, and State}{Camino
  et~al\mbox{.}}{2020}]%
        {camino2020data}
\bibfield{author}{\bibinfo{person}{Ramiro Camino},
  \bibinfo{person}{Christof~Ferreira Torres}, \bibinfo{person}{Mathis Baden},
  {and} \bibinfo{person}{Radu State}.} \bibinfo{year}{2020}\natexlab{}.
\newblock \showarticletitle{A data science approach for detecting honeypots in
  ethereum}. In \bibinfo{booktitle}{\emph{2020 IEEE International Conference on
  Blockchain and Cryptocurrency (ICBC)}}. IEEE, \bibinfo{pages}{1--9}.
\newblock


\bibitem[\protect\citeauthoryear{Cecchetti, Yao, Ni, and Myers}{Cecchetti
  et~al\mbox{.}}{[n.\,d.]}]%
        {cecchetti12compositional}
\bibfield{author}{\bibinfo{person}{Ethan Cecchetti}, \bibinfo{person}{Siqiu
  Yao}, \bibinfo{person}{Haobin Ni}, {and} \bibinfo{person}{Andrew~C Myers}.}
  \bibinfo{year}{[n.\,d.]}\natexlab{}.
\newblock \showarticletitle{Compositional Security for Reentrant Applications}.
\newblock \bibinfo{journal}{\emph{contract}} \bibinfo{volume}{12},
  \bibinfo{number}{13} (\bibinfo{year}{[n.\,d.]}), \bibinfo{pages}{14}.
\newblock


\bibitem[\protect\citeauthoryear{Cecchetti, Yao, Ni, and Myers}{Cecchetti
  et~al\mbox{.}}{2020}]%
        {cecchetti2020securing}
\bibfield{author}{\bibinfo{person}{Ethan Cecchetti}, \bibinfo{person}{Siqiu
  Yao}, \bibinfo{person}{Haobin Ni}, {and} \bibinfo{person}{Andrew~C Myers}.}
  \bibinfo{year}{2020}\natexlab{}.
\newblock \showarticletitle{Securing smart contracts with information flow}. In
  \bibinfo{booktitle}{\emph{International Symposium on Foundations and
  Applications of Blockchain}}.
\newblock


\bibitem[\protect\citeauthoryear{Chang, Gao, Xiao, Sun, Cai, and Yang}{Chang
  et~al\mbox{.}}{2019}]%
        {chang2019scompile}
\bibfield{author}{\bibinfo{person}{Jialiang Chang}, \bibinfo{person}{Bo Gao},
  \bibinfo{person}{Hao Xiao}, \bibinfo{person}{Jun Sun}, \bibinfo{person}{Yan
  Cai}, {and} \bibinfo{person}{Zijiang Yang}.} \bibinfo{year}{2019}\natexlab{}.
\newblock \showarticletitle{scompile: Critical path identification and analysis
  for smart contracts}. In \bibinfo{booktitle}{\emph{International Conference
  on Formal Engineering Methods}}. Springer, \bibinfo{pages}{286--304}.
\newblock


\bibitem[\protect\citeauthoryear{Chen, Pendleton, Njilla, and Xu}{Chen
  et~al\mbox{.}}{2020b}]%
        {chen2020survey}
\bibfield{author}{\bibinfo{person}{Huashan Chen}, \bibinfo{person}{Marcus
  Pendleton}, \bibinfo{person}{Laurent Njilla}, {and} \bibinfo{person}{Shouhuai
  Xu}.} \bibinfo{year}{2020}\natexlab{b}.
\newblock \showarticletitle{A survey on ethereum systems security:
  Vulnerabilities, attacks, and defenses}.
\newblock \bibinfo{journal}{\emph{ACM Computing Surveys (CSUR)}}
  \bibinfo{volume}{53}, \bibinfo{number}{3} (\bibinfo{year}{2020}),
  \bibinfo{pages}{1--43}.
\newblock


\bibitem[\protect\citeauthoryear{Chen, Xia, Lo, Grundy, Luo, and Chen}{Chen
  et~al\mbox{.}}{2020c}]%
        {chen2020defining}
\bibfield{author}{\bibinfo{person}{Jiachi Chen}, \bibinfo{person}{Xin Xia},
  \bibinfo{person}{David Lo}, \bibinfo{person}{John Grundy},
  \bibinfo{person}{Xiapu Luo}, {and} \bibinfo{person}{Ting Chen}.}
  \bibinfo{year}{2020}\natexlab{c}.
\newblock \showarticletitle{Defining smart contract defects on ethereum}.
\newblock \bibinfo{journal}{\emph{IEEE Transactions on Software Engineering}}
  (\bibinfo{year}{2020}).
\newblock


\bibitem[\protect\citeauthoryear{Chen, Xia, Lo, Grundy, Luo, and Chen}{Chen
  et~al\mbox{.}}{2021}]%
        {chen2021defectchecker}
\bibfield{author}{\bibinfo{person}{Jiachi Chen}, \bibinfo{person}{Xin Xia},
  \bibinfo{person}{David Lo}, \bibinfo{person}{John Grundy},
  \bibinfo{person}{Xiapu Luo}, {and} \bibinfo{person}{Ting Chen}.}
  \bibinfo{year}{2021}\natexlab{}.
\newblock \showarticletitle{Defectchecker: Automated smart contract defect
  detection by analyzing evm bytecode}.
\newblock \bibinfo{journal}{\emph{IEEE Transactions on Software Engineering}}
  (\bibinfo{year}{2021}).
\newblock


\bibitem[\protect\citeauthoryear{Chen, Cao, Li, Luo, Gu, Zhang, Liao, Zhu,
  Chen, He, et~al\mbox{.}}{Chen et~al\mbox{.}}{2020a}]%
        {chen2020soda}
\bibfield{author}{\bibinfo{person}{Ting Chen}, \bibinfo{person}{Rong Cao},
  \bibinfo{person}{Ting Li}, \bibinfo{person}{Xiapu Luo},
  \bibinfo{person}{Guofei Gu}, \bibinfo{person}{Yufei Zhang},
  \bibinfo{person}{Zhou Liao}, \bibinfo{person}{Hang Zhu},
  \bibinfo{person}{Gang Chen}, \bibinfo{person}{Zheyuan He}, {et~al\mbox{.}}}
  \bibinfo{year}{2020}\natexlab{a}.
\newblock \showarticletitle{SODA: A generic online detection framework for
  smart contracts}. In \bibinfo{booktitle}{\emph{27th Ann. Network and
  Distributed Systems Security Symp}}. The Internet Society.
\newblock


\bibitem[\protect\citeauthoryear{Chen, Li, Luo, and Zhang}{Chen
  et~al\mbox{.}}{2017}]%
        {chen2017under}
\bibfield{author}{\bibinfo{person}{Ting Chen}, \bibinfo{person}{Xiaoqi Li},
  \bibinfo{person}{Xiapu Luo}, {and} \bibinfo{person}{Xiaosong Zhang}.}
  \bibinfo{year}{2017}\natexlab{}.
\newblock \showarticletitle{Under-optimized smart contracts devour your money}.
  In \bibinfo{booktitle}{\emph{2017 IEEE 24th International Conference on
  Software Analysis, Evolution and Reengineering (SANER)}}. IEEE,
  \bibinfo{pages}{442--446}.
\newblock


\bibitem[\protect\citeauthoryear{Chen, Zhang, Li, Luo, Wang, Cao, Xiao, and
  Zhang}{Chen et~al\mbox{.}}{2019}]%
        {chen2019tokenscope}
\bibfield{author}{\bibinfo{person}{Ting Chen}, \bibinfo{person}{Yufei Zhang},
  \bibinfo{person}{Zihao Li}, \bibinfo{person}{Xiapu Luo},
  \bibinfo{person}{Ting Wang}, \bibinfo{person}{Rong Cao},
  \bibinfo{person}{Xiuzhuo Xiao}, {and} \bibinfo{person}{Xiaosong Zhang}.}
  \bibinfo{year}{2019}\natexlab{}.
\newblock \showarticletitle{Tokenscope: Automatically detecting inconsistent
  behaviors of cryptocurrency tokens in ethereum}. In
  \bibinfo{booktitle}{\emph{Proceedings of the 2019 ACM SIGSAC Conference on
  Computer and Communications Security}}. \bibinfo{pages}{1503--1520}.
\newblock


\bibitem[\protect\citeauthoryear{Chen, Zheng, Cui, Ngai, Zheng, and Zhou}{Chen
  et~al\mbox{.}}{2018}]%
        {chen2018detecting}
\bibfield{author}{\bibinfo{person}{Weili Chen}, \bibinfo{person}{Zibin Zheng},
  \bibinfo{person}{Jiahui Cui}, \bibinfo{person}{Edith Ngai},
  \bibinfo{person}{Peilin Zheng}, {and} \bibinfo{person}{Yuren Zhou}.}
  \bibinfo{year}{2018}\natexlab{}.
\newblock \showarticletitle{Detecting ponzi schemes on ethereum: Towards
  healthier blockchain technology}. In \bibinfo{booktitle}{\emph{Proceedings of
  the 2018 world wide web conference}}. \bibinfo{pages}{1409--1418}.
\newblock


\bibitem[\protect\citeauthoryear{Chinen, Yanai, Cruz, and Okamura}{Chinen
  et~al\mbox{.}}{2020}]%
        {chinen2020ra}
\bibfield{author}{\bibinfo{person}{Yuchiro Chinen}, \bibinfo{person}{Naoto
  Yanai}, \bibinfo{person}{Jason~Paul Cruz}, {and} \bibinfo{person}{Shingo
  Okamura}.} \bibinfo{year}{2020}\natexlab{}.
\newblock \showarticletitle{RA: Hunting for Re-Entrancy Attacks in Ethereum
  Smart Contracts via Static Analysis}. In \bibinfo{booktitle}{\emph{2020 IEEE
  International Conference on Blockchain (Blockchain)}}. IEEE,
  \bibinfo{pages}{327--336}.
\newblock


\bibitem[\protect\citeauthoryear{Daian, Goldfeder, Kell, Li, Zhao, Bentov,
  Breidenbach, and Juels}{Daian et~al\mbox{.}}{2020}]%
        {daian2020flash}
\bibfield{author}{\bibinfo{person}{Philip Daian}, \bibinfo{person}{Steven
  Goldfeder}, \bibinfo{person}{Tyler Kell}, \bibinfo{person}{Yunqi Li},
  \bibinfo{person}{Xueyuan Zhao}, \bibinfo{person}{Iddo Bentov},
  \bibinfo{person}{Lorenz Breidenbach}, {and} \bibinfo{person}{Ari Juels}.}
  \bibinfo{year}{2020}\natexlab{}.
\newblock \showarticletitle{Flash boys 2.0: Frontrunning in decentralized
  exchanges, miner extractable value, and consensus instability}. In
  \bibinfo{booktitle}{\emph{2020 IEEE Symposium on Security and Privacy (SP)}}.
  IEEE, \bibinfo{pages}{910--927}.
\newblock


\bibitem[\protect\citeauthoryear{Di~Angelo and Salzer}{Di~Angelo and
  Salzer}{2019}]%
        {di2019survey}
\bibfield{author}{\bibinfo{person}{Monika Di~Angelo} {and}
  \bibinfo{person}{Gernot Salzer}.} \bibinfo{year}{2019}\natexlab{}.
\newblock \showarticletitle{A survey of tools for analyzing Ethereum smart
  contracts}. In \bibinfo{booktitle}{\emph{2019 IEEE International Conference
  on Decentralized Applications and Infrastructures (DAPPCON)}}. IEEE,
  \bibinfo{pages}{69--78}.
\newblock


\bibitem[\protect\citeauthoryear{Ding, Li, Li, and Zhang}{Ding
  et~al\mbox{.}}{2021}]%
        {ding2021hfcontractfuzzer}
\bibfield{author}{\bibinfo{person}{Mengjie Ding}, \bibinfo{person}{Peiru Li},
  \bibinfo{person}{Shanshan Li}, {and} \bibinfo{person}{He Zhang}.}
  \bibinfo{year}{2021}\natexlab{}.
\newblock \showarticletitle{HFContractFuzzer: Fuzzing Hyperledger Fabric Smart
  Contracts for Vulnerability Detection}.
\newblock In \bibinfo{booktitle}{\emph{Evaluation and Assessment in Software
  Engineering}}. \bibinfo{pages}{321--328}.
\newblock


\bibitem[\protect\citeauthoryear{Duo, Xin, and Xiaofeng}{Duo
  et~al\mbox{.}}{2020}]%
        {duo2020formal}
\bibfield{author}{\bibinfo{person}{Wang Duo}, \bibinfo{person}{Huang Xin},
  {and} \bibinfo{person}{Ma Xiaofeng}.} \bibinfo{year}{2020}\natexlab{}.
\newblock \showarticletitle{Formal analysis of smart contract based on colored
  petri nets}.
\newblock \bibinfo{journal}{\emph{IEEE Intelligent Systems}}
  \bibinfo{volume}{35}, \bibinfo{number}{3} (\bibinfo{year}{2020}),
  \bibinfo{pages}{19--30}.
\newblock


\bibitem[\protect\citeauthoryear{Ellul and Pace}{Ellul and Pace}{2018}]%
        {ellul2018runtime}
\bibfield{author}{\bibinfo{person}{Joshua Ellul} {and}
  \bibinfo{person}{Gordon~J Pace}.} \bibinfo{year}{2018}\natexlab{}.
\newblock \showarticletitle{Runtime verification of ethereum smart contracts}.
  In \bibinfo{booktitle}{\emph{2018 14th European Dependable Computing
  Conference (EDCC)}}. IEEE, \bibinfo{pages}{158--163}.
\newblock


\bibitem[\protect\citeauthoryear{Feist, Grieco, and Groce}{Feist
  et~al\mbox{.}}{2019}]%
        {feist2019slither}
\bibfield{author}{\bibinfo{person}{Josselin Feist}, \bibinfo{person}{Gustavo
  Grieco}, {and} \bibinfo{person}{Alex Groce}.}
  \bibinfo{year}{2019}\natexlab{}.
\newblock \showarticletitle{Slither: a static analysis framework for smart
  contracts}. In \bibinfo{booktitle}{\emph{2019 IEEE/ACM 2nd International
  Workshop on Emerging Trends in Software Engineering for Blockchain
  (WETSEB)}}. IEEE, \bibinfo{pages}{8--15}.
\newblock


\bibitem[\protect\citeauthoryear{Feng, Torlak, and Bodik}{Feng
  et~al\mbox{.}}{2019}]%
        {feng2019precise}
\bibfield{author}{\bibinfo{person}{Yu Feng}, \bibinfo{person}{Emina Torlak},
  {and} \bibinfo{person}{Rastislav Bodik}.} \bibinfo{year}{2019}\natexlab{}.
\newblock \showarticletitle{Precise attack synthesis for smart contracts}.
\newblock \bibinfo{journal}{\emph{arXiv preprint arXiv:1902.06067}}
  (\bibinfo{year}{2019}).
\newblock


\bibitem[\protect\citeauthoryear{Feng, Torlak, and Bodik}{Feng
  et~al\mbox{.}}{2020}]%
        {feng2020summary}
\bibfield{author}{\bibinfo{person}{Yu Feng}, \bibinfo{person}{Emina Torlak},
  {and} \bibinfo{person}{Rastislav Bodik}.} \bibinfo{year}{2020}\natexlab{}.
\newblock \showarticletitle{Summary-based symbolic evaluation for smart
  contracts}. In \bibinfo{booktitle}{\emph{2020 35th IEEE/ACM International
  Conference on Automated Software Engineering (ASE)}}. IEEE,
  \bibinfo{pages}{1141--1152}.
\newblock


\bibitem[\protect\citeauthoryear{Ferreira~Torres, Baden, Norvill,
  Fiz~Pontiveros, Jonker, and Mauw}{Ferreira~Torres et~al\mbox{.}}{2020}]%
        {ferreira2020aegis}
\bibfield{author}{\bibinfo{person}{Christof Ferreira~Torres},
  \bibinfo{person}{Mathis Baden}, \bibinfo{person}{Robert Norvill},
  \bibinfo{person}{Beltran~Borja Fiz~Pontiveros}, \bibinfo{person}{Hugo
  Jonker}, {and} \bibinfo{person}{Sjouke Mauw}.}
  \bibinfo{year}{2020}\natexlab{}.
\newblock \showarticletitle{{\AE}gis: Shielding vulnerable smart contracts
  against attacks}. In \bibinfo{booktitle}{\emph{Proceedings of the 15th ACM
  Asia Conference on Computer and Communications Security}}.
  \bibinfo{pages}{584--597}.
\newblock


\bibitem[\protect\citeauthoryear{Ferreira~Torres, Iannillo, Gervais,
  et~al\mbox{.}}{Ferreira~Torres et~al\mbox{.}}{2021a}]%
        {ferreira2021confuzzius}
\bibfield{author}{\bibinfo{person}{Christof Ferreira~Torres},
  \bibinfo{person}{Antonio~Ken Iannillo}, \bibinfo{person}{Arthur Gervais},
  {et~al\mbox{.}}} \bibinfo{year}{2021}\natexlab{a}.
\newblock \showarticletitle{CONFUZZIUS: A Data Dependency-Aware Hybrid Fuzzer
  for Smart Contracts}.
\newblock  (\bibinfo{year}{2021}).
\newblock


\bibitem[\protect\citeauthoryear{Ferreira~Torres, Iannillo, Gervais,
  et~al\mbox{.}}{Ferreira~Torres et~al\mbox{.}}{2021b}]%
        {ferreira2021eye}
\bibfield{author}{\bibinfo{person}{Christof Ferreira~Torres},
  \bibinfo{person}{Antonio~Ken Iannillo}, \bibinfo{person}{Arthur Gervais},
  {et~al\mbox{.}}} \bibinfo{year}{2021}\natexlab{b}.
\newblock \showarticletitle{The Eye of Horus: Spotting and Analyzing Attacks on
  Ethereum Smart Contracts}. In \bibinfo{booktitle}{\emph{International
  Conference on Financial Cryptography and Data Security, Grenada 1-5 March
  2021}}.
\newblock


\bibitem[\protect\citeauthoryear{Frank, Aschermann, and Holz}{Frank
  et~al\mbox{.}}{2020}]%
        {frank2020ethbmc}
\bibfield{author}{\bibinfo{person}{Joel Frank}, \bibinfo{person}{Cornelius
  Aschermann}, {and} \bibinfo{person}{Thorsten Holz}.}
  \bibinfo{year}{2020}\natexlab{}.
\newblock \showarticletitle{{ETHBMC}: A Bounded Model Checker for Smart
  Contracts}. In \bibinfo{booktitle}{\emph{29th {USENIX} Security Symposium
  ({USENIX} Security 20)}}. \bibinfo{pages}{2757--2774}.
\newblock


\bibitem[\protect\citeauthoryear{Frontera}{Frontera}{[n.\,d.]}]%
        {thedao}
\bibfield{author}{\bibinfo{person}{Ernesto Frontera}.}
  \bibinfo{year}{[n.\,d.]}\natexlab{}.
\newblock \bibinfo{title}{{A History of The DAO Hack}}.
\newblock
  \bibinfo{howpublished}{{https://coinmarketcap.com/alexandria/article/a-history-of-the-dao-hack}}.
\newblock
\newblock
\shownote{Accessed: 2022-02-15}.


\bibitem[\protect\citeauthoryear{Fu, Ren, Ma, Jiang, Shi, and Sun}{Fu
  et~al\mbox{.}}{2019a}]%
        {fu2019evmfuzz}
\bibfield{author}{\bibinfo{person}{Ying Fu}, \bibinfo{person}{Meng Ren},
  \bibinfo{person}{Fuchen Ma}, \bibinfo{person}{Yu Jiang},
  \bibinfo{person}{Heyuan Shi}, {and} \bibinfo{person}{Jiaguang Sun}.}
  \bibinfo{year}{2019}\natexlab{a}.
\newblock \showarticletitle{Evmfuzz: Differential fuzz testing of ethereum
  virtual machine}.
\newblock \bibinfo{journal}{\emph{arXiv preprint arXiv:1903.08483}}
  (\bibinfo{year}{2019}).
\newblock


\bibitem[\protect\citeauthoryear{Fu, Ren, Ma, Shi, Yang, Jiang, Li, and Shi}{Fu
  et~al\mbox{.}}{2019b}]%
        {fu2019evmfuzzer}
\bibfield{author}{\bibinfo{person}{Ying Fu}, \bibinfo{person}{Meng Ren},
  \bibinfo{person}{Fuchen Ma}, \bibinfo{person}{Heyuan Shi},
  \bibinfo{person}{Xin Yang}, \bibinfo{person}{Yu Jiang},
  \bibinfo{person}{Huizhong Li}, {and} \bibinfo{person}{Xiang Shi}.}
  \bibinfo{year}{2019}\natexlab{b}.
\newblock \showarticletitle{Evmfuzzer: detect evm vulnerabilities via fuzz
  testing}. In \bibinfo{booktitle}{\emph{Proceedings of the 2019 27th ACM Joint
  Meeting on European Software Engineering Conference and Symposium on the
  Foundations of Software Engineering}}. \bibinfo{pages}{1110--1114}.
\newblock


\bibitem[\protect\citeauthoryear{Gao, Liu, Liu, Li, Guan, and Chen}{Gao
  et~al\mbox{.}}{2019}]%
        {gao2019easyflow}
\bibfield{author}{\bibinfo{person}{Jianbo Gao}, \bibinfo{person}{Han Liu},
  \bibinfo{person}{Chao Liu}, \bibinfo{person}{Qingshan Li},
  \bibinfo{person}{Zhi Guan}, {and} \bibinfo{person}{Zhong Chen}.}
  \bibinfo{year}{2019}\natexlab{}.
\newblock \showarticletitle{Easyflow: Keep ethereum away from overflow}. In
  \bibinfo{booktitle}{\emph{2019 IEEE/ACM 41st International Conference on
  Software Engineering: Companion Proceedings (ICSE-Companion)}}. IEEE,
  \bibinfo{pages}{23--26}.
\newblock


\bibitem[\protect\citeauthoryear{Goodin}{Goodin}{2021}]%
        {expl1}
\bibfield{author}{\bibinfo{person}{Dan Goodin}.}
  \bibinfo{year}{2021}\natexlab{}.
\newblock \bibinfo{title}{Really stupid “smart contract” bug let hackers
  steal \$31 million in digital coin}.
\newblock
  \bibinfo{howpublished}{{https://arstechnica.com/information-technology/2021/12/hackers-drain-31-million-from-cryptocurrency-service-monox-finance/}}.
\newblock


\bibitem[\protect\citeauthoryear{Grech, Kong, Jurisevic, Brent, Scholz, and
  Smaragdakis}{Grech et~al\mbox{.}}{2018}]%
        {grech2018madmax}
\bibfield{author}{\bibinfo{person}{Neville Grech}, \bibinfo{person}{Michael
  Kong}, \bibinfo{person}{Anton Jurisevic}, \bibinfo{person}{Lexi Brent},
  \bibinfo{person}{Bernhard Scholz}, {and} \bibinfo{person}{Yannis
  Smaragdakis}.} \bibinfo{year}{2018}\natexlab{}.
\newblock \showarticletitle{Madmax: Surviving out-of-gas conditions in ethereum
  smart contracts}.
\newblock \bibinfo{journal}{\emph{Proceedings of the ACM on Programming
  Languages}} \bibinfo{volume}{2}, \bibinfo{number}{OOPSLA}
  (\bibinfo{year}{2018}), \bibinfo{pages}{1--27}.
\newblock


\bibitem[\protect\citeauthoryear{Grieco, Song, Cygan, Feist, and Groce}{Grieco
  et~al\mbox{.}}{2020}]%
        {grieco2020echidna}
\bibfield{author}{\bibinfo{person}{Gustavo Grieco}, \bibinfo{person}{Will
  Song}, \bibinfo{person}{Artur Cygan}, \bibinfo{person}{Josselin Feist}, {and}
  \bibinfo{person}{Alex Groce}.} \bibinfo{year}{2020}\natexlab{}.
\newblock \showarticletitle{Echidna: effective, usable, and fast fuzzing for
  smart contracts}. In \bibinfo{booktitle}{\emph{Proceedings of the 29th ACM
  SIGSOFT International Symposium on Software Testing and Analysis}}.
  \bibinfo{pages}{557--560}.
\newblock


\bibitem[\protect\citeauthoryear{Grishchenko, Maffei, and
  Schneidewind}{Grishchenko et~al\mbox{.}}{2018a}]%
        {grishchenko2018ethertrust}
\bibfield{author}{\bibinfo{person}{Ilya Grishchenko}, \bibinfo{person}{Matteo
  Maffei}, {and} \bibinfo{person}{Clara Schneidewind}.}
  \bibinfo{year}{2018}\natexlab{a}.
\newblock \showarticletitle{Ethertrust: Sound static analysis of ethereum
  bytecode}.
\newblock \bibinfo{journal}{\emph{Technische Universit{\"a}t Wien, Tech. Rep}}
  (\bibinfo{year}{2018}).
\newblock


\bibitem[\protect\citeauthoryear{Grishchenko, Maffei, and
  Schneidewind}{Grishchenko et~al\mbox{.}}{2018b}]%
        {grishchenko2018semantic}
\bibfield{author}{\bibinfo{person}{Ilya Grishchenko}, \bibinfo{person}{Matteo
  Maffei}, {and} \bibinfo{person}{Clara Schneidewind}.}
  \bibinfo{year}{2018}\natexlab{b}.
\newblock \showarticletitle{A semantic framework for the security analysis of
  ethereum smart contracts}. In \bibinfo{booktitle}{\emph{International
  Conference on Principles of Security and Trust}}. Springer,
  \bibinfo{pages}{243--269}.
\newblock


\bibitem[\protect\citeauthoryear{Grossman, Abraham, Golan-Gueta, Michalevsky,
  Rinetzky, Sagiv, and Zohar}{Grossman et~al\mbox{.}}{2017}]%
        {grossman2017online}
\bibfield{author}{\bibinfo{person}{Shelly Grossman}, \bibinfo{person}{Ittai
  Abraham}, \bibinfo{person}{Guy Golan-Gueta}, \bibinfo{person}{Yan
  Michalevsky}, \bibinfo{person}{Noam Rinetzky}, \bibinfo{person}{Mooly Sagiv},
  {and} \bibinfo{person}{Yoni Zohar}.} \bibinfo{year}{2017}\natexlab{}.
\newblock \showarticletitle{Online detection of effectively callback free
  objects with applications to smart contracts}.
\newblock \bibinfo{journal}{\emph{Proceedings of the ACM on Programming
  Languages}} \bibinfo{volume}{2}, \bibinfo{number}{POPL}
  (\bibinfo{year}{2017}), \bibinfo{pages}{1--28}.
\newblock


\bibitem[\protect\citeauthoryear{Hajdu and Jovanovi{\'c}}{Hajdu and
  Jovanovi{\'c}}{2019}]%
        {hajdu2019solc}
\bibfield{author}{\bibinfo{person}{{\'A}kos Hajdu} {and} \bibinfo{person}{Dejan
  Jovanovi{\'c}}.} \bibinfo{year}{2019}\natexlab{}.
\newblock \showarticletitle{solc-verify: A modular verifier for Solidity smart
  contracts}.
\newblock \bibinfo{journal}{\emph{arXiv preprint arXiv:1907.04262}}
  (\bibinfo{year}{2019}).
\newblock


\bibitem[\protect\citeauthoryear{Hajdu, Jovanovi{\'c}, and Ciocarlie}{Hajdu
  et~al\mbox{.}}{2020}]%
        {hajdu2020formal}
\bibfield{author}{\bibinfo{person}{{\'A}kos Hajdu}, \bibinfo{person}{Dejan
  Jovanovi{\'c}}, {and} \bibinfo{person}{Gabriela Ciocarlie}.}
  \bibinfo{year}{2020}\natexlab{}.
\newblock \showarticletitle{Formal Specification and Verification of Solidity
  Contracts with Events (Short Paper)}. In \bibinfo{booktitle}{\emph{2nd
  Workshop on Formal Methods for Blockchains (FMBC 2020)}}. Schloss
  Dagstuhl-Leibniz-Zentrum f{\"u}r Informatik.
\newblock


\bibitem[\protect\citeauthoryear{Harz and Knottenbelt}{Harz and
  Knottenbelt}{2018}]%
        {harz2018towards}
\bibfield{author}{\bibinfo{person}{Dominik Harz} {and} \bibinfo{person}{William
  Knottenbelt}.} \bibinfo{year}{2018}\natexlab{}.
\newblock \showarticletitle{Towards safer smart contracts: A survey of
  languages and verification methods}.
\newblock \bibinfo{journal}{\emph{arXiv preprint arXiv:1809.09805}}
  (\bibinfo{year}{2018}).
\newblock


\bibitem[\protect\citeauthoryear{He, Balunovi{\'c}, Ambroladze, Tsankov, and
  Vechev}{He et~al\mbox{.}}{2019}]%
        {he2019learning}
\bibfield{author}{\bibinfo{person}{Jingxuan He}, \bibinfo{person}{Mislav
  Balunovi{\'c}}, \bibinfo{person}{Nodar Ambroladze}, \bibinfo{person}{Petar
  Tsankov}, {and} \bibinfo{person}{Martin Vechev}.}
  \bibinfo{year}{2019}\natexlab{}.
\newblock \showarticletitle{Learning to fuzz from symbolic execution with
  application to smart contracts}. In \bibinfo{booktitle}{\emph{Proceedings of
  the 2019 ACM SIGSAC Conference on Computer and Communications Security}}.
  \bibinfo{pages}{531--548}.
\newblock


\bibitem[\protect\citeauthoryear{He, Zhang, Wang, Wu, Luo, Guo, Yu, and
  Jiang}{He et~al\mbox{.}}{2021}]%
        {he2021eosafe}
\bibfield{author}{\bibinfo{person}{Ningyu He}, \bibinfo{person}{Ruiyi Zhang},
  \bibinfo{person}{Haoyu Wang}, \bibinfo{person}{Lei Wu},
  \bibinfo{person}{Xiapu Luo}, \bibinfo{person}{Yao Guo}, \bibinfo{person}{Ting
  Yu}, {and} \bibinfo{person}{Xuxian Jiang}.} \bibinfo{year}{2021}\natexlab{}.
\newblock \showarticletitle{$\{$EOSAFE$\}$: Security Analysis of $\{$EOSIO$\}$
  Smart Contracts}. In \bibinfo{booktitle}{\emph{30th USENIX Security Symposium
  (USENIX Security 21)}}. \bibinfo{pages}{1271--1288}.
\newblock


\bibitem[\protect\citeauthoryear{Hildenbrandt, Saxena, Rodrigues, Zhu, Daian,
  Guth, Moore, Park, Zhang, Stefanescu, et~al\mbox{.}}{Hildenbrandt
  et~al\mbox{.}}{2018}]%
        {hildenbrandt2018kevm}
\bibfield{author}{\bibinfo{person}{Everett Hildenbrandt},
  \bibinfo{person}{Manasvi Saxena}, \bibinfo{person}{Nishant Rodrigues},
  \bibinfo{person}{Xiaoran Zhu}, \bibinfo{person}{Philip Daian},
  \bibinfo{person}{Dwight Guth}, \bibinfo{person}{Brandon Moore},
  \bibinfo{person}{Daejun Park}, \bibinfo{person}{Yi Zhang},
  \bibinfo{person}{Andrei Stefanescu}, {et~al\mbox{.}}}
  \bibinfo{year}{2018}\natexlab{}.
\newblock \showarticletitle{Kevm: A complete formal semantics of the ethereum
  virtual machine}. In \bibinfo{booktitle}{\emph{2018 IEEE 31st Computer
  Security Foundations Symposium (CSF)}}. IEEE, \bibinfo{pages}{204--217}.
\newblock


\bibitem[\protect\citeauthoryear{Hu, Zhang, Liu, Liu, Yin, Lu, and Lin}{Hu
  et~al\mbox{.}}{2021b}]%
        {hu2021comprehensive}
\bibfield{author}{\bibinfo{person}{Bin Hu}, \bibinfo{person}{Zongyang Zhang},
  \bibinfo{person}{Jianwei Liu}, \bibinfo{person}{Yizhong Liu},
  \bibinfo{person}{Jiayuan Yin}, \bibinfo{person}{Rongxing Lu}, {and}
  \bibinfo{person}{Xiaodong Lin}.} \bibinfo{year}{2021}\natexlab{b}.
\newblock \showarticletitle{A comprehensive survey on smart contract
  construction and execution: paradigms, tools, and systems}.
\newblock \bibinfo{journal}{\emph{Patterns}} \bibinfo{volume}{2},
  \bibinfo{number}{2} (\bibinfo{year}{2021}), \bibinfo{pages}{100179}.
\newblock


\bibitem[\protect\citeauthoryear{Hu, Liu, Chen, Zhang, Huang, Niu, Lu, Zhou,
  and Liu}{Hu et~al\mbox{.}}{2021a}]%
        {hu2021transaction}
\bibfield{author}{\bibinfo{person}{Teng Hu}, \bibinfo{person}{Xiaolei Liu},
  \bibinfo{person}{Ting Chen}, \bibinfo{person}{Xiaosong Zhang},
  \bibinfo{person}{Xiaoming Huang}, \bibinfo{person}{Weina Niu},
  \bibinfo{person}{Jiazhong Lu}, \bibinfo{person}{Kun Zhou}, {and}
  \bibinfo{person}{Yuan Liu}.} \bibinfo{year}{2021}\natexlab{a}.
\newblock \showarticletitle{Transaction-based classification and detection
  approach for Ethereum smart contract}.
\newblock \bibinfo{journal}{\emph{Information Processing \& Management}}
  \bibinfo{volume}{58}, \bibinfo{number}{2} (\bibinfo{year}{2021}),
  \bibinfo{pages}{102462}.
\newblock


\bibitem[\protect\citeauthoryear{Hu, Zhuang, Lin, Zhang, Kan, and Cao}{Hu
  et~al\mbox{.}}{2021c}]%
        {hu2021security}
\bibfield{author}{\bibinfo{person}{Xinwen Hu}, \bibinfo{person}{Yi Zhuang},
  \bibinfo{person}{Shang-Wei Lin}, \bibinfo{person}{Fuyuan Zhang},
  \bibinfo{person}{Shuanglong Kan}, {and} \bibinfo{person}{Zining Cao}.}
  \bibinfo{year}{2021}\natexlab{c}.
\newblock \showarticletitle{A Security Type Verifier for Smart Contracts}.
\newblock \bibinfo{journal}{\emph{Computers \& Security}}
  (\bibinfo{year}{2021}), \bibinfo{pages}{102343}.
\newblock


\bibitem[\protect\citeauthoryear{Huang, Han, You, Shi, Liang, Wu, and Wu}{Huang
  et~al\mbox{.}}{2021}]%
        {huang2021hunting}
\bibfield{author}{\bibinfo{person}{Jianjun Huang}, \bibinfo{person}{Songming
  Han}, \bibinfo{person}{Wei You}, \bibinfo{person}{Wenchang Shi},
  \bibinfo{person}{Bin Liang}, \bibinfo{person}{Jingzheng Wu}, {and}
  \bibinfo{person}{Yanjun Wu}.} \bibinfo{year}{2021}\natexlab{}.
\newblock \showarticletitle{Hunting Vulnerable Smart Contracts via Graph
  Embedding Based Bytecode Matching}.
\newblock \bibinfo{journal}{\emph{IEEE Transactions on Information Forensics
  and Security}}  \bibinfo{volume}{16} (\bibinfo{year}{2021}),
  \bibinfo{pages}{2144--2156}.
\newblock


\bibitem[\protect\citeauthoryear{Ivanov, Guo, and Yan}{Ivanov
  et~al\mbox{.}}{2021a}]%
        {ivanov2021rectifying}
\bibfield{author}{\bibinfo{person}{Nikolay Ivanov}, \bibinfo{person}{Hanqing
  Guo}, {and} \bibinfo{person}{Qiben Yan}.} \bibinfo{year}{2021}\natexlab{a}.
\newblock \showarticletitle{Rectifying Administrated ERC20 Tokens}. In
  \bibinfo{booktitle}{\emph{International Conference on Information and
  Communications Security}}. Springer, \bibinfo{pages}{22--37}.
\newblock


\bibitem[\protect\citeauthoryear{Ivanov, Lou, Chen, Li, and Yan}{Ivanov
  et~al\mbox{.}}{2021b}]%
        {ivanov2021targeting}
\bibfield{author}{\bibinfo{person}{Nikolay Ivanov}, \bibinfo{person}{Jianzhi
  Lou}, \bibinfo{person}{Ting Chen}, \bibinfo{person}{Jin Li}, {and}
  \bibinfo{person}{Qiben Yan}.} \bibinfo{year}{2021}\natexlab{b}.
\newblock \showarticletitle{Targeting the Weakest Link: Social Engineering
  Attacks in Ethereum Smart Contracts}. In
  \bibinfo{booktitle}{\emph{Proceedings of the 2021 ACM Asia Conference on
  Computer and Communications Security}}. \bibinfo{pages}{787--801}.
\newblock


\bibitem[\protect\citeauthoryear{Ivanov and Yan}{Ivanov and Yan}{2021}]%
        {ivanov2021ethclipper}
\bibfield{author}{\bibinfo{person}{Nikolay Ivanov} {and} \bibinfo{person}{Qiben
  Yan}.} \bibinfo{year}{2021}\natexlab{}.
\newblock \showarticletitle{Ethclipper: a clipboard meddling attack on hardware
  wallets with address verification evasion}. In \bibinfo{booktitle}{\emph{2021
  IEEE Conference on Communications and Network Security (CNS)}}. IEEE,
  \bibinfo{pages}{191--199}.
\newblock


\bibitem[\protect\citeauthoryear{Ivanov, Yan, and Kompalli}{Ivanov
  et~al\mbox{.}}{2023}]%
        {ivanov2023txt}
\bibfield{author}{\bibinfo{person}{Nikolay Ivanov}, \bibinfo{person}{Qiben
  Yan}, {and} \bibinfo{person}{Anurag Kompalli}.}
  \bibinfo{year}{2023}\natexlab{}.
\newblock \showarticletitle{TxT: Real-time Transaction Encapsulation for
  Ethereum Smart Contracts}.
\newblock \bibinfo{journal}{\emph{IEEE Transactions on Information Forensics
  and Security}} (\bibinfo{year}{2023}).
\newblock


\bibitem[\protect\citeauthoryear{Jiang, Liu, and Chan}{Jiang
  et~al\mbox{.}}{2018}]%
        {jiang2018contractfuzzer}
\bibfield{author}{\bibinfo{person}{Bo Jiang}, \bibinfo{person}{Ye Liu}, {and}
  \bibinfo{person}{WK Chan}.} \bibinfo{year}{2018}\natexlab{}.
\newblock \showarticletitle{Contractfuzzer: Fuzzing smart contracts for
  vulnerability detection}. In \bibinfo{booktitle}{\emph{2018 33rd IEEE/ACM
  International Conference on Automated Software Engineering (ASE)}}. IEEE,
  \bibinfo{pages}{259--269}.
\newblock


\bibitem[\protect\citeauthoryear{Jiao, Kan, Lin, Sanan, Liu, and Sun}{Jiao
  et~al\mbox{.}}{2020}]%
        {jiao2020semantic}
\bibfield{author}{\bibinfo{person}{Jiao Jiao}, \bibinfo{person}{Shuanglong
  Kan}, \bibinfo{person}{Shang-Wei Lin}, \bibinfo{person}{David Sanan},
  \bibinfo{person}{Yang Liu}, {and} \bibinfo{person}{Jun Sun}.}
  \bibinfo{year}{2020}\natexlab{}.
\newblock \showarticletitle{Semantic understanding of smart contracts:
  executable operational semantics of Solidity}. In
  \bibinfo{booktitle}{\emph{2020 IEEE Symposium on Security and Privacy (SP)}}.
  IEEE, \bibinfo{pages}{1695--1712}.
\newblock


\bibitem[\protect\citeauthoryear{Jin, Cao, Chen, Zhang, and Campanoni}{Jin
  et~al\mbox{.}}{2022}]%
        {jin2022exgen}
\bibfield{author}{\bibinfo{person}{Ling Jin}, \bibinfo{person}{Yinzhi Cao},
  \bibinfo{person}{Yan Chen}, \bibinfo{person}{Di Zhang}, {and}
  \bibinfo{person}{Simone Campanoni}.} \bibinfo{year}{2022}\natexlab{}.
\newblock \showarticletitle{EXGEN: Cross-platform, Automated Exploit Generation
  for Smart Contract Vulnerabilities}.
\newblock \bibinfo{journal}{\emph{IEEE Transactions on Dependable and Secure
  Computing}} (\bibinfo{year}{2022}).
\newblock


\bibitem[\protect\citeauthoryear{Kalra, Goel, Dhawan, and Sharma}{Kalra
  et~al\mbox{.}}{2018}]%
        {kalra2018zeus}
\bibfield{author}{\bibinfo{person}{Sukrit Kalra}, \bibinfo{person}{Seep Goel},
  \bibinfo{person}{Mohan Dhawan}, {and} \bibinfo{person}{Subodh Sharma}.}
  \bibinfo{year}{2018}\natexlab{}.
\newblock \showarticletitle{ZEUS: Analyzing Safety of Smart Contracts.}. In
  \bibinfo{booktitle}{\emph{Ndss}}. \bibinfo{pages}{1--12}.
\newblock


\bibitem[\protect\citeauthoryear{King}{King}{1976}]%
        {king1976symbolic}
\bibfield{author}{\bibinfo{person}{James~C King}.}
  \bibinfo{year}{1976}\natexlab{}.
\newblock \showarticletitle{Symbolic execution and program testing}.
\newblock \bibinfo{journal}{\emph{Commun. ACM}} \bibinfo{volume}{19},
  \bibinfo{number}{7} (\bibinfo{year}{1976}), \bibinfo{pages}{385--394}.
\newblock


\bibitem[\protect\citeauthoryear{Kolluri, Nikolic, Sergey, Hobor, and
  Saxena}{Kolluri et~al\mbox{.}}{2019}]%
        {kolluri2019exploiting}
\bibfield{author}{\bibinfo{person}{Aashish Kolluri}, \bibinfo{person}{Ivica
  Nikolic}, \bibinfo{person}{Ilya Sergey}, \bibinfo{person}{Aquinas Hobor},
  {and} \bibinfo{person}{Prateek Saxena}.} \bibinfo{year}{2019}\natexlab{}.
\newblock \showarticletitle{Exploiting the laws of order in smart contracts}.
  In \bibinfo{booktitle}{\emph{Proceedings of the 28th ACM SIGSOFT
  international symposium on software testing and analysis}}.
  \bibinfo{pages}{363--373}.
\newblock


\bibitem[\protect\citeauthoryear{Kongmanee, Kijsanayothin, and
  Hewett}{Kongmanee et~al\mbox{.}}{2019}]%
        {kongmanee2019securing}
\bibfield{author}{\bibinfo{person}{Jaturong Kongmanee},
  \bibinfo{person}{Phongphun Kijsanayothin}, {and} \bibinfo{person}{Rattikorn
  Hewett}.} \bibinfo{year}{2019}\natexlab{}.
\newblock \showarticletitle{Securing smart contracts in blockchain}. In
  \bibinfo{booktitle}{\emph{2019 34th IEEE/ACM International Conference on
  Automated Software Engineering Workshop (ASEW)}}. IEEE,
  \bibinfo{pages}{69--76}.
\newblock


\bibitem[\protect\citeauthoryear{Krupp and Rossow}{Krupp and Rossow}{2018}]%
        {krupp2018teether}
\bibfield{author}{\bibinfo{person}{Johannes Krupp} {and}
  \bibinfo{person}{Christian Rossow}.} \bibinfo{year}{2018}\natexlab{}.
\newblock \showarticletitle{teether: Gnawing at ethereum to automatically
  exploit smart contracts}. In \bibinfo{booktitle}{\emph{27th {USENIX} Security
  Symposium ({USENIX} Security 18)}}. \bibinfo{pages}{1317--1333}.
\newblock


\bibitem[\protect\citeauthoryear{Le, Xu, Chen, and Shi}{Le
  et~al\mbox{.}}{2018}]%
        {le2018proving}
\bibfield{author}{\bibinfo{person}{Ton~Chanh Le}, \bibinfo{person}{Lei Xu},
  \bibinfo{person}{Lin Chen}, {and} \bibinfo{person}{Weidong Shi}.}
  \bibinfo{year}{2018}\natexlab{}.
\newblock \showarticletitle{Proving conditional termination for smart
  contracts}. In \bibinfo{booktitle}{\emph{Proceedings of the 2nd ACM Workshop
  on Blockchains, Cryptocurrencies, and Contracts}}. \bibinfo{pages}{57--59}.
\newblock


\bibitem[\protect\citeauthoryear{Li, Choi, and Long}{Li et~al\mbox{.}}{2020a}]%
        {li2020securing}
\bibfield{author}{\bibinfo{person}{Ao Li}, \bibinfo{person}{Jemin~Andrew Choi},
  {and} \bibinfo{person}{Fan Long}.} \bibinfo{year}{2020}\natexlab{a}.
\newblock \showarticletitle{Securing smart contract with runtime validation}.
  In \bibinfo{booktitle}{\emph{Proceedings of the 41st ACM SIGPLAN Conference
  on Programming Language Design and Implementation}}.
  \bibinfo{pages}{438--453}.
\newblock


\bibitem[\protect\citeauthoryear{Li, Jiang, Chen, Luo, and Wen}{Li
  et~al\mbox{.}}{2020b}]%
        {li2020survey}
\bibfield{author}{\bibinfo{person}{Xiaoqi Li}, \bibinfo{person}{Peng Jiang},
  \bibinfo{person}{Ting Chen}, \bibinfo{person}{Xiapu Luo}, {and}
  \bibinfo{person}{Qiaoyan Wen}.} \bibinfo{year}{2020}\natexlab{b}.
\newblock \showarticletitle{A survey on the security of blockchain systems}.
\newblock \bibinfo{journal}{\emph{Future Generation Computer Systems}}
  \bibinfo{volume}{107} (\bibinfo{year}{2020}), \bibinfo{pages}{841--853}.
\newblock


\bibitem[\protect\citeauthoryear{Li, Su, Xiong, Huang, and Wang}{Li
  et~al\mbox{.}}{2019}]%
        {li2019formal}
\bibfield{author}{\bibinfo{person}{Xiaoyu Li}, \bibinfo{person}{Cheng Su},
  \bibinfo{person}{Yan Xiong}, \bibinfo{person}{Wenchao Huang}, {and}
  \bibinfo{person}{Wansen Wang}.} \bibinfo{year}{2019}\natexlab{}.
\newblock \showarticletitle{Formal verification of BNB smart contract}. In
  \bibinfo{booktitle}{\emph{2019 5th International Conference on Big Data
  Computing and Communications (BIGCOM)}}. IEEE, \bibinfo{pages}{74--78}.
\newblock


\bibitem[\protect\citeauthoryear{Li}{Li}{2019}]%
        {li2019finding}
\bibfield{author}{\bibinfo{person}{Yue Li}.} \bibinfo{year}{2019}\natexlab{}.
\newblock \showarticletitle{Finding concurrency exploits on smart contracts}.
  In \bibinfo{booktitle}{\emph{2019 IEEE/ACM 41st International Conference on
  Software Engineering: Companion Proceedings (ICSE-Companion)}}. IEEE,
  \bibinfo{pages}{144--146}.
\newblock


\bibitem[\protect\citeauthoryear{Li, Liu, Yang, Ren, Wang, and Chen}{Li
  et~al\mbox{.}}{2020c}]%
        {li2020safepay}
\bibfield{author}{\bibinfo{person}{Yue Li}, \bibinfo{person}{Han Liu},
  \bibinfo{person}{Zhiqiang Yang}, \bibinfo{person}{Qian Ren},
  \bibinfo{person}{Lei Wang}, {and} \bibinfo{person}{Bangdao Chen}.}
  \bibinfo{year}{2020}\natexlab{c}.
\newblock \showarticletitle{SafePay on Ethereum: A Framework For Detecting
  Unfair Payments in Smart Contracts}. In \bibinfo{booktitle}{\emph{2020 IEEE
  40th International Conference on Distributed Computing Systems (ICDCS)}}.
  IEEE, \bibinfo{pages}{1219--1222}.
\newblock


\bibitem[\protect\citeauthoryear{Liao, Tsai, He, and Tien}{Liao
  et~al\mbox{.}}{2019}]%
        {liao2019soliaudit}
\bibfield{author}{\bibinfo{person}{Jian-Wei Liao}, \bibinfo{person}{Tsung-Ta
  Tsai}, \bibinfo{person}{Chia-Kang He}, {and} \bibinfo{person}{Chin-Wei
  Tien}.} \bibinfo{year}{2019}\natexlab{}.
\newblock \showarticletitle{Soliaudit: Smart contract vulnerability assessment
  based on machine learning and fuzz testing}. In
  \bibinfo{booktitle}{\emph{2019 Sixth International Conference on Internet of
  Things: Systems, Management and Security (IOTSMS)}}. IEEE,
  \bibinfo{pages}{458--465}.
\newblock


\bibitem[\protect\citeauthoryear{Linoy, Ray, and Stakhanova}{Linoy
  et~al\mbox{.}}{[n.\,d.]}]%
        {linoyetherprov}
\bibfield{author}{\bibinfo{person}{Shlomi Linoy}, \bibinfo{person}{Suprio Ray},
  {and} \bibinfo{person}{Natalia Stakhanova}.}
  \bibinfo{year}{[n.\,d.]}\natexlab{}.
\newblock \showarticletitle{EtherProv: provenance-aware detection, analysis,
  and mitigation of Ethereum smart contract security issues}.
\newblock  (\bibinfo{year}{[n.\,d.]}).
\newblock


\bibitem[\protect\citeauthoryear{Liu, Liu, Cao, Chen, Chen, and Roscoe}{Liu
  et~al\mbox{.}}{2018a}]%
        {liu2018reguard}
\bibfield{author}{\bibinfo{person}{Chao Liu}, \bibinfo{person}{Han Liu},
  \bibinfo{person}{Zhao Cao}, \bibinfo{person}{Zhong Chen},
  \bibinfo{person}{Bangdao Chen}, {and} \bibinfo{person}{Bill Roscoe}.}
  \bibinfo{year}{2018}\natexlab{a}.
\newblock \showarticletitle{Reguard: finding reentrancy bugs in smart
  contracts}. In \bibinfo{booktitle}{\emph{2018 IEEE/ACM 40th International
  Conference on Software Engineering: Companion (ICSE-Companion)}}. IEEE,
  \bibinfo{pages}{65--68}.
\newblock


\bibitem[\protect\citeauthoryear{Liu, Liu, Zhao, Jiang, and Sun}{Liu
  et~al\mbox{.}}{2018b}]%
        {liu2018s}
\bibfield{author}{\bibinfo{person}{Han Liu}, \bibinfo{person}{Chao Liu},
  \bibinfo{person}{Wenqi Zhao}, \bibinfo{person}{Yu Jiang}, {and}
  \bibinfo{person}{Jiaguang Sun}.} \bibinfo{year}{2018}\natexlab{b}.
\newblock \showarticletitle{S-gram: towards semantic-aware security auditing
  for ethereum smart contracts}. In \bibinfo{booktitle}{\emph{2018 33rd
  IEEE/ACM International Conference on Automated Software Engineering (ASE)}}.
  IEEE, \bibinfo{pages}{814--819}.
\newblock


\bibitem[\protect\citeauthoryear{Liu, Li, Lin, and Yan}{Liu
  et~al\mbox{.}}{2020}]%
        {liu2020modcon}
\bibfield{author}{\bibinfo{person}{Ye Liu}, \bibinfo{person}{Yi Li},
  \bibinfo{person}{Shang-Wei Lin}, {and} \bibinfo{person}{Qiang Yan}.}
  \bibinfo{year}{2020}\natexlab{}.
\newblock \showarticletitle{ModCon: A model-based testing platform for smart
  contracts}. In \bibinfo{booktitle}{\emph{Proceedings of the 28th ACM Joint
  Meeting on European Software Engineering Conference and Symposium on the
  Foundations of Software Engineering}}. \bibinfo{pages}{1601--1605}.
\newblock


\bibitem[\protect\citeauthoryear{Liu, Qian, Wang, Zhu, He, and Ji}{Liu
  et~al\mbox{.}}{2021}]%
        {liu2021smart}
\bibfield{author}{\bibinfo{person}{Zhenguang Liu}, \bibinfo{person}{Peng Qian},
  \bibinfo{person}{Xiang Wang}, \bibinfo{person}{Lei Zhu},
  \bibinfo{person}{Qinming He}, {and} \bibinfo{person}{Shouling Ji}.}
  \bibinfo{year}{2021}\natexlab{}.
\newblock \showarticletitle{Smart Contract Vulnerability Detection: From Pure
  Neural Network to Interpretable Graph Feature and Expert Pattern Fusion}.
\newblock \bibinfo{journal}{\emph{arXiv preprint arXiv:2106.09282}}
  (\bibinfo{year}{2021}).
\newblock


\bibitem[\protect\citeauthoryear{Lu, Wang, Zhang, Shi, and Esposito}{Lu
  et~al\mbox{.}}{2019}]%
        {lu2019neucheck}
\bibfield{author}{\bibinfo{person}{Ning Lu}, \bibinfo{person}{Bin Wang},
  \bibinfo{person}{Yongxin Zhang}, \bibinfo{person}{Wenbo Shi}, {and}
  \bibinfo{person}{Christian Esposito}.} \bibinfo{year}{2019}\natexlab{}.
\newblock \showarticletitle{NeuCheck: A more practical Ethereum smart contract
  security analysis tool}.
\newblock \bibinfo{journal}{\emph{Software: Practice and Experience}}
  (\bibinfo{year}{2019}).
\newblock


\bibitem[\protect\citeauthoryear{Lutz, Chen, Fereidooni, Sendner, Dmitrienko,
  Sadeghi, and Koushanfar}{Lutz et~al\mbox{.}}{2021}]%
        {lutz2021escort}
\bibfield{author}{\bibinfo{person}{Oliver Lutz}, \bibinfo{person}{Huili Chen},
  \bibinfo{person}{Hossein Fereidooni}, \bibinfo{person}{Christoph Sendner},
  \bibinfo{person}{Alexandra Dmitrienko}, \bibinfo{person}{Ahmad~Reza Sadeghi},
  {and} \bibinfo{person}{Farinaz Koushanfar}.} \bibinfo{year}{2021}\natexlab{}.
\newblock \showarticletitle{ESCORT: Ethereum Smart COntRacTs Vulnerability
  Detection using Deep Neural Network and Transfer Learning}.
\newblock \bibinfo{journal}{\emph{arXiv preprint arXiv:2103.12607}}
  (\bibinfo{year}{2021}).
\newblock


\bibitem[\protect\citeauthoryear{Luu, Chu, Olickel, Saxena, and Hobor}{Luu
  et~al\mbox{.}}{2016}]%
        {luu2016making}
\bibfield{author}{\bibinfo{person}{Loi Luu}, \bibinfo{person}{Duc-Hiep Chu},
  \bibinfo{person}{Hrishi Olickel}, \bibinfo{person}{Prateek Saxena}, {and}
  \bibinfo{person}{Aquinas Hobor}.} \bibinfo{year}{2016}\natexlab{}.
\newblock \showarticletitle{Making smart contracts smarter}. In
  \bibinfo{booktitle}{\emph{Proceedings of the 2016 ACM SIGSAC conference on
  computer and communications security}}. \bibinfo{pages}{254--269}.
\newblock


\bibitem[\protect\citeauthoryear{Ma, Fu, Ren, Sun, Liu, Jiang, Sun, and Sun}{Ma
  et~al\mbox{.}}{2019a}]%
        {ma2019gasfuzz}
\bibfield{author}{\bibinfo{person}{Fuchen Ma}, \bibinfo{person}{Ying Fu},
  \bibinfo{person}{Meng Ren}, \bibinfo{person}{Wanting Sun},
  \bibinfo{person}{Zhe Liu}, \bibinfo{person}{Yu Jiang}, \bibinfo{person}{Jun
  Sun}, {and} \bibinfo{person}{Jiaguang Sun}.}
  \bibinfo{year}{2019}\natexlab{a}.
\newblock \showarticletitle{Gasfuzz: Generating high gas consumption inputs to
  avoid out-of-gas vulnerability}.
\newblock \bibinfo{journal}{\emph{arXiv preprint arXiv:1910.02945}}
  (\bibinfo{year}{2019}).
\newblock


\bibitem[\protect\citeauthoryear{Ma, Fu, Ren, Wang, Jiang, Zhang, Li, and
  Shi}{Ma et~al\mbox{.}}{2019b}]%
        {ma2019evm}
\bibfield{author}{\bibinfo{person}{Fuchen Ma}, \bibinfo{person}{Ying Fu},
  \bibinfo{person}{Meng Ren}, \bibinfo{person}{Mingzhe Wang},
  \bibinfo{person}{Yu Jiang}, \bibinfo{person}{Kaixiang Zhang},
  \bibinfo{person}{Huizhong Li}, {and} \bibinfo{person}{Xiang Shi}.}
  \bibinfo{year}{2019}\natexlab{b}.
\newblock \showarticletitle{EVM*: from offline detection to online
  reinforcement for ethereum virtual machine}. In
  \bibinfo{booktitle}{\emph{2019 IEEE 26th International Conference on Software
  Analysis, Evolution and Reengineering (SANER)}}. IEEE,
  \bibinfo{pages}{554--558}.
\newblock


\bibitem[\protect\citeauthoryear{Marescotti, Otoni, Alt, Eugster,
  Hyv{\"a}rinen, and Sharygina}{Marescotti et~al\mbox{.}}{2020}]%
        {marescotti2020accurate}
\bibfield{author}{\bibinfo{person}{Matteo Marescotti}, \bibinfo{person}{Rodrigo
  Otoni}, \bibinfo{person}{Leonardo Alt}, \bibinfo{person}{Patrick Eugster},
  \bibinfo{person}{Antti~EJ Hyv{\"a}rinen}, {and} \bibinfo{person}{Natasha
  Sharygina}.} \bibinfo{year}{2020}\natexlab{}.
\newblock \showarticletitle{Accurate smart contract verification through direct
  modelling}. In \bibinfo{booktitle}{\emph{International Symposium on
  Leveraging Applications of Formal Methods}}. Springer,
  \bibinfo{pages}{178--194}.
\newblock


\bibitem[\protect\citeauthoryear{Mavridou and Laszka}{Mavridou and
  Laszka}{2018}]%
        {mavridou2018designing}
\bibfield{author}{\bibinfo{person}{Anastasia Mavridou} {and}
  \bibinfo{person}{Aron Laszka}.} \bibinfo{year}{2018}\natexlab{}.
\newblock \showarticletitle{Designing secure ethereum smart contracts: A finite
  state machine based approach}. In \bibinfo{booktitle}{\emph{International
  Conference on Financial Cryptography and Data Security}}. Springer,
  \bibinfo{pages}{523--540}.
\newblock


\bibitem[\protect\citeauthoryear{Mavridou, Laszka, Stachtiari, and
  Dubey}{Mavridou et~al\mbox{.}}{2019}]%
        {mavridou2019verisolid}
\bibfield{author}{\bibinfo{person}{Anastasia Mavridou}, \bibinfo{person}{Aron
  Laszka}, \bibinfo{person}{Emmanouela Stachtiari}, {and}
  \bibinfo{person}{Abhishek Dubey}.} \bibinfo{year}{2019}\natexlab{}.
\newblock \showarticletitle{VeriSolid: Correct-by-design smart contracts for
  Ethereum}. In \bibinfo{booktitle}{\emph{International Conference on Financial
  Cryptography and Data Security}}. Springer, \bibinfo{pages}{446--465}.
\newblock


\bibitem[\protect\citeauthoryear{Meier, Schmidt, Cremers, and Basin}{Meier
  et~al\mbox{.}}{2013}]%
        {meier2013tamarin}
\bibfield{author}{\bibinfo{person}{Simon Meier}, \bibinfo{person}{Benedikt
  Schmidt}, \bibinfo{person}{Cas Cremers}, {and} \bibinfo{person}{David
  Basin}.} \bibinfo{year}{2013}\natexlab{}.
\newblock \showarticletitle{The TAMARIN prover for the symbolic analysis of
  security protocols}. In \bibinfo{booktitle}{\emph{International conference on
  computer aided verification}}. Springer, \bibinfo{pages}{696--701}.
\newblock


\bibitem[\protect\citeauthoryear{Momeni, Wang, and Samavi}{Momeni
  et~al\mbox{.}}{2019}]%
        {momeni2019machine}
\bibfield{author}{\bibinfo{person}{Pouyan Momeni}, \bibinfo{person}{Yu Wang},
  {and} \bibinfo{person}{Reza Samavi}.} \bibinfo{year}{2019}\natexlab{}.
\newblock \showarticletitle{Machine learning model for smart contracts security
  analysis}. In \bibinfo{booktitle}{\emph{2019 17th International Conference on
  Privacy, Security and Trust (PST)}}. IEEE, \bibinfo{pages}{1--6}.
\newblock


\bibitem[\protect\citeauthoryear{Mossberg, Manzano, Hennenfent, Groce, Grieco,
  Feist, Brunson, and Dinaburg}{Mossberg et~al\mbox{.}}{2019}]%
        {mossberg2019manticore}
\bibfield{author}{\bibinfo{person}{Mark Mossberg}, \bibinfo{person}{Felipe
  Manzano}, \bibinfo{person}{Eric Hennenfent}, \bibinfo{person}{Alex Groce},
  \bibinfo{person}{Gustavo Grieco}, \bibinfo{person}{Josselin Feist},
  \bibinfo{person}{Trent Brunson}, {and} \bibinfo{person}{Artem Dinaburg}.}
  \bibinfo{year}{2019}\natexlab{}.
\newblock \showarticletitle{Manticore: A user-friendly symbolic execution
  framework for binaries and smart contracts}. In
  \bibinfo{booktitle}{\emph{2019 34th IEEE/ACM International Conference on
  Automated Software Engineering (ASE)}}. IEEE, \bibinfo{pages}{1186--1189}.
\newblock


\bibitem[\protect\citeauthoryear{Mueller}{Mueller}{2018}]%
        {mueller2018smashing}
\bibfield{author}{\bibinfo{person}{Bernhard Mueller}.}
  \bibinfo{year}{2018}\natexlab{}.
\newblock \showarticletitle{Smashing ethereum smart contracts for fun and real
  profit}. In \bibinfo{booktitle}{\emph{9th Annual HITB Security Conference
  (HITBSecConf)}}, Vol.~\bibinfo{volume}{54}.
\newblock


\bibitem[\protect\citeauthoryear{Nehai, Piriou, and Daumas}{Nehai
  et~al\mbox{.}}{2018}]%
        {nehai2018model}
\bibfield{author}{\bibinfo{person}{Zeinab Nehai}, \bibinfo{person}{Pierre-Yves
  Piriou}, {and} \bibinfo{person}{Fr{\'e}d{\'e}ric Daumas}.}
  \bibinfo{year}{2018}\natexlab{}.
\newblock \showarticletitle{Model-Checking of Smart Contracts}. In
  \bibinfo{booktitle}{\emph{IEEE International Conference on Blockchain}}.
\newblock


\bibitem[\protect\citeauthoryear{Nguyen, Pham, and Sun}{Nguyen
  et~al\mbox{.}}{[n.\,d.]}]%
        {nguyen2021sguard}
\bibfield{author}{\bibinfo{person}{Tai~D Nguyen}, \bibinfo{person}{Long~H
  Pham}, {and} \bibinfo{person}{Jun Sun}.} \bibinfo{year}{[n.\,d.]}\natexlab{}.
\newblock \showarticletitle{sGUARD: Towards Fixing Vulnerable Smart Contracts
  Automatically}.
\newblock \bibinfo{journal}{\emph{arXiv preprint arXiv:2101.01917}}
  (\bibinfo{year}{[n.\,d.]}).
\newblock


\bibitem[\protect\citeauthoryear{Nguyen, Pham, Sun, Lin, and Minh}{Nguyen
  et~al\mbox{.}}{2020}]%
        {nguyen2020sfuzz}
\bibfield{author}{\bibinfo{person}{Tai~D Nguyen}, \bibinfo{person}{Long~H
  Pham}, \bibinfo{person}{Jun Sun}, \bibinfo{person}{Yun Lin}, {and}
  \bibinfo{person}{Quang~Tran Minh}.} \bibinfo{year}{2020}\natexlab{}.
\newblock \showarticletitle{sfuzz: An efficient adaptive fuzzer for solidity
  smart contracts}. In \bibinfo{booktitle}{\emph{Proceedings of the ACM/IEEE
  42nd International Conference on Software Engineering}}.
  \bibinfo{pages}{778--788}.
\newblock


\bibitem[\protect\citeauthoryear{NI, ZHANG, and YIN}{NI et~al\mbox{.}}{2020}]%
        {ni2020survey}
\bibfield{author}{\bibinfo{person}{Yuandong NI}, \bibinfo{person}{Chao ZHANG},
  {and} \bibinfo{person}{Tingting YIN}.} \bibinfo{year}{2020}\natexlab{}.
\newblock \showarticletitle{A Survey of Smart Contract Vulnerability Research}.
\newblock \bibinfo{journal}{\emph{Journal of Cyber Security}}
  \bibinfo{volume}{5}, \bibinfo{number}{3} (\bibinfo{year}{2020}),
  \bibinfo{pages}{78--99}.
\newblock


\bibitem[\protect\citeauthoryear{Nikoli{\'c}, Kolluri, Sergey, Saxena, and
  Hobor}{Nikoli{\'c} et~al\mbox{.}}{2018}]%
        {nikolic2018finding}
\bibfield{author}{\bibinfo{person}{Ivica Nikoli{\'c}}, \bibinfo{person}{Aashish
  Kolluri}, \bibinfo{person}{Ilya Sergey}, \bibinfo{person}{Prateek Saxena},
  {and} \bibinfo{person}{Aquinas Hobor}.} \bibinfo{year}{2018}\natexlab{}.
\newblock \showarticletitle{Finding the greedy, prodigal, and suicidal
  contracts at scale}. In \bibinfo{booktitle}{\emph{Proceedings of the 34th
  Annual Computer Security Applications Conference}}.
  \bibinfo{pages}{653--663}.
\newblock


\bibitem[\protect\citeauthoryear{Norvill, Pontiveros, State, and
  Cullen}{Norvill et~al\mbox{.}}{2018}]%
        {norvill2018visual}
\bibfield{author}{\bibinfo{person}{Robert Norvill}, \bibinfo{person}{Beltran
  Borja~Fiz Pontiveros}, \bibinfo{person}{Radu State}, {and}
  \bibinfo{person}{Andrea Cullen}.} \bibinfo{year}{2018}\natexlab{}.
\newblock \showarticletitle{Visual emulation for Ethereum's virtual machine}.
  In \bibinfo{booktitle}{\emph{NOMS 2018-2018 IEEE/IFIP Network Operations and
  Management Symposium}}. IEEE, \bibinfo{pages}{1--4}.
\newblock


\bibitem[\protect\citeauthoryear{Perez and Livshits}{Perez and
  Livshits}{2019}]%
        {perez2019smart}
\bibfield{author}{\bibinfo{person}{Daniel Perez} {and}
  \bibinfo{person}{Benjamin Livshits}.} \bibinfo{year}{2019}\natexlab{}.
\newblock \showarticletitle{Smart contract vulnerabilities: Does anyone care?}
\newblock \bibinfo{journal}{\emph{arXiv preprint arXiv:1902.06710}}
  (\bibinfo{year}{2019}), \bibinfo{pages}{1--15}.
\newblock


\bibitem[\protect\citeauthoryear{Perez and Livshits}{Perez and
  Livshits}{2021}]%
        {perez2021smart}
\bibfield{author}{\bibinfo{person}{Daniel Perez} {and} \bibinfo{person}{Ben
  Livshits}.} \bibinfo{year}{2021}\natexlab{}.
\newblock \showarticletitle{Smart contract vulnerabilities: Vulnerable does not
  imply exploited}. In \bibinfo{booktitle}{\emph{30th {USENIX} Security
  Symposium ({USENIX} Security 21)}}.
\newblock


\bibitem[\protect\citeauthoryear{Permenev, Dimitrov, Tsankov, Drachsler-Cohen,
  and Vechev}{Permenev et~al\mbox{.}}{2020}]%
        {permenev2020verx}
\bibfield{author}{\bibinfo{person}{Anton Permenev}, \bibinfo{person}{Dimitar
  Dimitrov}, \bibinfo{person}{Petar Tsankov}, \bibinfo{person}{Dana
  Drachsler-Cohen}, {and} \bibinfo{person}{Martin Vechev}.}
  \bibinfo{year}{2020}\natexlab{}.
\newblock \showarticletitle{Verx: Safety verification of smart contracts}. In
  \bibinfo{booktitle}{\emph{2020 IEEE Symposium on Security and Privacy (SP)}}.
  IEEE, \bibinfo{pages}{1661--1677}.
\newblock


\bibitem[\protect\citeauthoryear{Praitheeshan, Pan, Yu, Liu, and
  Doss}{Praitheeshan et~al\mbox{.}}{2019}]%
        {praitheeshan2019security}
\bibfield{author}{\bibinfo{person}{Purathani Praitheeshan},
  \bibinfo{person}{Lei Pan}, \bibinfo{person}{Jiangshan Yu},
  \bibinfo{person}{Joseph Liu}, {and} \bibinfo{person}{Robin Doss}.}
  \bibinfo{year}{2019}\natexlab{}.
\newblock \showarticletitle{Security analysis methods on ethereum smart
  contract vulnerabilities: a survey}.
\newblock \bibinfo{journal}{\emph{arXiv preprint arXiv:1908.08605}}
  (\bibinfo{year}{2019}).
\newblock


\bibitem[\protect\citeauthoryear{Qin, Zhou, and Gervais}{Qin
  et~al\mbox{.}}{2022}]%
        {qin2022quantifying}
\bibfield{author}{\bibinfo{person}{Kaihua Qin}, \bibinfo{person}{Liyi Zhou},
  {and} \bibinfo{person}{Arthur Gervais}.} \bibinfo{year}{2022}\natexlab{}.
\newblock \showarticletitle{Quantifying blockchain extractable value: How dark
  is the forest?}. In \bibinfo{booktitle}{\emph{2022 IEEE Symposium on Security
  and Privacy (SP)}}. IEEE, \bibinfo{pages}{198--214}.
\newblock


\bibitem[\protect\citeauthoryear{Quan, Wu, and Wang}{Quan
  et~al\mbox{.}}{2019}]%
        {quan2019evulhunter}
\bibfield{author}{\bibinfo{person}{Lijin Quan}, \bibinfo{person}{Lei Wu}, {and}
  \bibinfo{person}{Haoyu Wang}.} \bibinfo{year}{2019}\natexlab{}.
\newblock \showarticletitle{EVulHunter: detecting fake transfer vulnerabilities
  for EOSIO's smart contracts at Webassembly-level}.
\newblock \bibinfo{journal}{\emph{arXiv preprint arXiv:1906.10362}}
  (\bibinfo{year}{2019}).
\newblock


\bibitem[\protect\citeauthoryear{Rodler, Li, Karame, and Davi}{Rodler
  et~al\mbox{.}}{2018}]%
        {rodler2018sereum}
\bibfield{author}{\bibinfo{person}{Michael Rodler}, \bibinfo{person}{Wenting
  Li}, \bibinfo{person}{Ghassan~O Karame}, {and} \bibinfo{person}{Lucas Davi}.}
  \bibinfo{year}{2018}\natexlab{}.
\newblock \showarticletitle{Sereum: Protecting existing smart contracts against
  re-entrancy attacks}.
\newblock \bibinfo{journal}{\emph{arXiv preprint arXiv:1812.05934}}
  (\bibinfo{year}{2018}).
\newblock


\bibitem[\protect\citeauthoryear{Rodler, Li, Karame, and Davi}{Rodler
  et~al\mbox{.}}{2021}]%
        {rodler2021evmpatch}
\bibfield{author}{\bibinfo{person}{Michael Rodler}, \bibinfo{person}{Wenting
  Li}, \bibinfo{person}{Ghassan~O Karame}, {and} \bibinfo{person}{Lucas Davi}.}
  \bibinfo{year}{2021}\natexlab{}.
\newblock \showarticletitle{EVMPatch: timely and automated patching of ethereum
  smart contracts}. In \bibinfo{booktitle}{\emph{30th {USENIX} Security
  Symposium ({USENIX} Security 21)}}.
\newblock


\bibitem[\protect\citeauthoryear{Sahs and Khan}{Sahs and Khan}{2012}]%
        {sahs2012machine}
\bibfield{author}{\bibinfo{person}{Justin Sahs} {and} \bibinfo{person}{Latifur
  Khan}.} \bibinfo{year}{2012}\natexlab{}.
\newblock \showarticletitle{A machine learning approach to android malware
  detection}. In \bibinfo{booktitle}{\emph{2012 European Intelligence and
  Security Informatics Conference}}. IEEE, \bibinfo{pages}{141--147}.
\newblock


\bibitem[\protect\citeauthoryear{Samreen and Alalfi}{Samreen and
  Alalfi}{2021}]%
        {samreen2021survey}
\bibfield{author}{\bibinfo{person}{Noama~Fatima Samreen} {and}
  \bibinfo{person}{Manar~H. Alalfi}.} \bibinfo{year}{2021}\natexlab{}.
\newblock \showarticletitle{A Survey of Security Vulnerabilities in Ethereum
  Smart Contracts}.
\newblock \bibinfo{journal}{\emph{CoRR}}  \bibinfo{volume}{abs/2105.06974}
  (\bibinfo{year}{2021}).
\newblock
\showeprint[arXiv]{2105.06974}
\urldef\tempurl%
\url{https://arxiv.org/abs/2105.06974}
\showURL{%
\tempurl}


\bibitem[\protect\citeauthoryear{Schneidewind, Grishchenko, Scherer, and
  Maffei}{Schneidewind et~al\mbox{.}}{2020}]%
        {schneidewind2020ethor}
\bibfield{author}{\bibinfo{person}{Clara Schneidewind}, \bibinfo{person}{Ilya
  Grishchenko}, \bibinfo{person}{Markus Scherer}, {and} \bibinfo{person}{Matteo
  Maffei}.} \bibinfo{year}{2020}\natexlab{}.
\newblock \showarticletitle{eThor: Practical and provably sound static analysis
  of Ethereum smart contracts}. In \bibinfo{booktitle}{\emph{Proceedings of the
  2020 ACM SIGSAC Conference on Computer and Communications Security}}.
  \bibinfo{pages}{621--640}.
\newblock


\bibitem[\protect\citeauthoryear{Schrans, Eisenbach, and Drossopoulou}{Schrans
  et~al\mbox{.}}{2018}]%
        {schrans2018writing}
\bibfield{author}{\bibinfo{person}{Franklin Schrans}, \bibinfo{person}{Susan
  Eisenbach}, {and} \bibinfo{person}{Sophia Drossopoulou}.}
  \bibinfo{year}{2018}\natexlab{}.
\newblock \showarticletitle{Writing safe smart contracts in flint}. In
  \bibinfo{booktitle}{\emph{Conference companion of the 2nd international
  conference on art, science, and engineering of programming}}.
  \bibinfo{pages}{218--219}.
\newblock


\bibitem[\protect\citeauthoryear{Schwarz-Schilling, Neu, Monnot, Asgaonkar,
  Tas, and Tse}{Schwarz-Schilling et~al\mbox{.}}{2022}]%
        {schwarz2022three}
\bibfield{author}{\bibinfo{person}{Caspar Schwarz-Schilling},
  \bibinfo{person}{Joachim Neu}, \bibinfo{person}{Barnab{\'e} Monnot},
  \bibinfo{person}{Aditya Asgaonkar}, \bibinfo{person}{Ertem~Nusret Tas}, {and}
  \bibinfo{person}{David Tse}.} \bibinfo{year}{2022}\natexlab{}.
\newblock \showarticletitle{Three attacks on proof-of-stake ethereum}. In
  \bibinfo{booktitle}{\emph{Financial Cryptography and Data Security: 26th
  International Conference, FC 2022, Grenada, May 2--6, 2022, Revised Selected
  Papers}}. Springer, \bibinfo{pages}{560--576}.
\newblock


\bibitem[\protect\citeauthoryear{Sigalos}{Sigalos}{2021}]%
        {expl2}
\bibfield{author}{\bibinfo{person}{MacKenzie Sigalos}.}
  \bibinfo{year}{2021}\natexlab{}.
\newblock \bibinfo{title}{Bug puts \$162 million up for grabs, says founder of
  DeFi platform Compound}.
\newblock
  \bibinfo{howpublished}{{https://www.cnbc.com/2021/10/03/162-million-up-for-grabs-after-bug-in-defi-protocol-compound-.html}}.
\newblock


\bibitem[\protect\citeauthoryear{Singh, Parizi, Zhang, Choo, and
  Dehghantanha}{Singh et~al\mbox{.}}{2020}]%
        {singh2020blockchain}
\bibfield{author}{\bibinfo{person}{Amritraj Singh}, \bibinfo{person}{Reza~M
  Parizi}, \bibinfo{person}{Qi Zhang}, \bibinfo{person}{Kim-Kwang~Raymond
  Choo}, {and} \bibinfo{person}{Ali Dehghantanha}.}
  \bibinfo{year}{2020}\natexlab{}.
\newblock \showarticletitle{Blockchain smart contracts formalization:
  Approaches and challenges to address vulnerabilities}.
\newblock \bibinfo{journal}{\emph{Computers \& Security}}  \bibinfo{volume}{88}
  (\bibinfo{year}{2020}), \bibinfo{pages}{101654}.
\newblock


\bibitem[\protect\citeauthoryear{So, Hong, and Oh}{So et~al\mbox{.}}{2021}]%
        {so2021smartest}
\bibfield{author}{\bibinfo{person}{Sunbeom So}, \bibinfo{person}{Seongjoon
  Hong}, {and} \bibinfo{person}{Hakjoo Oh}.} \bibinfo{year}{2021}\natexlab{}.
\newblock \showarticletitle{SMARTEST: Effectively Hunting Vulnerable
  Transaction Sequences in Smart Contracts through Language Model-Guided
  Symbolic Execution}. In \bibinfo{booktitle}{\emph{30th {USENIX} Security
  Symposium ({USENIX} Security 21)}}.
\newblock


\bibitem[\protect\citeauthoryear{So, Lee, Park, Lee, and Oh}{So
  et~al\mbox{.}}{2020}]%
        {so2020verismart}
\bibfield{author}{\bibinfo{person}{Sunbeom So}, \bibinfo{person}{Myungho Lee},
  \bibinfo{person}{Jisu Park}, \bibinfo{person}{Heejo Lee}, {and}
  \bibinfo{person}{Hakjoo Oh}.} \bibinfo{year}{2020}\natexlab{}.
\newblock \showarticletitle{VeriSmart: A highly precise safety verifier for
  Ethereum smart contracts}. In \bibinfo{booktitle}{\emph{2020 IEEE Symposium
  on Security and Privacy (SP)}}. IEEE, \bibinfo{pages}{1678--1694}.
\newblock


\bibitem[\protect\citeauthoryear{Stephens, Ferles, Mariano, Lahiri, and
  Dillig}{Stephens et~al\mbox{.}}{2021}]%
        {stephens2021smartpulse}
\bibfield{author}{\bibinfo{person}{Jon Stephens}, \bibinfo{person}{Kostas
  Ferles}, \bibinfo{person}{Benjamin Mariano}, \bibinfo{person}{Shuvendu
  Lahiri}, {and} \bibinfo{person}{Isil Dillig}.}
  \bibinfo{year}{2021}\natexlab{}.
\newblock \showarticletitle{SmartPulse: Automated Checking of Temporal
  Properties in Smart Contracts}. In \bibinfo{booktitle}{\emph{IEEE S\&P}}.
\newblock


\bibitem[\protect\citeauthoryear{Su, Shen, Du, Liao, Wang, Xing, and Liu}{Su
  et~al\mbox{.}}{2021}]%
        {su2021evil}
\bibfield{author}{\bibinfo{person}{Liya Su}, \bibinfo{person}{Xinyue Shen},
  \bibinfo{person}{Xiangyu Du}, \bibinfo{person}{Xiaojing Liao},
  \bibinfo{person}{XiaoFeng Wang}, \bibinfo{person}{Luyi Xing}, {and}
  \bibinfo{person}{Baoxu Liu}.} \bibinfo{year}{2021}\natexlab{}.
\newblock \showarticletitle{Evil Under the Sun: Understanding and Discovering
  Attacks on Ethereum Decentralized Applications}. In
  \bibinfo{booktitle}{\emph{30th {USENIX} Security Symposium ({USENIX} Security
  21)}}.
\newblock


\bibitem[\protect\citeauthoryear{Sun, Luo, and Zhang}{Sun
  et~al\mbox{.}}{2023}]%
        {panda2023}
\bibfield{author}{\bibinfo{person}{Zhiyuan Sun}, \bibinfo{person}{Xiapu Luo},
  {and} \bibinfo{person}{Yinqian Zhang}.} \bibinfo{year}{2023}\natexlab{}.
\newblock \showarticletitle{Panda: Security Analysis of Algorand Smart
  Contracts}. In \bibinfo{booktitle}{\emph{USENIX Security Symposium}}.
\newblock


\bibitem[\protect\citeauthoryear{Szabo}{Szabo}{1996}]%
        {szabo1996smart}
\bibfield{author}{\bibinfo{person}{Nick Szabo}.}
  \bibinfo{year}{1996}\natexlab{}.
\newblock \showarticletitle{Smart contracts: building blocks for digital
  markets}.
\newblock \bibinfo{journal}{\emph{EXTROPY: The Journal of Transhumanist
  Thought,(16)}}  \bibinfo{volume}{18} (\bibinfo{year}{1996}),
  \bibinfo{pages}{2}.
\newblock


\bibitem[\protect\citeauthoryear{Tikhomirov, Voskresenskaya, Ivanitskiy,
  Takhaviev, Marchenko, and Alexandrov}{Tikhomirov et~al\mbox{.}}{2018}]%
        {tikhomirov2018smartcheck}
\bibfield{author}{\bibinfo{person}{Sergei Tikhomirov},
  \bibinfo{person}{Ekaterina Voskresenskaya}, \bibinfo{person}{Ivan
  Ivanitskiy}, \bibinfo{person}{Ramil Takhaviev}, \bibinfo{person}{Evgeny
  Marchenko}, {and} \bibinfo{person}{Yaroslav Alexandrov}.}
  \bibinfo{year}{2018}\natexlab{}.
\newblock \showarticletitle{Smartcheck: Static analysis of ethereum smart
  contracts}. In \bibinfo{booktitle}{\emph{Proceedings of the 1st International
  Workshop on Emerging Trends in Software Engineering for Blockchain}}.
  \bibinfo{pages}{9--16}.
\newblock


\bibitem[\protect\citeauthoryear{Tolmach, Li, Lin, Liu, and Li}{Tolmach
  et~al\mbox{.}}{2021}]%
        {tolmach2021survey}
\bibfield{author}{\bibinfo{person}{Palina Tolmach}, \bibinfo{person}{Yi Li},
  \bibinfo{person}{Shang-Wei Lin}, \bibinfo{person}{Yang Liu}, {and}
  \bibinfo{person}{Zengxiang Li}.} \bibinfo{year}{2021}\natexlab{}.
\newblock \showarticletitle{A survey of smart contract formal specification and
  verification}.
\newblock \bibinfo{journal}{\emph{ACM Computing Surveys (CSUR)}}
  \bibinfo{volume}{54}, \bibinfo{number}{7} (\bibinfo{year}{2021}),
  \bibinfo{pages}{1--38}.
\newblock


\bibitem[\protect\citeauthoryear{Torres, Sch{\"u}tte, and State}{Torres
  et~al\mbox{.}}{2018}]%
        {torres2018osiris}
\bibfield{author}{\bibinfo{person}{Christof~Ferreira Torres},
  \bibinfo{person}{Julian Sch{\"u}tte}, {and} \bibinfo{person}{Radu State}.}
  \bibinfo{year}{2018}\natexlab{}.
\newblock \showarticletitle{Osiris: Hunting for integer bugs in ethereum smart
  contracts}. In \bibinfo{booktitle}{\emph{Proceedings of the 34th Annual
  Computer Security Applications Conference}}. \bibinfo{pages}{664--676}.
\newblock


\bibitem[\protect\citeauthoryear{Torres, Steichen, et~al\mbox{.}}{Torres
  et~al\mbox{.}}{2019}]%
        {torres2019art}
\bibfield{author}{\bibinfo{person}{Christof~Ferreira Torres},
  \bibinfo{person}{Mathis Steichen}, {et~al\mbox{.}}}
  \bibinfo{year}{2019}\natexlab{}.
\newblock \showarticletitle{The art of the scam: Demystifying honeypots in
  ethereum smart contracts}. In \bibinfo{booktitle}{\emph{28th {USENIX}
  Security Symposium ({USENIX} Security 19)}}. \bibinfo{pages}{1591--1607}.
\newblock


\bibitem[\protect\citeauthoryear{Tsankov, Dan, Drachsler-Cohen, Gervais,
  Buenzli, and Vechev}{Tsankov et~al\mbox{.}}{2018}]%
        {tsankov2018securify}
\bibfield{author}{\bibinfo{person}{Petar Tsankov}, \bibinfo{person}{Andrei
  Dan}, \bibinfo{person}{Dana Drachsler-Cohen}, \bibinfo{person}{Arthur
  Gervais}, \bibinfo{person}{Florian Buenzli}, {and} \bibinfo{person}{Martin
  Vechev}.} \bibinfo{year}{2018}\natexlab{}.
\newblock \showarticletitle{Securify: Practical security analysis of smart
  contracts}. In \bibinfo{booktitle}{\emph{Proceedings of the 2018 ACM SIGSAC
  Conference on Computer and Communications Security}}.
  \bibinfo{pages}{67--82}.
\newblock


\bibitem[\protect\citeauthoryear{Vacca, {Di Sorbo}, Visaggio, and
  Canfora}{Vacca et~al\mbox{.}}{2021}]%
        {VACCA2021110891}
\bibfield{author}{\bibinfo{person}{Anna Vacca}, \bibinfo{person}{Andrea {Di
  Sorbo}}, \bibinfo{person}{Corrado~A. Visaggio}, {and}
  \bibinfo{person}{Gerardo Canfora}.} \bibinfo{year}{2021}\natexlab{}.
\newblock \showarticletitle{A systematic literature review of blockchain and
  smart contract development: Techniques, tools, and open challenges}.
\newblock \bibinfo{journal}{\emph{Journal of Systems and Software}}
  \bibinfo{volume}{174} (\bibinfo{year}{2021}), \bibinfo{pages}{110891}.
\newblock
\showISSN{0164-1212}
\urldef\tempurl%
\url{https://doi.org/10.1016/j.jss.2020.110891}
\showDOI{\tempurl}


\bibitem[\protect\citeauthoryear{Wan, Xia, Lo, Chen, Luo, and Yang}{Wan
  et~al\mbox{.}}{2021}]%
        {wan2021smart}
\bibfield{author}{\bibinfo{person}{Zhiyuan Wan}, \bibinfo{person}{Xin Xia},
  \bibinfo{person}{David Lo}, \bibinfo{person}{Jiachi Chen},
  \bibinfo{person}{Xiapu Luo}, {and} \bibinfo{person}{Xiaohu Yang}.}
  \bibinfo{year}{2021}\natexlab{}.
\newblock \showarticletitle{Smart contract security: a practitioners'
  perspective}. In \bibinfo{booktitle}{\emph{2021 IEEE/ACM 43rd International
  Conference on Software Engineering (ICSE)}}. IEEE,
  \bibinfo{pages}{1410--1422}.
\newblock


\bibitem[\protect\citeauthoryear{Wang, Wang, Jiang, and Chan}{Wang
  et~al\mbox{.}}{2020c}]%
        {wang2020artemis}
\bibfield{author}{\bibinfo{person}{Anqi Wang}, \bibinfo{person}{Hao Wang},
  \bibinfo{person}{Bo Jiang}, {and} \bibinfo{person}{Wing~Kwong Chan}.}
  \bibinfo{year}{2020}\natexlab{c}.
\newblock \showarticletitle{Artemis: An improved smart contract verification
  tool for vulnerability detection}. In \bibinfo{booktitle}{\emph{2020 7th
  International Conference on Dependable Systems and Their Applications
  (DSA)}}. IEEE, \bibinfo{pages}{173--181}.
\newblock


\bibitem[\protect\citeauthoryear{Wang, Liu, Liu, Yang, Ren, Zheng, and
  Lei}{Wang et~al\mbox{.}}{2021c}]%
        {wang2021blockeye}
\bibfield{author}{\bibinfo{person}{Bin Wang}, \bibinfo{person}{Han Liu},
  \bibinfo{person}{Chao Liu}, \bibinfo{person}{Zhiqiang Yang},
  \bibinfo{person}{Qian Ren}, \bibinfo{person}{Huixuan Zheng}, {and}
  \bibinfo{person}{Hong Lei}.} \bibinfo{year}{2021}\natexlab{c}.
\newblock \showarticletitle{BLOCKEYE: Hunting For DeFi Attacks on Blockchain}.
  In \bibinfo{booktitle}{\emph{2021 IEEE/ACM 43rd International Conference on
  Software Engineering: Companion Proceedings (ICSE-Companion)}}. IEEE,
  \bibinfo{pages}{17--20}.
\newblock


\bibitem[\protect\citeauthoryear{Wang, Jiang, and Chan}{Wang
  et~al\mbox{.}}{2020a}]%
        {wang2020wana}
\bibfield{author}{\bibinfo{person}{Dong Wang}, \bibinfo{person}{Bo Jiang},
  {and} \bibinfo{person}{WK Chan}.} \bibinfo{year}{2020}\natexlab{a}.
\newblock \showarticletitle{WANA: Symbolic Execution of Wasm Bytecode for
  Cross-Platform Smart Contract Vulnerability Detection}.
\newblock \bibinfo{journal}{\emph{arXiv preprint arXiv:2007.15510}}
  (\bibinfo{year}{2020}).
\newblock


\bibitem[\protect\citeauthoryear{Wang, Li, Lin, Ma, and Liu}{Wang
  et~al\mbox{.}}{2019b}]%
        {wang2019vultron}
\bibfield{author}{\bibinfo{person}{Haijun Wang}, \bibinfo{person}{Yi Li},
  \bibinfo{person}{Shang-Wei Lin}, \bibinfo{person}{Lei Ma}, {and}
  \bibinfo{person}{Yang Liu}.} \bibinfo{year}{2019}\natexlab{b}.
\newblock \showarticletitle{VULTRON: catching vulnerable smart contracts once
  and for all}. In \bibinfo{booktitle}{\emph{2019 IEEE/ACM 41st International
  Conference on Software Engineering: New Ideas and Emerging Results
  (ICSE-NIER)}}. IEEE, \bibinfo{pages}{1--4}.
\newblock


\bibitem[\protect\citeauthoryear{Wang, Zhang, and Su}{Wang
  et~al\mbox{.}}{2019c}]%
        {wang2019detecting}
\bibfield{author}{\bibinfo{person}{Shuai Wang}, \bibinfo{person}{Chengyu
  Zhang}, {and} \bibinfo{person}{Zhendong Su}.}
  \bibinfo{year}{2019}\natexlab{c}.
\newblock \showarticletitle{Detecting nondeterministic payment bugs in Ethereum
  smart contracts}.
\newblock \bibinfo{journal}{\emph{Proceedings of the ACM on Programming
  Languages}} \bibinfo{volume}{3}, \bibinfo{number}{OOPSLA}
  (\bibinfo{year}{2019}), \bibinfo{pages}{1--29}.
\newblock


\bibitem[\protect\citeauthoryear{Wang, Song, Xu, Li, Wang, and Su}{Wang
  et~al\mbox{.}}{2020b}]%
        {wang2020contractward}
\bibfield{author}{\bibinfo{person}{Wei Wang}, \bibinfo{person}{Jingjing Song},
  \bibinfo{person}{Guangquan Xu}, \bibinfo{person}{Yidong Li},
  \bibinfo{person}{Hao Wang}, {and} \bibinfo{person}{Chunhua Su}.}
  \bibinfo{year}{2020}\natexlab{b}.
\newblock \showarticletitle{Contractward: Automated vulnerability detection
  models for ethereum smart contracts}.
\newblock \bibinfo{journal}{\emph{IEEE Transactions on Network Science and
  Engineering}} (\bibinfo{year}{2020}).
\newblock


\bibitem[\protect\citeauthoryear{Wang, He, Zhu, Yi, Zhang, Song, and Xue}{Wang
  et~al\mbox{.}}{2021a}]%
        {wang2021securityenhancement}
\bibfield{author}{\bibinfo{person}{Yajing Wang}, \bibinfo{person}{Jingsha He},
  \bibinfo{person}{Nafei Zhu}, \bibinfo{person}{Yuzi Yi},
  \bibinfo{person}{Qingqing Zhang}, \bibinfo{person}{Hongyu Song}, {and}
  \bibinfo{person}{Ruixin Xue}.} \bibinfo{year}{2021}\natexlab{a}.
\newblock \showarticletitle{Security enhancement technologies for smart
  contracts in the blockchain: A survey}.
\newblock \bibinfo{journal}{\emph{Transactions on Emerging Telecommunications
  Technologies}} \bibinfo{volume}{32}, \bibinfo{number}{12}
  (\bibinfo{year}{2021}), \bibinfo{pages}{e4341}.
\newblock
\urldef\tempurl%
\url{https://doi.org/10.1002/ett.4341}
\showDOI{\tempurl}
\showeprint{https://onlinelibrary.wiley.com/doi/pdf/10.1002/ett.4341}


\bibitem[\protect\citeauthoryear{Wang, Lahiri, Chen, Pan, Dillig, Born, Naseer,
  and Ferles}{Wang et~al\mbox{.}}{2019a}]%
        {wang2019formal}
\bibfield{author}{\bibinfo{person}{Yuepeng Wang}, \bibinfo{person}{Shuvendu~K
  Lahiri}, \bibinfo{person}{Shuo Chen}, \bibinfo{person}{Rong Pan},
  \bibinfo{person}{Isil Dillig}, \bibinfo{person}{Cody Born},
  \bibinfo{person}{Immad Naseer}, {and} \bibinfo{person}{Kostas Ferles}.}
  \bibinfo{year}{2019}\natexlab{a}.
\newblock \showarticletitle{Formal verification of workflow policies for smart
  contracts in azure blockchain}. In \bibinfo{booktitle}{\emph{Working
  Conference on Verified Software: Theories, Tools, and Experiments}}.
  Springer, \bibinfo{pages}{87--106}.
\newblock


\bibitem[\protect\citeauthoryear{Wang, Jin, Dai, Choo, and Zou}{Wang
  et~al\mbox{.}}{2021b}]%
        {wang2021ethereum}
\bibfield{author}{\bibinfo{person}{Zeli Wang}, \bibinfo{person}{Hai Jin},
  \bibinfo{person}{Weiqi Dai}, \bibinfo{person}{Kim-Kwang~Raymond Choo}, {and}
  \bibinfo{person}{Deqing Zou}.} \bibinfo{year}{2021}\natexlab{b}.
\newblock \showarticletitle{Ethereum smart contract security research: survey
  and future research opportunities}.
\newblock \bibinfo{journal}{\emph{Frontiers of Computer Science}}
  \bibinfo{volume}{15}, \bibinfo{number}{2} (\bibinfo{year}{2021}),
  \bibinfo{pages}{1--18}.
\newblock


\bibitem[\protect\citeauthoryear{Wood et~al\mbox{.}}{Wood
  et~al\mbox{.}}{2014}]%
        {wood2014ethereum}
\bibfield{author}{\bibinfo{person}{Gavin Wood} {et~al\mbox{.}}}
  \bibinfo{year}{2014}\natexlab{}.
\newblock \showarticletitle{Ethereum: A secure decentralised generalised
  transaction ledger}.
\newblock \bibinfo{journal}{\emph{Ethereum project yellow paper}}
  \bibinfo{volume}{151}, \bibinfo{number}{2014} (\bibinfo{year}{2014}),
  \bibinfo{pages}{1--32}.
\newblock


\bibitem[\protect\citeauthoryear{Wu, Wu, Zhou, Li, Wang, Luo, Wang, and Ren}{Wu
  et~al\mbox{.}}{[n.\,d.]}]%
        {wu2020ethscope}
\bibfield{author}{\bibinfo{person}{Lei Wu}, \bibinfo{person}{Siwei Wu},
  \bibinfo{person}{Yajin Zhou}, \bibinfo{person}{Runhuai Li},
  \bibinfo{person}{Zhi Wang}, \bibinfo{person}{Xiapu Luo},
  \bibinfo{person}{Cong Wang}, {and} \bibinfo{person}{Kui Ren}.}
  \bibinfo{year}{[n.\,d.]}\natexlab{}.
\newblock \showarticletitle{Ethscope: A transaction-centric security analytics
  framework to detect malicious smart contracts on ethereum}.
\newblock  (\bibinfo{year}{[n.\,d.]}).
\newblock


\bibitem[\protect\citeauthoryear{Wu, Wang, He, Zhou, Wu, Yuan, He, and Ren}{Wu
  et~al\mbox{.}}{2021}]%
        {wu2021defiranger}
\bibfield{author}{\bibinfo{person}{Siwei Wu}, \bibinfo{person}{Dabao Wang},
  \bibinfo{person}{Jianting He}, \bibinfo{person}{Yajin Zhou},
  \bibinfo{person}{Lei Wu}, \bibinfo{person}{Xingliang Yuan},
  \bibinfo{person}{Qinming He}, {and} \bibinfo{person}{Kui Ren}.}
  \bibinfo{year}{2021}\natexlab{}.
\newblock \showarticletitle{DeFiRanger: Detecting Price Manipulation Attacks on
  DeFi Applications}.
\newblock \bibinfo{journal}{\emph{arXiv preprint arXiv:2104.15068}}
  (\bibinfo{year}{2021}).
\newblock


\bibitem[\protect\citeauthoryear{W{\"u}stholz and Christakis}{W{\"u}stholz and
  Christakis}{2020}]%
        {wustholz2020harvey}
\bibfield{author}{\bibinfo{person}{Valentin W{\"u}stholz} {and}
  \bibinfo{person}{Maria Christakis}.} \bibinfo{year}{2020}\natexlab{}.
\newblock \showarticletitle{Harvey: A greybox fuzzer for smart contracts}. In
  \bibinfo{booktitle}{\emph{Proceedings of the 28th ACM Joint Meeting on
  European Software Engineering Conference and Symposium on the Foundations of
  Software Engineering}}. \bibinfo{pages}{1398--1409}.
\newblock


\bibitem[\protect\citeauthoryear{Yan, Gao, Wu, Li, Guan, Li, and Chen}{Yan
  et~al\mbox{.}}{2020}]%
        {yan2020eshield}
\bibfield{author}{\bibinfo{person}{Wentian Yan}, \bibinfo{person}{Jianbo Gao},
  \bibinfo{person}{Zhenhao Wu}, \bibinfo{person}{Yue Li}, \bibinfo{person}{Zhi
  Guan}, \bibinfo{person}{Qingshan Li}, {and} \bibinfo{person}{Zhong Chen}.}
  \bibinfo{year}{2020}\natexlab{}.
\newblock \showarticletitle{Eshield: protect smart contracts against reverse
  engineering}. In \bibinfo{booktitle}{\emph{Proceedings of the 29th ACM
  SIGSOFT International Symposium on Software Testing and Analysis}}.
  \bibinfo{pages}{553--556}.
\newblock


\bibitem[\protect\citeauthoryear{Yang and Lei}{Yang and Lei}{2018}]%
        {yang2018lolisa}
\bibfield{author}{\bibinfo{person}{Zheng Yang} {and} \bibinfo{person}{Hang
  Lei}.} \bibinfo{year}{2018}\natexlab{}.
\newblock \showarticletitle{Lolisa: Formal syntax and semantics for a subset of
  the solidity programming language}.
\newblock \bibinfo{journal}{\emph{arXiv e-prints}} (\bibinfo{year}{2018}),
  \bibinfo{pages}{arXiv--1803}.
\newblock


\bibitem[\protect\citeauthoryear{Yang, Liu, Li, Zheng, Wang, and Chen}{Yang
  et~al\mbox{.}}{2020}]%
        {yang2020seraph}
\bibfield{author}{\bibinfo{person}{Zhiqiang Yang}, \bibinfo{person}{Han Liu},
  \bibinfo{person}{Yue Li}, \bibinfo{person}{Huixuan Zheng},
  \bibinfo{person}{Lei Wang}, {and} \bibinfo{person}{Bangdao Chen}.}
  \bibinfo{year}{2020}\natexlab{}.
\newblock \showarticletitle{Seraph: enabling cross-platform security analysis
  for evm and wasm smart contracts}. In \bibinfo{booktitle}{\emph{2020 IEEE/ACM
  42nd International Conference on Software Engineering: Companion Proceedings
  (ICSE-Companion)}}. IEEE, \bibinfo{pages}{21--24}.
\newblock


\bibitem[\protect\citeauthoryear{Ye, Ma, Lin, Sui, and Xue}{Ye
  et~al\mbox{.}}{2020}]%
        {ye2020clairvoyance}
\bibfield{author}{\bibinfo{person}{Jiaming Ye}, \bibinfo{person}{Mingliang Ma},
  \bibinfo{person}{Yun Lin}, \bibinfo{person}{Yulei Sui}, {and}
  \bibinfo{person}{Yinxing Xue}.} \bibinfo{year}{2020}\natexlab{}.
\newblock \showarticletitle{Clairvoyance: Cross-contract static analysis for
  detecting practical reentrancy vulnerabilities in smart contracts}. In
  \bibinfo{booktitle}{\emph{2020 IEEE/ACM 42nd International Conference on
  Software Engineering: Companion Proceedings (ICSE-Companion)}}. IEEE,
  \bibinfo{pages}{274--275}.
\newblock


\bibitem[\protect\citeauthoryear{Yi, Yang, Kelarev, Lam, and Tari}{Yi
  et~al\mbox{.}}{2022}]%
        {yi2022blockchain}
\bibfield{author}{\bibinfo{person}{Xun Yi}, \bibinfo{person}{Xuechao Yang},
  \bibinfo{person}{Andrei Kelarev}, \bibinfo{person}{Kwok Lam}, {and}
  \bibinfo{person}{Zahir Tari}.} \bibinfo{year}{2022}\natexlab{}.
\newblock \bibinfo{booktitle}{\emph{Blockchain Foundations and Applications}}.
\newblock \bibinfo{publisher}{Springer Nature}.
\newblock


\bibitem[\protect\citeauthoryear{Zhang, Zhang, Zhang, and Lin}{Zhang
  et~al\mbox{.}}{2020d}]%
        {zhang2020txspector}
\bibfield{author}{\bibinfo{person}{Mengya Zhang}, \bibinfo{person}{Xiaokuan
  Zhang}, \bibinfo{person}{Yinqian Zhang}, {and} \bibinfo{person}{Zhiqiang
  Lin}.} \bibinfo{year}{2020}\natexlab{d}.
\newblock \showarticletitle{{TXSPECTOR}: Uncovering Attacks in Ethereum from
  Transactions}. In \bibinfo{booktitle}{\emph{29th {USENIX} Security Symposium
  ({USENIX} Security 20)}}. \bibinfo{pages}{2775--2792}.
\newblock


\bibitem[\protect\citeauthoryear{Zhang, Xiao, and Luo}{Zhang
  et~al\mbox{.}}{2019b}]%
        {zhang2019soliditycheck}
\bibfield{author}{\bibinfo{person}{Pengcheng Zhang}, \bibinfo{person}{Feng
  Xiao}, {and} \bibinfo{person}{Xiapu Luo}.} \bibinfo{year}{2019}\natexlab{b}.
\newblock \showarticletitle{SolidityCheck: Quickly detecting smart contract
  problems through regular expressions}.
\newblock \bibinfo{journal}{\emph{arXiv preprint arXiv:1911.09425}}
  (\bibinfo{year}{2019}).
\newblock


\bibitem[\protect\citeauthoryear{Zhang, Xiao, and Luo}{Zhang
  et~al\mbox{.}}{2020c}]%
        {zhang2020framework}
\bibfield{author}{\bibinfo{person}{Pengcheng Zhang}, \bibinfo{person}{Feng
  Xiao}, {and} \bibinfo{person}{Xiapu Luo}.} \bibinfo{year}{2020}\natexlab{c}.
\newblock \showarticletitle{A framework and dataset for bugs in ethereum smart
  contracts}. In \bibinfo{booktitle}{\emph{2020 IEEE International Conference
  on Software Maintenance and Evolution (ICSME)}}. IEEE,
  \bibinfo{pages}{139--150}.
\newblock


\bibitem[\protect\citeauthoryear{Zhang, Wang, Li, and Ma}{Zhang
  et~al\mbox{.}}{2020b}]%
        {zhang2020ethploit}
\bibfield{author}{\bibinfo{person}{Qingzhao Zhang}, \bibinfo{person}{Yizhuo
  Wang}, \bibinfo{person}{Juanru Li}, {and} \bibinfo{person}{Siqi Ma}.}
  \bibinfo{year}{2020}\natexlab{b}.
\newblock \showarticletitle{Ethploit: From fuzzing to efficient exploit
  generation against smart contracts}. In \bibinfo{booktitle}{\emph{2020 IEEE
  27th International Conference on Software Analysis, Evolution and
  Reengineering (SANER)}}. IEEE, \bibinfo{pages}{116--126}.
\newblock


\bibitem[\protect\citeauthoryear{Zhang, Banescu, Pasos, Stewart, and
  Ganesh}{Zhang et~al\mbox{.}}{2019a}]%
        {zhang2019mpro}
\bibfield{author}{\bibinfo{person}{William Zhang}, \bibinfo{person}{Sebastian
  Banescu}, \bibinfo{person}{Leonardo Pasos}, \bibinfo{person}{Steven Stewart},
  {and} \bibinfo{person}{Vijay Ganesh}.} \bibinfo{year}{2019}\natexlab{a}.
\newblock \showarticletitle{Mpro: Combining static and symbolic analysis for
  scalable testing of smart contract}. In \bibinfo{booktitle}{\emph{2019 IEEE
  30th International Symposium on Software Reliability Engineering (ISSRE)}}.
  IEEE, \bibinfo{pages}{456--462}.
\newblock


\bibitem[\protect\citeauthoryear{Zhang, Ma, Li, Li, Nepal, and Gu}{Zhang
  et~al\mbox{.}}{2020a}]%
        {zhang2020smartshield}
\bibfield{author}{\bibinfo{person}{Yuyao Zhang}, \bibinfo{person}{Siqi Ma},
  \bibinfo{person}{Juanru Li}, \bibinfo{person}{Kailai Li},
  \bibinfo{person}{Surya Nepal}, {and} \bibinfo{person}{Dawu Gu}.}
  \bibinfo{year}{2020}\natexlab{a}.
\newblock \showarticletitle{Smartshield: Automatic smart contract protection
  made easy}. In \bibinfo{booktitle}{\emph{2020 IEEE 27th International
  Conference on Software Analysis, Evolution and Reengineering (SANER)}}. IEEE,
  \bibinfo{pages}{23--34}.
\newblock


\bibitem[\protect\citeauthoryear{Zhou, Hua, Pi, Sun, Nomura, Yamashita, and
  Kurihara}{Zhou et~al\mbox{.}}{2018a}]%
        {zhou2018security}
\bibfield{author}{\bibinfo{person}{Ence Zhou}, \bibinfo{person}{Song Hua},
  \bibinfo{person}{Bingfeng Pi}, \bibinfo{person}{Jun Sun},
  \bibinfo{person}{Yashihide Nomura}, \bibinfo{person}{Kazuhiro Yamashita},
  {and} \bibinfo{person}{Hidetoshi Kurihara}.}
  \bibinfo{year}{2018}\natexlab{a}.
\newblock \showarticletitle{Security assurance for smart contract}. In
  \bibinfo{booktitle}{\emph{2018 9th IFIP International Conference on New
  Technologies, Mobility and Security (NTMS)}}. IEEE, \bibinfo{pages}{1--5}.
\newblock


\bibitem[\protect\citeauthoryear{Zhou, M{\"o}ser, Yang, Adida, Holz, Xiang,
  Goldfeder, Cao, Plattner, Qin, et~al\mbox{.}}{Zhou et~al\mbox{.}}{2020}]%
        {zhou2020ever}
\bibfield{author}{\bibinfo{person}{Shunfan Zhou}, \bibinfo{person}{Malte
  M{\"o}ser}, \bibinfo{person}{Zhemin Yang}, \bibinfo{person}{Ben Adida},
  \bibinfo{person}{Thorsten Holz}, \bibinfo{person}{Jie Xiang},
  \bibinfo{person}{Steven Goldfeder}, \bibinfo{person}{Yinzhi Cao},
  \bibinfo{person}{Martin Plattner}, \bibinfo{person}{Xiaojun Qin},
  {et~al\mbox{.}}} \bibinfo{year}{2020}\natexlab{}.
\newblock \showarticletitle{An ever-evolving game: Evaluation of real-world
  attacks and defenses in ethereum ecosystem}. In
  \bibinfo{booktitle}{\emph{29th {USENIX} Security Symposium ({USENIX} Security
  20)}}. \bibinfo{pages}{2793--2810}.
\newblock


\bibitem[\protect\citeauthoryear{Zhou, Kumar, Bakshi, Mason, Miller, and
  Bailey}{Zhou et~al\mbox{.}}{2018b}]%
        {zhou2018erays}
\bibfield{author}{\bibinfo{person}{Yi Zhou}, \bibinfo{person}{Deepak Kumar},
  \bibinfo{person}{Surya Bakshi}, \bibinfo{person}{Joshua Mason},
  \bibinfo{person}{Andrew Miller}, {and} \bibinfo{person}{Michael Bailey}.}
  \bibinfo{year}{2018}\natexlab{b}.
\newblock \showarticletitle{Erays: reverse engineering ethereum's opaque smart
  contracts}. In \bibinfo{booktitle}{\emph{27th {USENIX} Security Symposium
  ({USENIX} Security 18)}}. \bibinfo{pages}{1371--1385}.
\newblock


\bibitem[\protect\citeauthoryear{Zou, Lo, Kochhar, Le, Xia, Feng, Chen, and
  Xu}{Zou et~al\mbox{.}}{2019}]%
        {zou2019smart}
\bibfield{author}{\bibinfo{person}{Weiqin Zou}, \bibinfo{person}{David Lo},
  \bibinfo{person}{Pavneet~Singh Kochhar}, \bibinfo{person}{Xuan-Bach~D Le},
  \bibinfo{person}{Xin Xia}, \bibinfo{person}{Yang Feng},
  \bibinfo{person}{Zhenyu Chen}, {and} \bibinfo{person}{Baowen Xu}.}
  \bibinfo{year}{2019}\natexlab{}.
\newblock \showarticletitle{Smart contract development: Challenges and
  opportunities}.
\newblock \bibinfo{journal}{\emph{IEEE Transactions on Software Engineering}}
  (\bibinfo{year}{2019}).
\newblock


\end{thebibliography}


% %%
% %% If your work has an appendix, this is the place to put it.
% \appendix




\end{document}